\documentclass[natbib]{emulateapj}
\usepackage{natbib}
\input epsf
\def\s2n{S^{\prime}/N}

\usepackage{amssymb,amsmath}
\def\bs{\boldsymbol}

\slugcomment{Submitted to ApJ, \today}
\shorttitle{Supernova Driving. I. The Origin of Molecular Cloud Turbulence}
\shortauthors{Padoan et al.}

\begin{document}
\title{Supernova Driving. I. The Origin of Molecular Cloud Turbulence}

\author{Paolo Padoan,}
\affil{ICREA \& Institut de Ci\`{e}ncies del Cosmos, Universitat de Barcelona, IEEC-UB, Mart\'{i} Franqu\`{e}s 1, E08028 Barcelona, Spain; ppadoan@icc.ub.edu}
\author{Liubin Pan}
\affil{Harvard-Smithsonian Center for Astrophysics,
60 Garden Street, Cambridge, MA 02138, USA;
lpan@cfa.harvard.edu}
\author{Troels Haugb{\o}lle,}
\affil{Centre for Star and Planet Formation, Niels Bohr Institute and Natural History Museum of Denmark, University of Copenhagen, {\O}ster Voldgade 5-7, DK-1350 Copenhagen K, Denmark; haugboel@nbi.ku.dk}
\author{{\AA}ke Nordlund,}
\affil{Centre for Star and Planet Formation, Niels Bohr Institute and Natural History Museum of Denmark, University of Copenhagen, {\O}ster Voldgade 5-7, DK-1350 Copenhagen K, Denmark; aake@nbi.ku.dk}

\begin{abstract}

Turbulence is ubiquitous in molecular clouds (MCs), but its origin is still unclear because MCs are usually assumed to live longer
than the turbulence dissipation time. Interstellar medium (ISM) turbulence is likely driven by SN explosions, but it has
never been demonstrated that SN explosions can establish and maintain a turbulent cascade inside MCs consistent with the observations.
In this work, we carry out a simulation of SN-driven turbulence in a volume of (250 pc)$^3$, specifically designed to test if SN driving alone
can be responsible for the observed turbulence inside MCs. We find that SN driving establishes a velocity scaling consistent with the usual
scaling laws of supersonic turbulence, suggesting that previous idealized simulations of MC turbulence, driven with a random, large-scale
volume force, were correctly adopted as appropriate models for MC turbulence, despite the artificial driving.
We also find that the same scaling laws extend to the interior of MCs, and
that the velocity-size relation of the MCs selected from our simulation
is consistent with that of MCs from the Outer-Galaxy Survey, the largest MC sample available. The mass-size relation and the
mass and size probability distributions also compare successfully with those of the Outer Galaxy Survey.
Finally, we show that MC turbulence is super-Alfv\'{e}nic with respect to both the mean and rms magnetic-field strength.
We conclude that MC structure and dynamics are the natural result of SN-driven turbulence.

\end{abstract}

\keywords{
ISM: kinematics and dynamics -- MHD -- stars: formation -- turbulence
}

\section{Introduction}

Understanding molecular cloud (MC) turbulence is key to understanding the star formation process, because supersonic
turbulence is ubiquitous in MCs and drives their fragmentation into stars. Supersonic turbulence has been studied
extensively in the context of star formation, and its statistical properties are at the core of recent models of the star formation rate
\citep{Krumholz+McKee05sfr,Padoan+Nordlund11sfr,Hennebelle+Chabrier11sfr,Federrath+Klessen12} and the stellar
initial mass function \citep{Padoan+Nordlund97imf,Padoan+Nordlund02imf,Hennebelle+Chabrier08,Hopkins12imf}.
Most numerical studies of supersonic turbulence have used a random large-scale force to drive the turbulence,
as customary in the turbulence literature. It remains to be shown that such an idealized external force is a good approximation of the actual large-scale
processes driving the interstellar-medium (ISM) turbulence. While the simulations use a volume force that penetrates the interior of the fluid, as long-range
forces do (e.g. gravity and magnetic fields), the real ISM driving forces may be surface forces, such as large-scale shocks from spiral arms or supernova (SN) bubbles.
The effect of different types of surface forces, or of a combination of volume and surface forces on the turbulence and, thus, on star formation, has not
been studied systematically.

The turbulence in MCs is also key to understanding their origin. The generation and maintenance of MC turbulence must be an integral
part of the cloud formation process, because most of the energy of an MC is in the form of turbulent kinetic energy. For example, the great majority
of small and intermediate-mass MCs are known to have rather large virial parameters \citep[see \S 10 and][]{Heyer+01,Heyer+09}, thus their turbulent
energy is large enough to form and disperse them in a few dynamical times. The same may be true also for the
most massive MCs, even if their virial parameter tends to be of order unity. The spatial and velocity structures of MCs follow
power laws that span all scales from the smallest to the most massive clouds, suggesting a universal origin of clouds and cloud turbulence.

In this work, we adopt the viewpoint that MC turbulence is just one component of the general ISM turbulence. The question of the origin of MC turbulence is
thus turned into the more general question of how the ISM turbulent energy is shared among its different gas phases. If we demonstrate that
the total large-scale turbulent energy of the ISM is dynamically consistent with the turbulent energy in its dense phase, no extra energy source specific to
MCs is needed.

It is generally accepted that SN explosions dominate the energy budget of star-forming galaxies at MC scales, although large-scale gravitational instabilities
in galactic disks \citep[e.g.][]{Elmegreen+2003,Bournaud+10} and gas compression in spiral density waves \citep[e.g.][]{Semenov+15} may also contribute
to the turbulence. The Kennicut-Schmidt star formation
law of disk galaxies, despite giving a large gas-consumption timescale of order 1 Gyr, corresponds to an energy input from SN explosions that exceeds the
turbulence dissipation rate of the ISM of those galaxies. The analysis of HI maps of nearby face-on galaxies also leads to the conclusion that SN feedback is
responsible for the observed HI line width, except in the disk outskirts \citep{Tamburro+09,Stilp+13_HI,Ianjamasimanana+15_HI}. Detailed modeling of the
disk vertical balance has also shown that SN feedback can maintain the ISM turbulence that determines the disk scale height, resulting in self-regulated star
formation \citep{Ostriker+10sfr,Ostriker+Shetty11,Faucher-Giguere+13}, a picture that may apply also to galaxies with high star formation rate at
redshift $z=1-3$ \citep{Lehnert+13}.

Prior to these studies, galactic-fountain simulations had already demonstrated that SN explosions can drive the observed ISM turbulence
\citep{deAvillez05,Joung+09}. However, these simulations barely resolved the formation of the most massive clouds, and could not resolve their
internal structure and dynamics in order to compare with observed MC properties. Furthermore, the evolution of individual SN explosions was
not resolved with high-enough spatial resolution to study their interaction with individual MCs.

Recent numerical works have studied the momentum injection by individual SN explosions into the ambient medium, assuming realistic ISM
density and temperature fluctuations \citep{Walch+Naab15SN,Martizzi+15SN,Iffrig+Hennebelle15SN,Kim+Ostriker15SN}. These studies are
useful for the derivation of feedback models to be implemented as sub-grid physics in galaxy formation simulations, but do not address the
problem of the generation and maintenance of MC turbulence by SN explosions.

In this work, we focus on a smaller-scale region than in galactic-fountain simulations, and use high-enough numerical resolution to model the MCs, their internal
structure and dynamics, and the evolution of individual SN bubbles with sub-pc resolution (see \S 2). We also select clouds from the simulation,
as connected regions above a threshold density (\S 3), to study their turbulence and to compare with observed MCs. First we study statistical properties
over the whole computational volume, such as total energies (\S 4), power spectra (\S 5) and velocity structure (\S 6.1). Then we present properties of
individual clouds selected from the simulation, such as the velocity structure (\S 6.2), the virial parameter (\S 7), the cloud lifetime (\S 8) and the
magnetic field strength (\S 9). Finally, we compare the properties of the clouds selected from the simulation with those of MCs from the Five College 
Radio Astronomy Observatory (FCRAO) Outer Galaxy Survey (\S 10).

\section{Numerical Setup}

The simulation is carried out with the Ramses AMR code \citep{Teyssier02}. We refer to \citet{Padoan+14accr} for a brief description of our version of
Ramses. What is new here, relative to the simulation in \citet{Padoan+14accr}, is the use of the full energy equation (instead of assuming an isothermal
equation of state), and the inclusion of SN explosions and tracer particles, besides the much larger physical size of the computational volume.

We simulate a cubic region of size $L_{\rm box}=250$ pc (large enough to contain a few turbulence correlation lengths, while small enough to allow sub-pc resolution),
with a mean density of 5 cm$^{-3}$ (corresponding to a column density of 30 M$_{\odot}$pc$^{-2}$ and a total mass of $1.9\times 10^6$ M$_{\odot}$)
and a Type II SN rate of 6.25 Myr$^{-1}$ (or a galactic rate of 100 Myr$^{-1}$kpc$^{-2}$, if all SN explosions occurred within the vertical extent of our box).
We do not consider Type Ia SNe, because of their lower rate and higher scale height, and because we distribute SN explosions randomly, so our SN rate
could also be interpreted as the sum of the Type II and Type Ia rates.

These rate and column density values are consistent with the Kennicutt-Schmidt relation, and our computational volume may be viewed as a dense section of
a spiral arm. For example, the total column density in the Perseus arm of the Milky Way is 23 M$_{\odot}$pc$^{-2}$ \citep{Heyer+Terebey98}. However, we do not
include a galactic gravitational field, and adopt periodic boundary conditions
in all directions, so vertical stratification and outflows of hot gas are neglected. We have chosen this idealized setup because one of the motivations of this work is
to relate the statistical properties of SN-driven turbulence to previous studies of randomly-driven, supersonic, isothermal turbulence that
were carried out on periodic boxes without stratification. Furthermore, we relate the velocity scaling to theoretical predictions that also
neglect complications such as gravity and stratification.\footnote{Star-formation simulations with gravity have shown that, under reasonable conditions
(e.g. MCs not collapsing as a whole), gravity does not affect the velocity scaling \citep{Federrath+Klessen13}}

Besides the pdV work, and the thermal energy introduced to model SN explosions, our energy equation adopts uniform photoelectric heating as in
\citet{Wolfire+95}, with efficiency $\epsilon=0.05$ and the FUV radiation field of \citet{Habing68} with coefficient $G_0=0.6$, chosen to obtain temperature
distributions consistent with those from the comprehensive simulations by \citet{Walch+15}.
Because the code conserves total energy, kinetic and magnetic
dissipations are included self-consistently as energy sources (this dissipation is purely numerical, as we do not include viscosity or resistivity explicitly).
We use a tabulated optically thin cooling function constructed from the extensive compilation by \citet{Gnedin+Hollon12}, based on 75 million runs of the Cloudy
code \citep{Ferland+98} to sample a large range of conditions, and from which the results have been made publicly available as a Fortran
code with accompanying database. All relevant atomic transitions are included in the Cloudy runs. Although available in Cloudy, molecular cooling is not included 
because the runs are restricted to a single computational zone, with a negligible column density, to enforce the optically thin case. Above a temperature of 100 K, 
atomic cooling is dominant up to densities of 10$^6$ cm$^{-3}$ \citep[e.g.][]{Neufeld+95}. At lower temperature and high densities, molecular cooling should be 
included. However, molecular cooling and cosmic-ray heating are neglected, and their thermal balance in very dense gas is emulated by clamping the resulting 
drop in temperature at 10 K. As pointed out by \citet{Gnedin+Hollon12}, including the balance of 
molecular cooling and cosmic-ray heating in such a treatment does not make much physical sense, since the balance of these processes at high densities 
crucially depends on radiative transfer effects; in particular the absorption of UV radiation by small-scale high-density cloud structures. 
In the absence of radiative transfer, and given the optically thin assumption, we approximate the 
UV shielding in MCs by tapering off the photoelectric heating exponentially above
a number density of 200 cm$^{-3}$ \citep[assuming a characteristic size of 1 pc for MC structures at our critical density, corresponding to a critical visual extinction
of 0.3 mag  --][]{Franco+Cox86}.

\begin{figure}[t]
\includegraphics[width=\columnwidth]{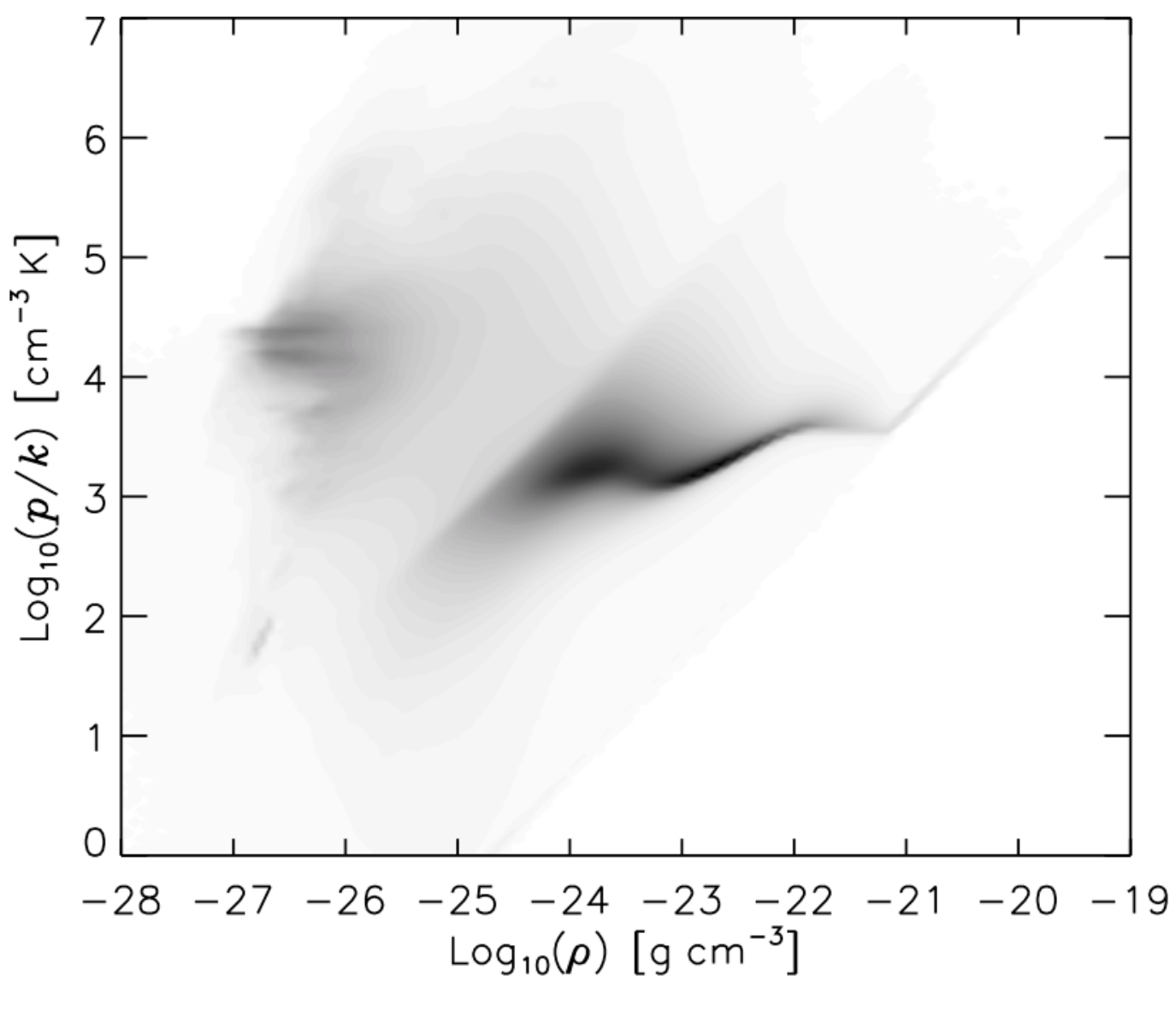}
\caption[]{Phase diagram of gas pressure versus density based on 11 snapshots covering uniformly the time period when gravity is included in the simulations,
$t=45$ to 56 Myr. The gray scale represents the square root of the volume fraction at a given pressure and density.}
\label{phase}
\end{figure}

Figure \ref{phase} shows the phase diagram of gas pressure versus density sampled over the last 11 Myr of the simulation.  The horizontal feature
around densities of a few $10^{-22}$~g\,cm$^{-3}$ is a consequence of the approximation of self-shielding.  In a real MC, or a model where the absorption
of UV-radiation by the filamentary structure of dense gas and dust is taken into account more realistically, the transition to opaqueness would take
place at different densities at different locations, and similar local phase diagram features would be washed out.  Comparing our current phase
diagram with the ones in \citet{Walch+15}, we actually find the largest discrepancy not to be near that horizontal feature, but rather
at higher densities where, for some reason, their balance of heating and cooling results in a temperature of about 30 K, instead of our assumed value of 10 K.

The simulation is started with zero velocity, a uniform density $n_{\rm H,0}=5$ cm$^{-3}$, a uniform magnetic field $B_0=4.6$ $\mu$G and a
uniform temperature $T_0=10^4$ K.
The first few SN explosions rapidly bring the mean thermal, magnetic and kinetic energy to approximately steady-state values, with the magnetic field
amplified to an rms value of 7.2 $\mu$G. The value of the mean magnetic field is chosen to achieve near equipartition with the kinetic energy at large
scales, as shown in \S 4. It also yields an average value of $|\bs B|$ of 6.0 $\mu$G, consistent with the value of $6.0 \pm 1.8$ $\mu$G derived
from the `Millennium Arecibo 21-cm Absorption-Line Survey' by \citet{Heiles+Troland05_B}.

The simulation is run for 56 Myr (with a total of 359 SN explosions), initially without tracer particles and self-gravity. At $t=33$ Myr we include 150 million
passively-advected tracer particles following the mass distribution in the computational volume (each particle represents a fluid element of 0.013 M$_{\odot}$).
Tracer particles are advected with the same symplectic Kick-Drift-Kick scheme used for dark matter particles in Ramses, but the kicks are ignored
-- they are passive -- and velocities in the drift are instead sampled using CIC interpolation of fluid velocities.

At $t=45$ Myr we include self-gravity. Several clumps of order 1,000 M$_{\odot}$ start to collapse. Larger resolution
and sink particles are required at that point, which will be the topic of a followup work. For the purpose of this work, we stop the simulation after approximately
11 Myr of evolution with self-gravity. This is also motivated by the need to trace the exact location of SN explosions when clouds have been forming massive
stars for a time of order 10 Myr, as explained in the next section.

\begin{figure*}[ht]
\includegraphics[width=\textwidth]{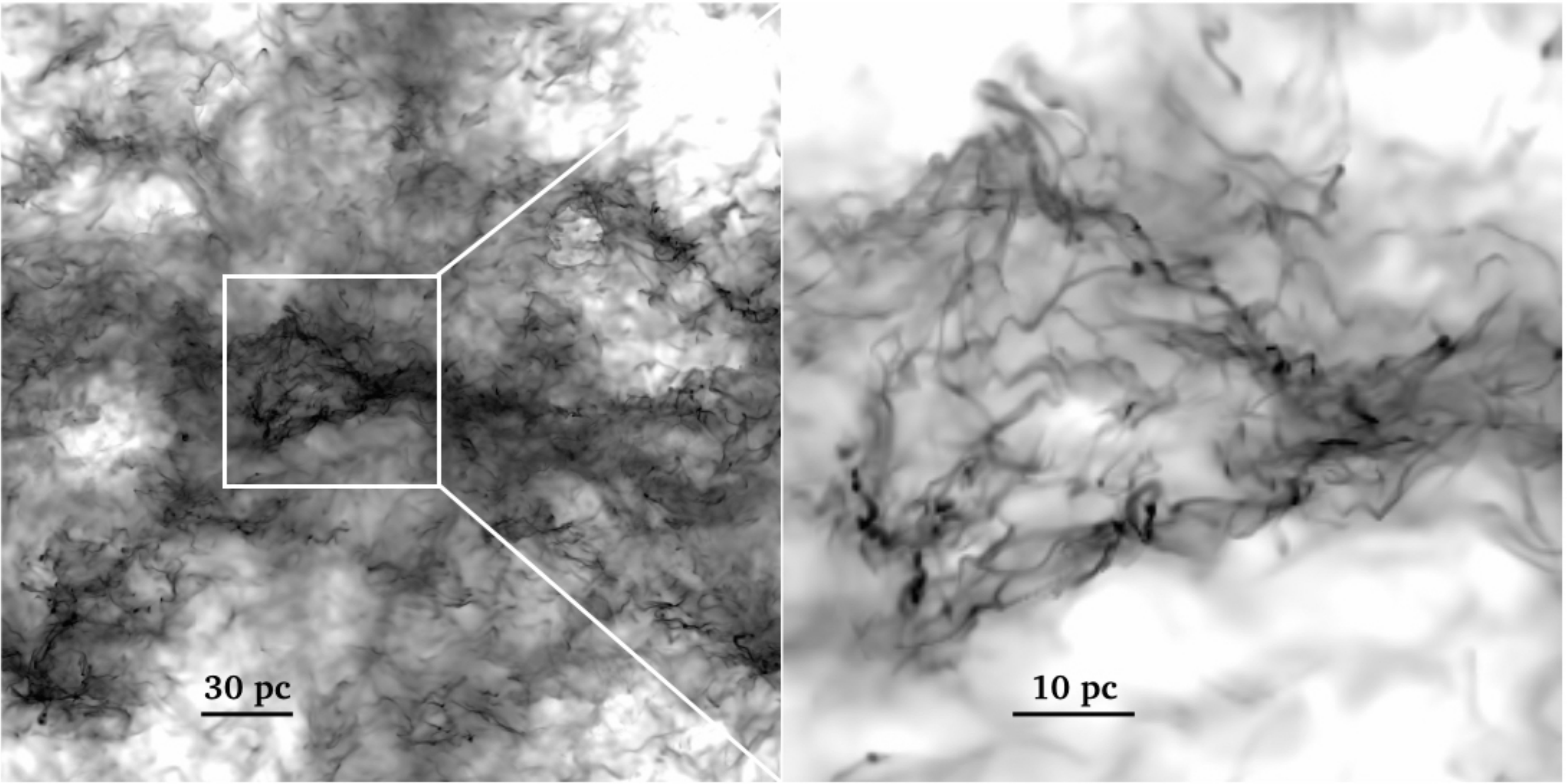}
\caption[]{Left Panel: Logarithm of projected density along the x-axis of the simulation volume. The mean magnetic field direction is horizontal on the image.
The column density value is larger in darker regions, with black set to a maximum value of $5\times10^{22}$ cm$^{-2}$, and white to a minimum value of
$6\times10^{20}$ cm$^{-2}$. The actual values of column density span a much larger range, and these limits have been chosen to optimize the contrast of
the density structure. The image includes the whole computational volume, that is a size of 250 pc. Right Panel: Same as in the left panel, but for a region
four times smaller (62.5 pc) shown by the white squared in the left panel. The color-table limit values have been increased by a factor of 4 and 5,
$2\times10^{23}$ cm$^{-2}$ (black) and $3\times10^{21}$ cm$^{-2}$ (white), to match the larger mean density in this region.}
\label{fig_zoom}
\end{figure*}

To enforce a sub-pc resolution of the evolution of SN bubbles and of their interaction with the ISM, we adopt AMR criteria based on density, density gradients
and pressure gradients. Although our root grid contains only 128$^3$ cells, our AMR criteria result in rather large volume filling factors of high-resolution cells:
75\% of our computational domain is covered at a resolution equivalent to 256$^3$, 22\% at a resolution of 512$^3$, and 2\% at 1024$^3$. Because the volume
filling factor of the clouds selected in this work is approximately 0.5\%, our clouds are all resolved at the maximum resolution.

The left panel of Figure \ref{fig_zoom} shows the logarithm of the projected density of the whole computational volume, at the final time of the simulation. The structure
is highly filamentary and appears to be self-similar. The right panel of Figure \ref{fig_zoom} shows a sub-region magnified by a factor of four; the same type of filamentary
structure is seen at this smaller scale. This structure is very similar to that previously found in supersonic simulations of isothermal turbulence, and consistent with
the appearance of nearby GMCs mapped with the Herschel satellite.

Apart from SN explosions, the simulation neglects any other energy source, such as winds and radiation feedback from massive stars, or external forces, such as
spiral arm shocks and fluctuations of the large-scale gravitational potential, that may affect MC turbulence in real galaxies. This choice allows us to test if
SN turbulence {\it alone} can explain the origin and maintenance of GMC turbulence, while also providing a significant reduction of the computational cost.
We also neglect the chemical evolution of the ISM. Although we do not expect that the dynamics would be strongly affected by a more precise computation
of cooling and heating based on dynamically evolved chemical abundances, the selection of MCs and their comparison with observations would certainly
benefit from a dynamical computation of the H2 and CO abundances \citep{Glover+12,Walch+14}. The neglect of chemistry in this work is solely motivated
by considerations of computational cost.

\subsection{SN Explosions}

Individual SN explosions are implemented with an instantaneous addition of 10$^{51}$ erg of thermal energy and 15 M$_{\odot}$ of gas,
distributed according to an exponential profile on a spherical region of radius $r_{\rm SN}=3 dx=0.73$ pc, where $dx$ is our smallest cell size,
$dx=L_{\rm box}/1024=0.24$ pc. \citet{Kim+Ostriker15SN} have derived a useful condition for numerical convergence of the evolution of SN remnants,
which states that both the grid size and the initial SN radius must be smaller than one third of the shell-formation radius, $dx, r_{\rm SN} < r_{\rm sf}/3$.
Because in our case $r_{\rm SN} > dx$, and given the expression for $r_{\rm sf}$ in \citet{Kim+Ostriker15SN}, the condition for our simulation is
$r_{\rm SN} < 10 \, {\rm pc} \, n_0^{-0.46}$, where $n_0$ is the ambient density. Given our value of $r_{\rm SN}=0.73$ pc, the condition becomes
$n_0 \lesssim 300$ cm$^{-3}$. The volume filling factor of gas at density above that value varies in the approximate range between 0.0003 and 0.0009,
with the largest value reached only at the end of the simulation, while the total number of SN explosions in the simulation is 359. Because the locations
of SN explosions are randomly distributed, there is only a small probability that one or more SN explosions violate the condition for numerical convergence.

SN explosions are generated at random positions and times, neglecting the possibility of spatial and temporal clustering.
In the galactic fountain simulations by \citet{deAvillez05} and \citet{Joung+09} it was assumed
that 60\% of the SN explosions occurred in clusters, and only the remaining 40\% at random locations.
\citet{Joung+09} assumed that the clustered SN population is distributed in clusters with up to 40 stars more massive than
8 M$_{\odot}$, $N_{\rm SN}= 40$, with a probability of cluster size following a power law, $\sim N_{\rm SN}^{-2}$.
All massive stars from a given cluster were assumed to explode at the same location, though at different times distributed
in an interval of 40 Myr, the approximate lifetimes of the least massive stars to explode.

However, assuming that clusters have a typical velocity dispersion of the cold gas of order 10 km/s, and considering the average value of
$N_{\rm SN}= 10$, the typical cluster would contribute on average 1 SN explosion every 4 Myr, with a separation of approximately 40 pc
between consecutive SN explosions, due to the cluster motion not accounted for by \citet{Joung+09}. The 10 explosions would cover a
distance of 400 pc over 40 Myr. Thus, even assuming the reasonable cluster statistics of \citet{Joung+09}, the explosions would not appear
to be clustered. A realistic prediction of spatial and temporal clustering of SN explosions requires that the formation and kinematics of individual
stars is numerically resolved through the adoption of very high spatial resolution and sink particles. In the absence of that, a random
SN distribution is a reasonable assumption.

Another potentially important issue is the correlation between the location of SN explosions and the position of their parent clouds.
Recent simulations have shown that the large-scale structure of the ISM is sensitive to the amount of correlation between SN positions
and density peaks \citep{Gatto+15,Walch+14}. \citet{Walch+14} have concluded that the random distribution case or the case of
clustering in space and time, but not correlated with density peaks, are favored by the observations. On the contrary, in \citet{Dobbs+Pringle13}
and \citet{Dobbs15} all SN explosions are assumed to occur inside MCs. As an MC (or any density peak above a certain threshold) is created,
SN feedback is turned on in its interior. This unrealistic SN feedback does not allow to address the question of the origin of MC turbulence, as
this is driven directly by SNe by design.

\citet{Iffrig+Hennebelle15SN} have addressed this issue by focusing on the effect of a single SN on a single MC of $10^4$ M$_{\odot}$, showing
that the effect of a SN explosion on a MC is much stronger if the explosion occurs inside the cloud than outside of it. Their results may be affected
by their specific choice of the ambient density, 1.2, 20 and 700 cm$^{-3}$, for the three SN positions they tested. The probability of high density is
certainly increased inside a MC than outside of it. However, the probability of a SN explosion at high density must be quite small, because the
volume filling fraction of dense filaments and clumps is low even within the volume enclosing a MC. Furthermore, HII regions and winds would
probably prevent SN explosions in dense gas in general. Nevertheless, it seems reasonable that SN remnants expanding from within a MC would
be more effective at bringing material above the escape velocity than SN remnants pushing on a MC from the outside.

With a Salpeter IMF \citep{Salpeter55}, the mean stellar mass between 8 and 100 M$_{\odot}$ is approximately
19 M$_{\odot}$, corresponding to a stellar lifetime of approximately 9 Myr. Assuming that it takes at least 1 Myr
to initiate star formation in a young cloud, and to accrete the stellar mass, the SN feedback would thus be important
in a MC after a characteristic time of 10
Myr (assuming that massive stars can remain in the general region of their
parent cloud for a parent-cloud crossing time).
We will show in \S 8 that our most massive clouds of order
$10^5$ M$_{\odot}$ selected towards the end of the simulation were formed approximately 20 Myr earlier
($t_{\rm form} \approx 20$ Myr in Figure \ref{t_form_death}). For these clouds, thus, the SN feedback from locally-formed
massive stars should start to play an important role, depending on the delay between cloud formation and the formation of
the first massive stars.

The same is true for all clouds with lifetime of order of 10 Myr or larger. Because we find in \S 8 that the cloud lifetime is
on average four times longer than the cloud dynamical time, and using the expression (\ref{tdyn_Mcl}) for the dynamical time derived
in \S 10.4, the condition is that the dynamical time
is at least 5 Myr or longer, or, equivalently, that the cloud mass is larger than
500 M$_{\odot}$. As the formation of massive stars most likely requires clouds more massive than 500 M$_{\odot}$ as well,
we conclude that SN feedback from locally-formed massive stars should generally play a role for most clouds forming massive stars.
As mentioned in \S 2, this is in fact the main reason why the simulation
was stopped approximately 11 Myr after introducing self-gravity, as neglecting the locally-formed massive stars beyond that
time would be unrealistic. The effect of the correlation of SN explosions with their parent clouds will be considered in a future
work where the formation and kinematics of individual massive stars will be numerically resolved in order to model self-consistently
the precise time and location of SN explosions.

\begin{figure*}[t]
\centering
\includegraphics[width=18cm]{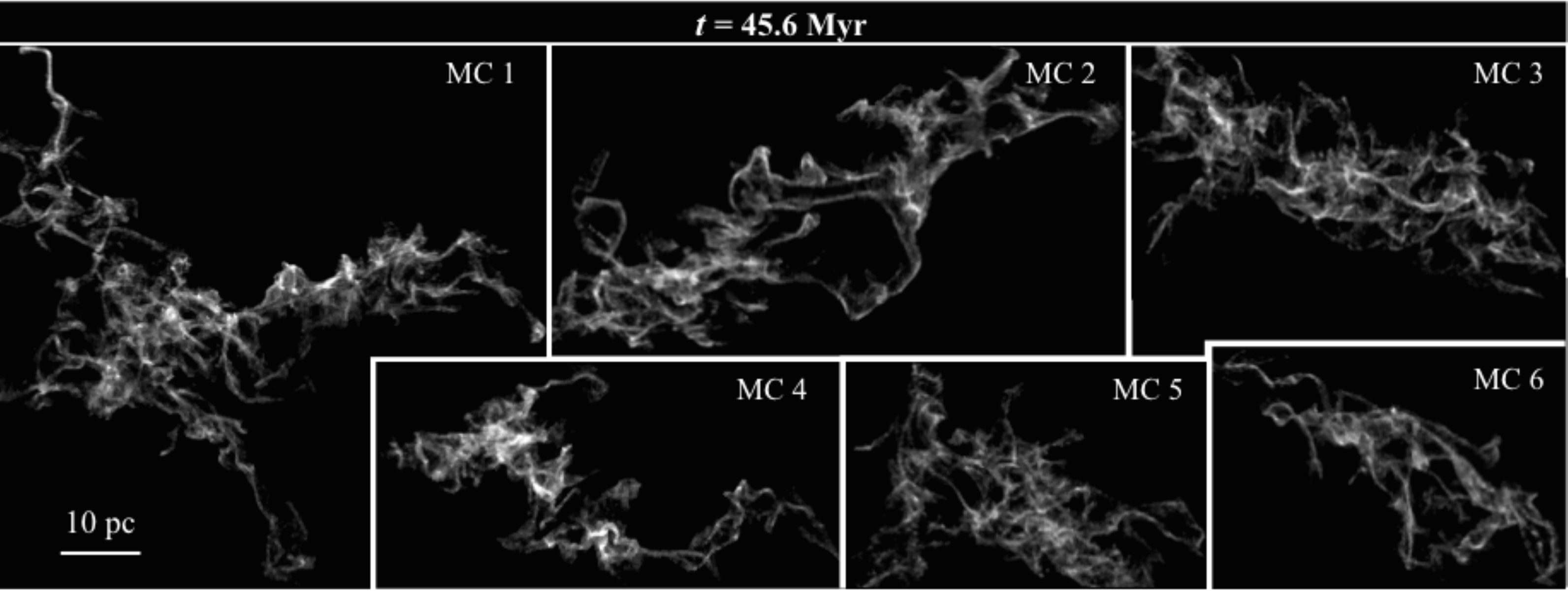}
\includegraphics[width=18cm]{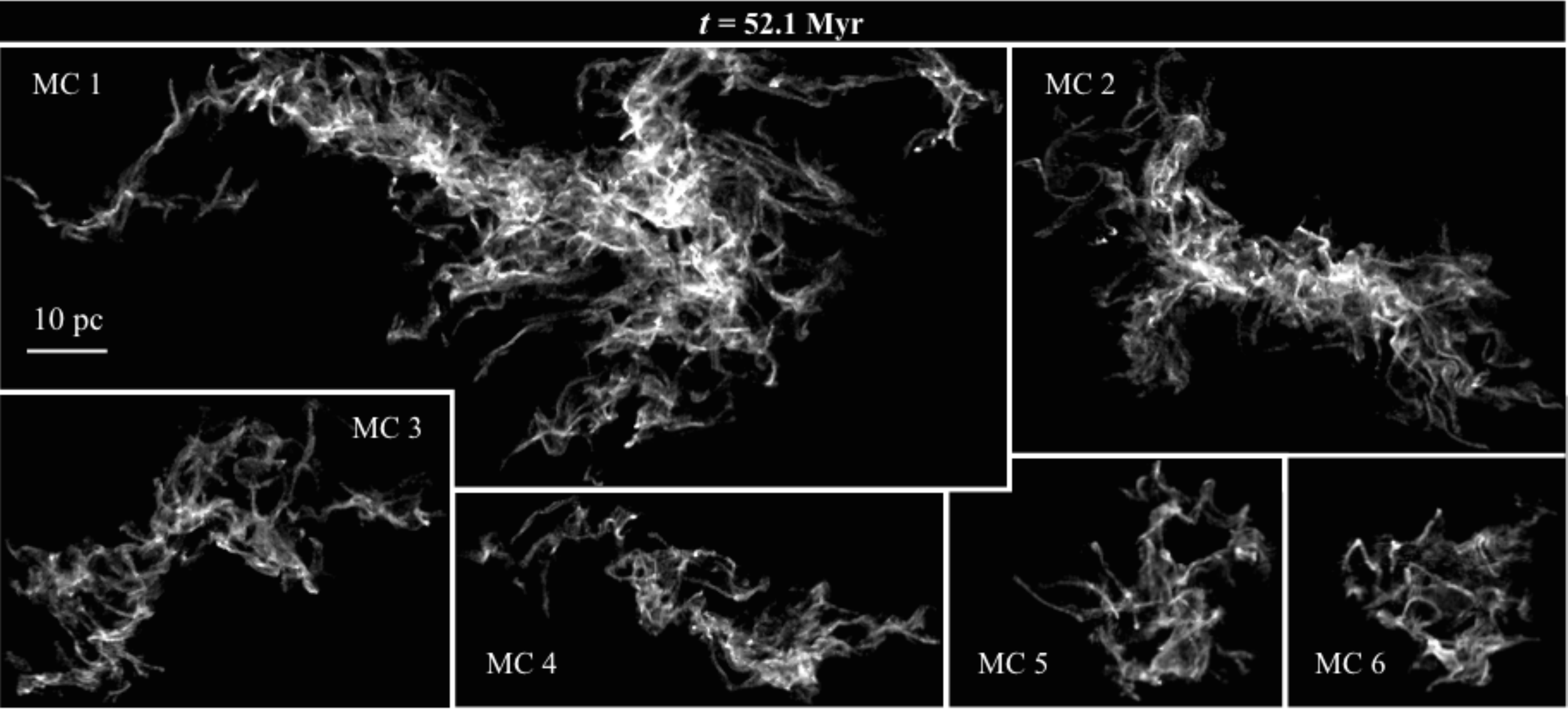}
\includegraphics[width=18cm]{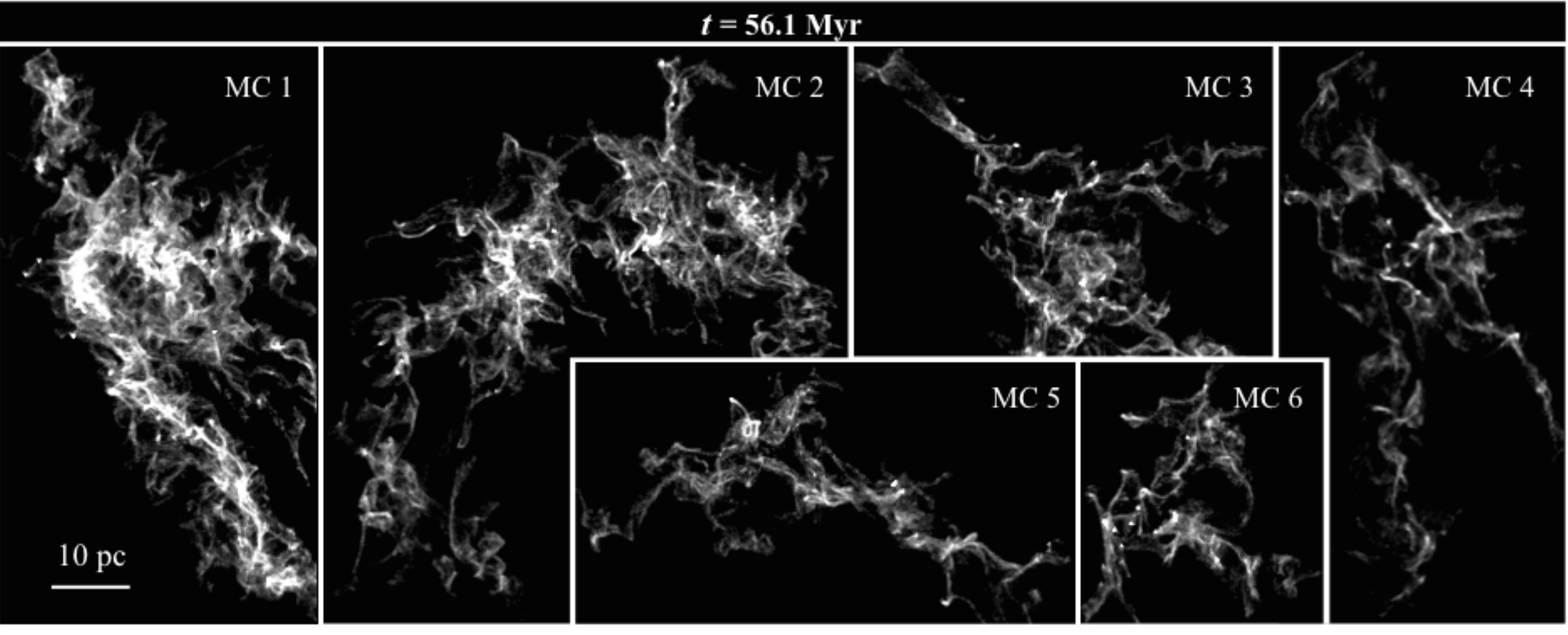}

\caption[]{Square root of the projected density of 18 clouds extracted with $n_{\rm H,min}=100$ cm$^{-3}$ and $512^3$ resolution from
three different snapshots near the beginning, the middle, and the end of the time interval with self-gravity. The clouds are numbered in
order of decreasing mass. To save space in the panels, the most massive cloud from each snapshot (cloud 0) is not shown. The grey
color table covers a range of column densities from 0 M$_{\odot}$pc$^{-2}$ (black) to 200 M$_{\odot}$pc$^{-2}$ (white). The column
density of gas below the threshold density of 100 cm$^{-3}$ is not included. Cloud 6 of the top panel and cloud 4 of the bottom one
are the same as in Figures \ref{vel_structure_gmc6} and \ref{vel_structure_gmc4} respectively. The complex filamentary structure
is very similar to that observed in MCs.}
\label{clouds}
\end{figure*}

\section{MC Selection}

The main question addressed by this work, whether SN explosions can drive and maintain the observed MC turbulence, can be
partly answered independently of the definition of MCs, by computing the velocity structure functions for the dense and cold gas
throughout the computational volume. As long as most of such gas is in clouds, its global structure functions should
be equivalent to the average of the structure functions of MCs. Nevertheless, it is important to characterize the turbulence within
individual clouds and to also compute other cloud properties that can be compared with observations.

Because chemistry is not included in this work, we can only define MCs as cold over-densities in the SN-driven ISM turbulence.
A detailed comparison with the observations is beyond the scope of this work. It would require synthetic observations, hence
molecular abundances from chemical-network calculations. In the absence of synthetic observations, we prefer to avoid the
selection of MCs in position-position-velocity (PPV) space, and instead define MCs in 3D space (PPP) in the simplest possible way,
as connected regions above a single threshold gas density, $n_{\rm H,min}$. In order to test if our results depend on spatial resolution or
threshold density, we extract clouds using three different mesh sizes, $2dx$, $4dx$ and $8dx$ (we create uniform grids of $512^3$,
$256^3$ and $128^3$ cells, respectively), and four different threshold densities, $n_{\rm H,min}=100$, 200, 400 and 800 cm$^{-3}$,
generating 12 cloud catalogs. Examples of clouds from one such catalog ($n_{\rm H,min}=100$ cm$^{-3}$ and $512^3$ resolution) are
shown in Figure \ref{clouds}. The images show the projected density of 18 clouds, the 2nd to the 7th most massive ones in each of three
snapshots.

The analysis of MC properties derived with the full three-dimensional information, such as the results on MC velocity scaling,
the discussion on the virial parameter and cloud structure, the evaluation of cloud lifetimes and the study of the cloud magnetic field,
is based on clouds extracted with $n_{\rm H,min}=100$ cm$^{-3}$ and $128^3$ resolution. For this analysis, the resolution only affects the definition
of cloud boundaries, as all results are derived from the position, velocity, density and magnetic field of the tracer particles, thus taking
advantage of the highest resolution.

When we compare with observational data, deriving projected quantities such as surface density,
equivalent radius and line-of-sight velocity dispersion, we verify the results on all 12 catalogs. For the mass and size distributions,
we also report quantitatively on the dependence of the slope of their power-law tails on cloud extraction density and resolution. However,
all the plots we show in the comparison with the observations are based on the highest-resolution catalog ($512^3$ cells, or 0.49 pc) and
on the threshold density that best matches the observed mass-size relation, $n_{\rm H,min}=200$ cm$^{-3}$.
Furthermore, velocity dispersions are based on tracer particles and so take advantage of the highest available resolution, $dx=0.24$ pc.\footnote{The
number density of tracer particles is very large in dense gas, where the mesh is refined to the highest resolution.} This resolution matches well the
highest one in the observational survey we consider \citep{Heyer+01}. After selecting clouds with circular velocity $v_{c}< 20$ km s$^{-1}$ and
mass $M_{\rm cl} > 100$ M$_{\odot}$, we are left with 3,228 observed clouds with measured distances corresponding to a range of spatial resolutions
of $0.24-3.0$ pc.

We select from the simulations only clouds more massive than 100 M$_{\odot}$ to guarantee that, even at the lowest value of $n_{\rm H,min}=100$
cm$^{-3}$, the smallest clouds contain more than 1,000 computational cells (our largest cloud of nearly $3\times 10^5$ M$_{\odot}$ contains more than
3$\times 10^6$ cells). Because we use 150 million tracer particles, our smallest and lowest-threshold-density clouds contain a minimum of
approximately 7,000 particles, while our most massive cloud of $3\times 10^5$ M$_{\odot}$ contains more than 20 million particles.
With this minimum mass, each simulation snapshot yields over 200 clouds. In the comparison with the observations (see \S 10) we use 7 snapshots from
approximately the last 6 Myr of the simulation, to include the effect of self-gravity, resulting in sample sizes ranging from 595
clouds in the smallest catalog ($128^3$ resolution and $n_{\rm H,min}=800$ cm$^{-3}$), to 1,615 in the largest one ($512^3$ resolution and
$n_{\rm H,min}=100$ cm$^{-3}$). The plots we show in \S 10, based on $512^3$ resolution and $n_{\rm H,min}=200$ cm$^{-3}$, use
measurements from 1,547 clouds. The catalog sizes could be considered three times larger, as projected quantities are computed in the three
orthogonal directions. However, the clouds (and cloud masses) would be the same, so the three samples would not be completely independent.
Thus, we compare with the observations using only one of the directions, after verifying that the results are independent of
direction.

\begin{figure}[t]
\includegraphics[width=\columnwidth]{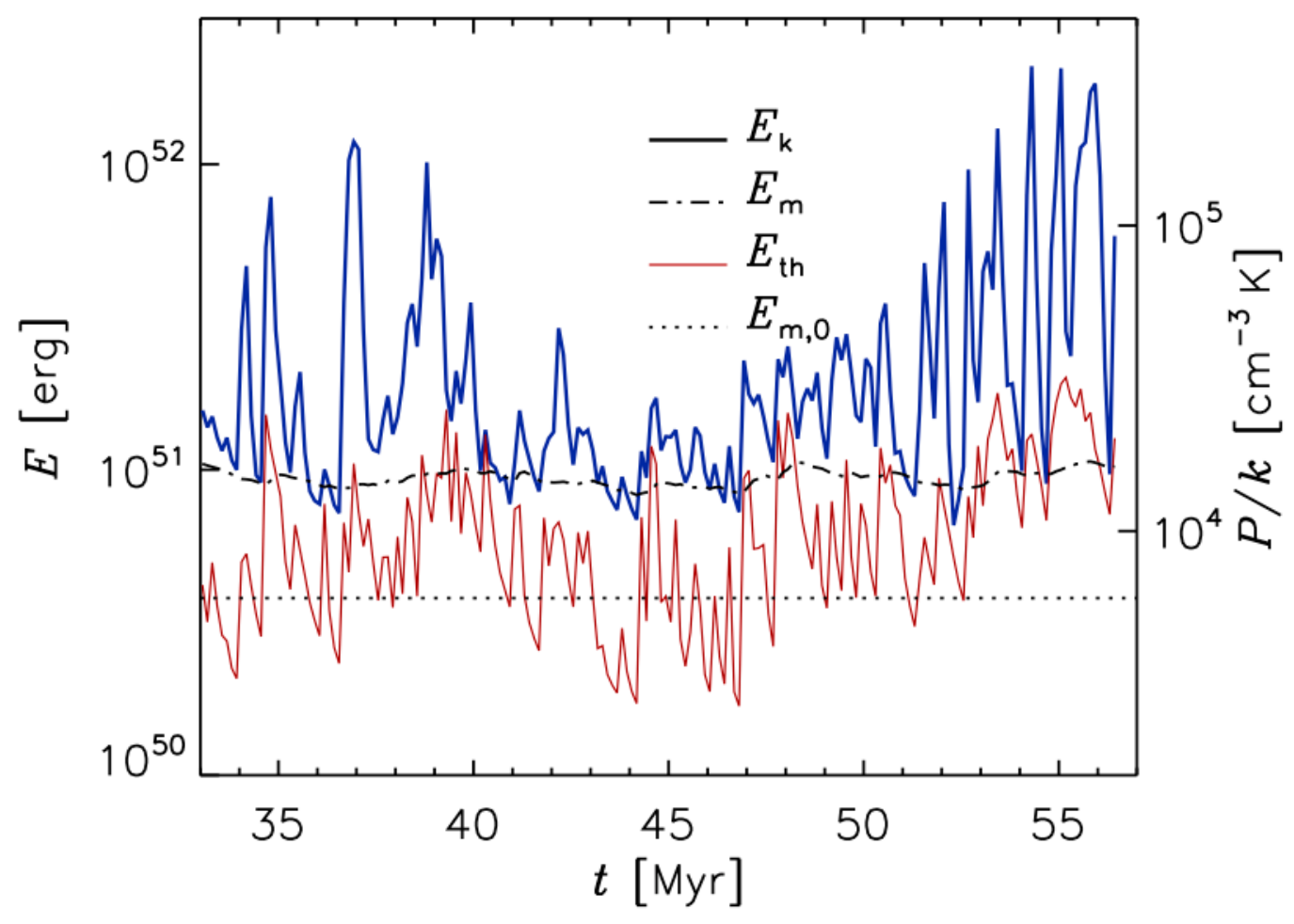}
\caption[]{Total kinetic, magnetic and thermal energies in the simulation versus time, during the period analyzed in this work, between
$t=33.06$ Myr and $t=56.43$ Myr. The energies are measured in 188 snapshots (8 snapshots per Myr), integrating over the whole
simulation volume, independent of density. The horizontal dotted line shows the initial magnetic energy (also the magnetic energy
corresponding to the mean magnetic field in the simulation, which is a conserved quantity). There is near equipartition between the
lowest values of kinetic energy, the mean value of the magnetic energy, and the highest values of the thermal energy.}
\label{fig_energies1}
\end{figure}

Despite the mass range of over three orders of magnitude of our clouds, we refer to all of them as MCs, instead of following the common
nomenclature that classifies clouds as cores, clumps, molecular clouds, giant molecular clouds, and giant molecular cloud complexes, in order
of increasing mass. As we view all clouds as cold density enhancements of the ISM turbulence, and because of the scale-free nature of
the turbulence, that nomenclature is not useful for this work.

\section{Total ISM Energies}

We analyze our simulation in a time interval starting after the turbulence has been fully developed until the end of the simulation,
between $t=33.06$ Myr and $t=56.43$ Myr. This is also the time interval for which we have tracer particles in the simulation. The
evolution of the total kinetic, magnetic and thermal energies in that time interval is shown in Figure \ref{fig_energies1}. The three
energies are all around 10$^{51}$ erg. Both the thermal ($E_{\rm th}$) and kinetic ($E_{\rm k}$) energies show strong oscillations,
while the magnetic energy ($E_{\rm m}$) is almost constant with time. The kinetic energy oscillates mostly above 10$^{51}$ erg,
and, interestingly, at its valleys, the magnetic energy is almost in equipartition with $E_{\rm k}$. The thermal energy is the smallest,
mostly below both the magnetic and kinetic energies.

\begin{figure}[t]
\includegraphics[width=\columnwidth]{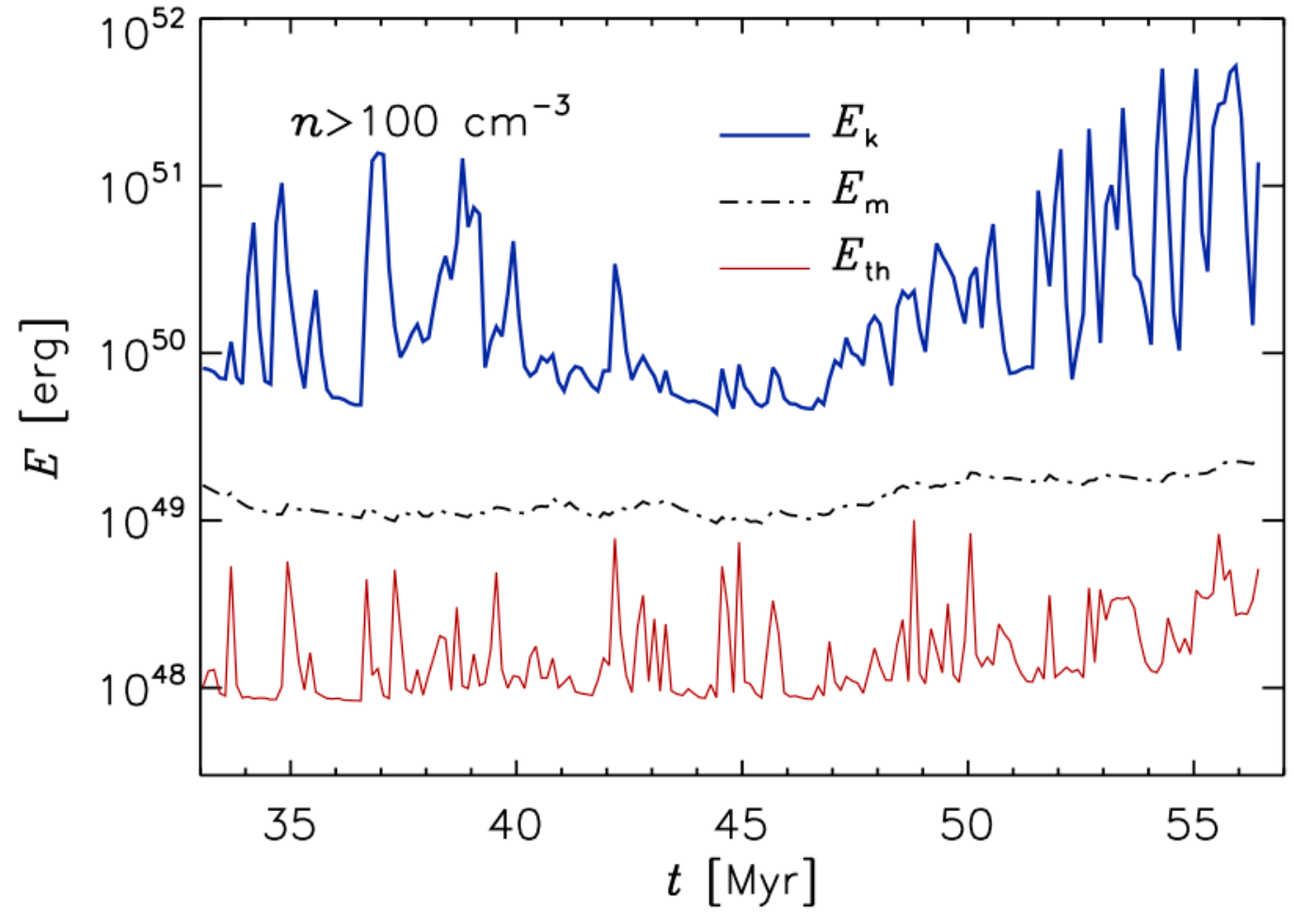}
\caption[]{Same as Figure \ref{fig_energies1}, but with the energies computed only for gas with density above 100 cm$^{-3}$, the same threshold density
adopted for the cloud selection. There is a clear energy separation in the dense gas, with the lowest kinetic energy values being approximately an order
of magnitude larger than the mean magnetic energy, and two orders of magnitude larger than the lowest values of the thermal energy. Thus, the turbulence
in the dense gas is both supersonic and super-Alfv\'{e}nic.}
\label{fig_energies2}
\end{figure}

Given the strong temporal oscillations in kinetic energy, it may seem surprising that the magnetic energy does not experience
significant fluctuations at all. The reason is that the kinetic energy peaks correspond to the energy injected by SN explosions
mainly in the form of expansions and shocks. While the magnetic field can be strongly compressed and amplified by the
shocks, the volume filling factor of the postshock gas is tiny, and thus the total magnetic energy of the computational volume
is barely changed, as illustrated by the following estimate.

Consider a SN remnant of radius $R_{\rm SN}$ and a preshock magnetic field $B_1$. Compression by the SN shock would
amplify the magnetic field strength by a factor equal to the density-jump factor, $\Gamma$, of the shock. In the adiabatic phase,
$\Gamma = 4$ and in the radiative phase, $\Gamma \simeq {\cal M}_{\rm A}$ with ${\cal M}_{\rm A}=v_{\rm sh}/(B_1/\sqrt{4\pi\rho_1})$
the Alfv\'{e}nic Mach number of the shock. As the remnant expands, solenoidal turbulent motions
also develop (see \S 5.1), which can amplify the magnetic field as well. The timescale for the
magnetic energy amplification by solenoidal motions is roughly the turnover time of the
largest eddies in the postshock region \citep[e.g.][]{Federrath+11PRL}.  We assume that
the turbulent velocity in the postshock region is the postshock velocity, $v_{\rm sh}/\Gamma$,
and the large eddy size is the thickness of the postshock region of the remnant, which is $\simeq R_{\rm SN}/(3 \Gamma)$
(due to the compression by the shock). The large eddy turnover time is then estimated to be $\simeq R_{\rm SN}/(3 v_{\rm sh})$.
Considering the age of the remnant is $\simeq \frac{2}{5} R_{\rm SN}/v_{\rm sh}$ and $\simeq \frac{2}{7} R_{\rm SN}/v_{\rm sh}$
in the adiabatic and radiative phases, respectively, solenoidal motions may amplify the magnetic energy
by one e-fold or so, meaning an amplification factor of  $A \simeq 3$ for the magnetic energy.
Including both the effects of shocks and solenoidal motions, the magnetic energy density in
the postshock region would be $ \simeq \frac{1}{8 \pi} A \Gamma^2 B_1^2$.

The above description of the magnetic field amplification only applies to the compressed layer behind the shock.
Due to the compression by a factor of $\Gamma$, the width of the compressed layer is given by $R_{\rm SN}/(3\Gamma)$,
so the total magnetic energy within the compressed layer is estimated to be $\frac{1}{6} A\Gamma B_1^2R_{\rm SN}^3$.
On the other hand, the magnetic energy in the hot cavity interior to the compressed layer is small due to the expansion and
can be neglected. Considering that the magnetic strength is still $\simeq B_1$ in regions not reached by the SN remnant,
the total magnetic energy in the simulation box is
$\simeq \frac{1}{8\pi}B_1^2  (4\pi A \Gamma  N_{\rm SN} R_{\rm SN}^3/3 + V_{\rm box} - 4 \pi N_{\rm SN}R_{\rm SN}^3/3)$,
where $N_{\rm SN}$ is the number of SNe exploded around the same time and $V_{\rm box}$ is the volume of the
simulation box. Because the highest kinetic-energy peaks are of the order $10^{52}$ erg, $N_{\rm SN}$ may be as
large as $\simeq 10$. If we define a filling factor of each SN remnant as $f =  4 \pi R_{\rm SN}^3/(3V_{\rm box})$,
the total magnetic energy can be written as $\simeq  \frac{1}{8\pi}  [N_{\rm SN}(A \Gamma -1) f +1] B_1^2 V_{\rm box}$,
meaning that $E_{\rm m}$ is amplified by a factor of $N_{\rm SN}(A \Gamma -1) f +1$.
In the Sedov phase,  $\Gamma$ is constant, and the amplification factor for the magnetic energy increases
with the filling factor, $f$.
Applying the physical conditions in our simulation, we find that, when the
Sedov phase ends due to radiative cooling, $f$ increases to $\simeq 3\times 10^{-4}$.
Thus, with $\Gamma =4$  and $A \simeq 3$, $(A \Gamma -1) f$ is only $\simeq 0.003$.
Therefore, due to the tiny filling factor of the SN remnant, the total magnetic energy is
amplified only by a negligible amount during its early evolution.

When the remnant evolution enters the radiative phase, the jumping factor, $\Gamma \simeq {\cal M}_{\rm A}$,
which can be significantly larger than 4. Using the pressure-driven snow-plough solution to account for the deceleration
of the shock velocity (and hence the decrease of ${\cal M}_{\rm A}$ with time)
and the increase of $R_{\rm SN}$ (and the filling factor, $f$), we find that the maximum of the amplification factor, $(A^2 \Gamma -1) f$,
by a single SN in the radiative phase is $\simeq 0.05$.
Therefore, even if at a given time there are 10 SNRs that happen to simultaneously amplify the magnetic energy by the maximum
possible amount, the total amplification of the magnetic energy is only
50\%. Note that, in the above estimate, we have ignored the dynamical effect
of the magnetic field and the dissipation of the magnetic energy, and thus
the realistic amplification due to the SN explosions may be considerably
smaller than 50\%.  This explains the near constancy of the total magnetic
energy despite the strong oscillations in the total kinetic energy of the flow.

\begin{figure}[t]
\includegraphics[width=\columnwidth]{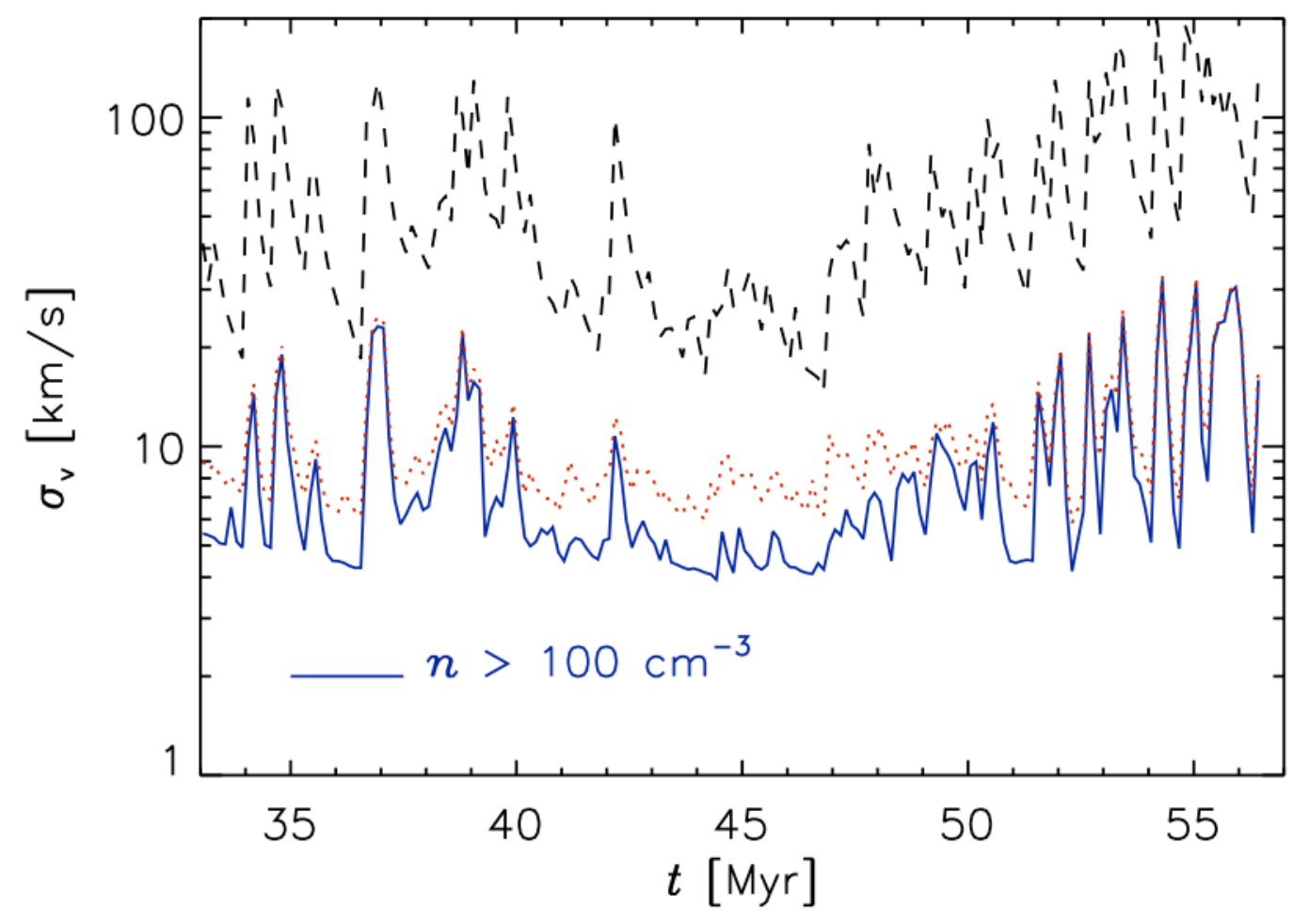}
\caption[]{Velocity dispersion versus time. The rms velocity is computed over the whole computational volume (dashed line) or only for gas with density
larger than 100 cm$^{-3}$ (solid line). The dotted line shows the mass-weighted rms velocity, averaged over the whole volume.}
\label{fig_vrms}
\end{figure}

Although the SN energy is introduced purely as thermal energy in the simulation, the thermal energy is efficiently converted into kinetic
energy, and is also partly radiated, so the resulting turbulence is mildly supersonic, with a time-averaged kinetic to thermal energy ratio of
$\langle E_{\rm k}/E_{\rm th}\rangle=3.75$. A crucial question for this work is what fraction of that total kinetic energy is
given to the dense gas and, thus, is available as turbulent kinetic energy for the MC turbulence.
The answer is given in Figure \ref{fig_energies2}, showing the time evolution of the total energies in the gas with density larger than
100 cm$^{-3}$. The time-averaged ratio of dense-gas kinetic energy and total kinetic energy is $\langle E_{\rm k,d}/E_{\rm k}\rangle=0.11$.
The ratio of the kinetic energies per unit mass is $\langle (E_{\rm k,d}/M_{\rm d})/(E_{\rm k}/M_{\rm box})\rangle=0.55$ (the dense-gas mass
fraction is $\langle M_{\rm d}/M_{\rm box}\rangle=0.18$), and oscillates in time between 0.19 and 0.97, with the largest values achieved during
the peaks of total kinetic energy. This large ratio means that, per unit mass, the kinetic energy is only a factor of $\sim2$ less than equally
distributed between the dense gas and the rest of the gas. This high efficiency of kinetic energy transfer to the dense gas results in a realistic
velocity dispersion of dense gas,  $\langle \sigma_{\rm v,d} \rangle=8.5$ km/s, as shown in Figure \ref{fig_vrms}.

Figure \ref{fig_energies2} also shows a clear separation of energies in the dense gas, with $\langle E_{\rm k,d}/E_{\rm m,d} \rangle=27.4$
and $\langle E_{\rm m,d}/E_{\rm th,d}\rangle=9.7$. Thus, the turbulence in the dense gas is both supersonic and super-Alfv\'{e}nic, as
further confirmed below for individual clouds selected from our simulation (see \S 10) and first suggested by
\citet{Padoan+Nordlund97MHD,Padoan+Nordlund99MHD}.

\section{Power Spectra and Driving Scale}

\begin{figure}[t]
\includegraphics[width=\columnwidth]{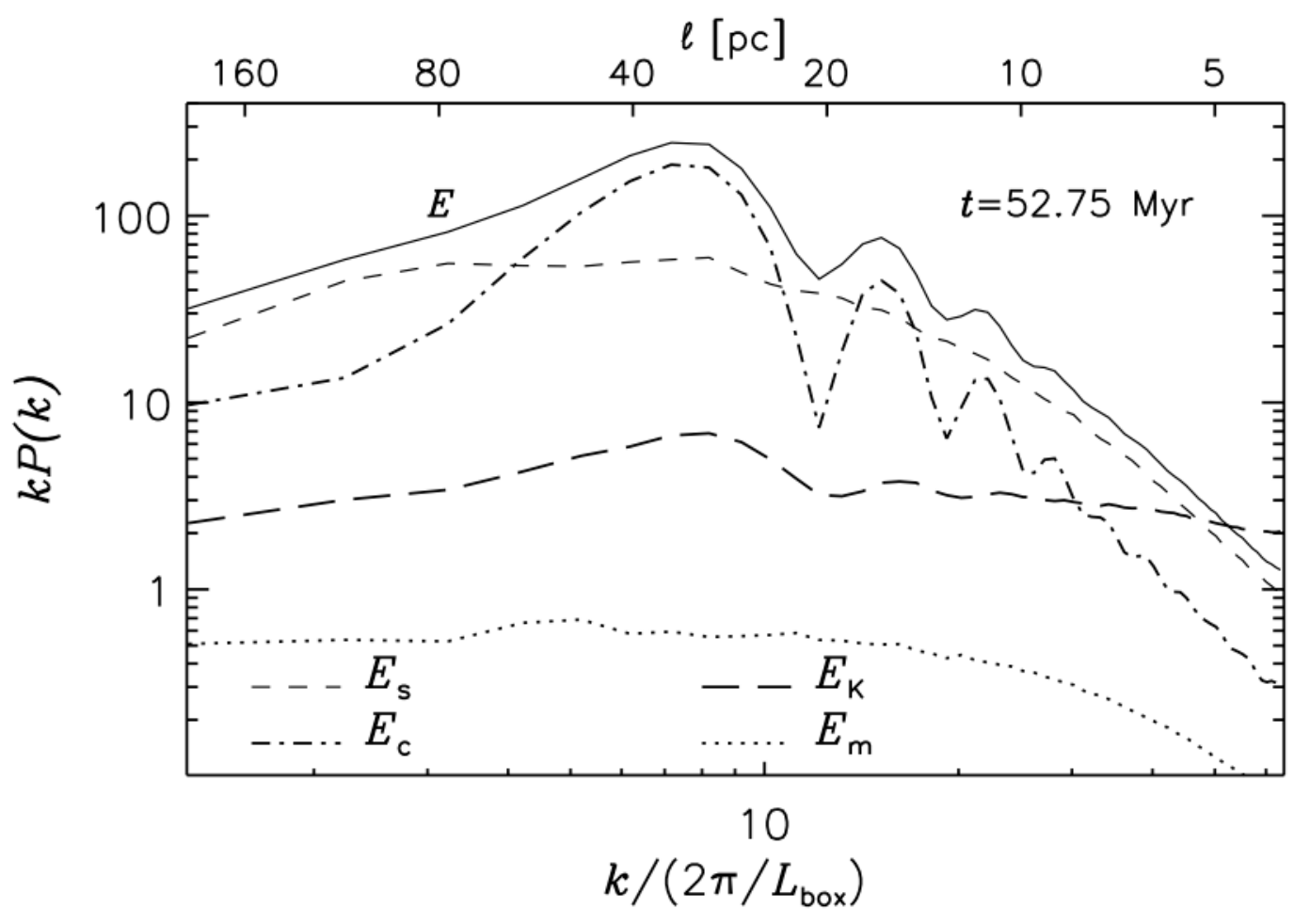}
\caption[]{Power spectra of velocity, $E(k)$, solenoidal velocity component, $E_{\rm s}(k)$, compressive velocity component, $E_{\rm c}(k)$,
square root of kinetic energy, $E_{\rm K}(k)$, and magnetic field, $E_{\rm m}(k)$, obtained from the average of the power spectra of the $x$, $y$
and $z$ components of these fields in the root grid ($128^3$ computational cells) of a single snapshot at time $t=52.75$ Myr. This time captures
the early expansion of a SN remnant to a diameter of approximately 30 pc, as indicated by the peak of the velocity power spectrum. The wavy
appearance of $E_{\rm c}(k)$ is explained in the text. The fact that $E_{\rm s}$ does not show similar fluctuations, and $E_{\rm s} < E_{\rm c}$
at the energy peak, indicates that this remnant has so far expanded into a relatively uniform hot medium, where the baroclinic effect is negligible
and solenoidal modes are not efficiently generated yet.}
\label{wavypowers}
\end{figure}

The expansion of SN remnants deposit energy on a broad range of scales, affecting the velocity scaling of the turbulence. We
observe this in the form of wave-like features in the velocity power spectrum or deviations in the velocity structure functions
(see \S 6) at small and intermediate scales in nearly one fourth of the snapshots of our simulation, as expected for our rate
of approximately 6 SN explosions per Myr, and a characteristic expansion time to 20-40 pc of $\sim 4 \times 10^4$ yr.

Figure \ref{wavypowers} shows the power spectra from a single snapshot that captures the early expansion phase of
a single SN remnant (here it has reached a diameter of approximately 30 pc) that has not had any major interaction with dense gas
yet, so most of the freshly-injected energy is still in the compressive modes (see \S 5.1)\footnote{This is a relatively rare event, as most
SN remnants experience some interaction with dense gas before they reach a size of 30 pc, and thus the compressive modes are usually
not dominant, as explained in \S 5.1.}. The Figure shows the power of the velocity, $E(k)$,
of the solenoidal (divergence-free) velocity component from a Helmholtz decomposition, $E_{\rm s}(k)$, of the compressive (curl-free)
velocity component from the same Helmholtz decomposition, $E_{\rm c}(k)$, of the square root of kinetic energy ($\rho^{1/2}u_{\rm i}$),
$E_{\rm K}(k)$, and of the magnetic field, $E_{\rm m}(k)$. They are computed from the root grid ($128^3$ computational cells) of a single
snapshot at time $t=52.75$ Myr. The velocity power spectra, especially the compressive one, exhibit significant
fluctuations at large and intermediate $k$. The wavy behavior is a signature of strong SN shocks and can be understood with a
one-dimensional illustration. Consider two SN shocks moving in opposite directions away from the origin (the explosion center).
If the velocity profile in between the two shocks is assumed to be roughly linear with the distance from the origin, it is
straightforward to show that the power spectrum is $\propto k^{-2} (\cos(kR) - \sin(kR)/kR)^2$, where $R$ is the radius of the
SN shock. Clearly, this spectrum oscillates at $k \gtrsim R^{-1}$, which explains the wavy behavior of the velocity spectra shown
in Figure \ref{wavypowers}.

\begin{figure}[t]
\includegraphics[width=\columnwidth]{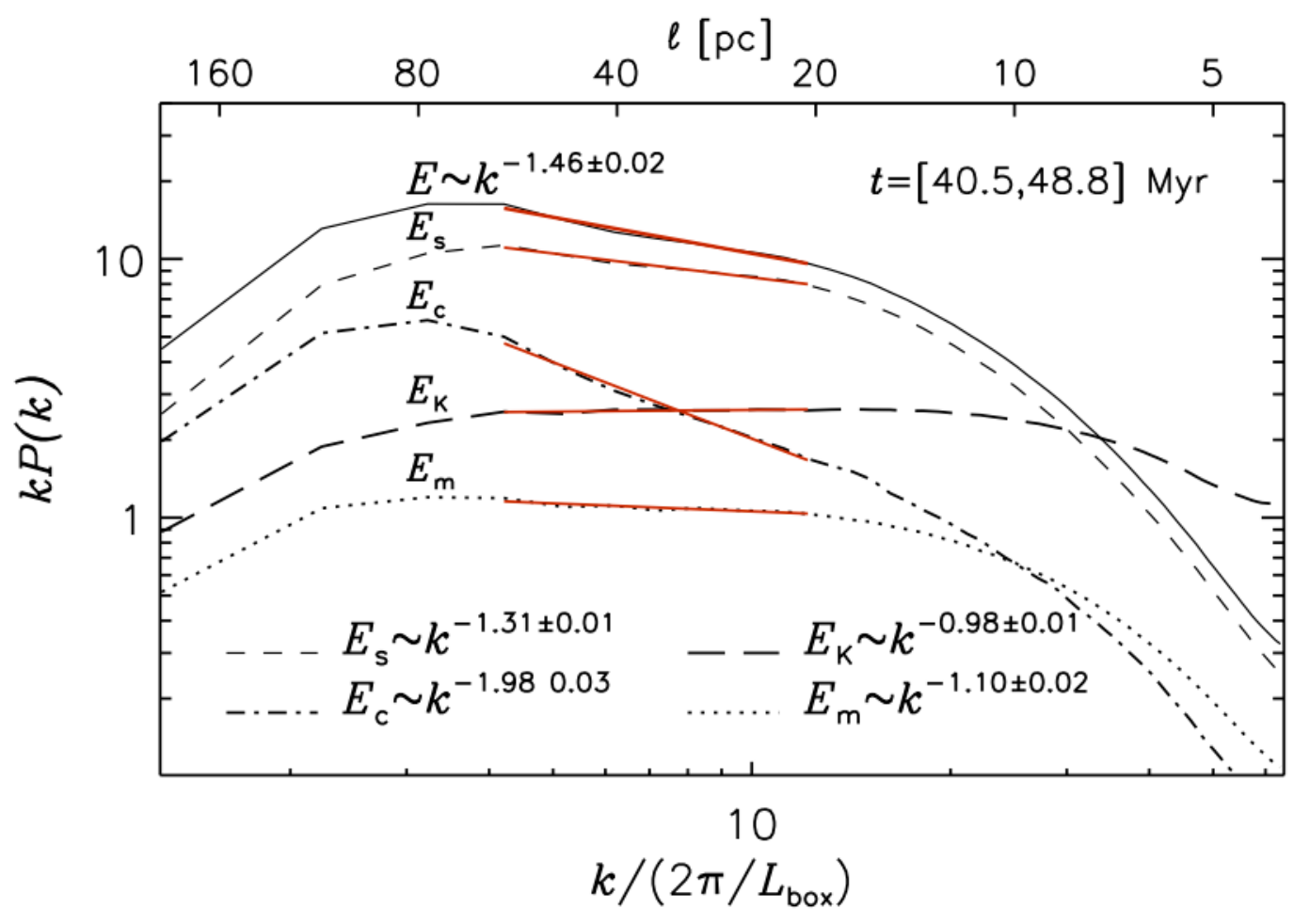}
\caption[]{Power spectra as in Figure \ref{wavypowers}, but averaged over 66 snapshots within the time interval $t=40.5-48.8$ Myr. During this time
interval the random SN rate is a bit lower than during the first and last third of the simulation, so there is a reduced probability that a snapshot captures the
early expansion of SN remnants with the corresponding perturbations to the power spectra. As a result, the time average is representative of the
statistically relaxed power spectra. The power-law slopes are obtained from a least-square fit in the range of wave numbers $4.2 \le k\,L_{\rm box}/2\pi \le 12.1$.}
\label{powers}
\end{figure}

In between SN explosions, or in regions not directly affected by a rapidly expanding SN remnant, the flow should have time to relax
from the transient state with direct SN impact, and we expect the velocity spectra to be smoother. To test this, we consider a time
range $t=40.5-48.8$ Myr, around the middle of our integration time. This choice avoids the larger SN rate towards the beginning
and the end of the simulation (see Figure \ref{fig_energies1}), and thus minimizes the number of snapshots with direct impact from
very recent SN explosions.  Figure \ref{powers} shows the average three-dimensional power spectra computed from the
root grid of 66 snapshots in the chosen time range. For each snapshot, the 3D spectra are obtained by averaging the
three components (e.g., the average of the power spectra of ${u}_{x}$, ${u}_{y}$ and ${u}_{z}$). The average spectra we
obtain are qualitatively similar to those of the most relaxed snapshots, and they exhibit a clear (though short) inertial range.
The slopes have been computed in the wavenumber interval $4.2 \le k\,L_{\rm box}/2\pi \le 12.1$, corresponding to the scale
interval $59.3 pc \ge \ell \ge 20.6 pc$, where all spectra are very well described by power laws. The measured inertial-range slopes
are given in Figure \ref{powers}. To our knowledge, this is the first time that inertial-range slopes are identified in SN-driven
turbulence. Because of the importance of this result, a full discussion of the power spectra will be presented elsewhere.
Here, we focus only on two points of direct interest for this work, the ratio of compressive to solenoidal modes and the
driving scale.

\subsection{Ratio of Compressive to Solenoidal Modes}

\begin{figure*}[t]
\includegraphics[width=\columnwidth]{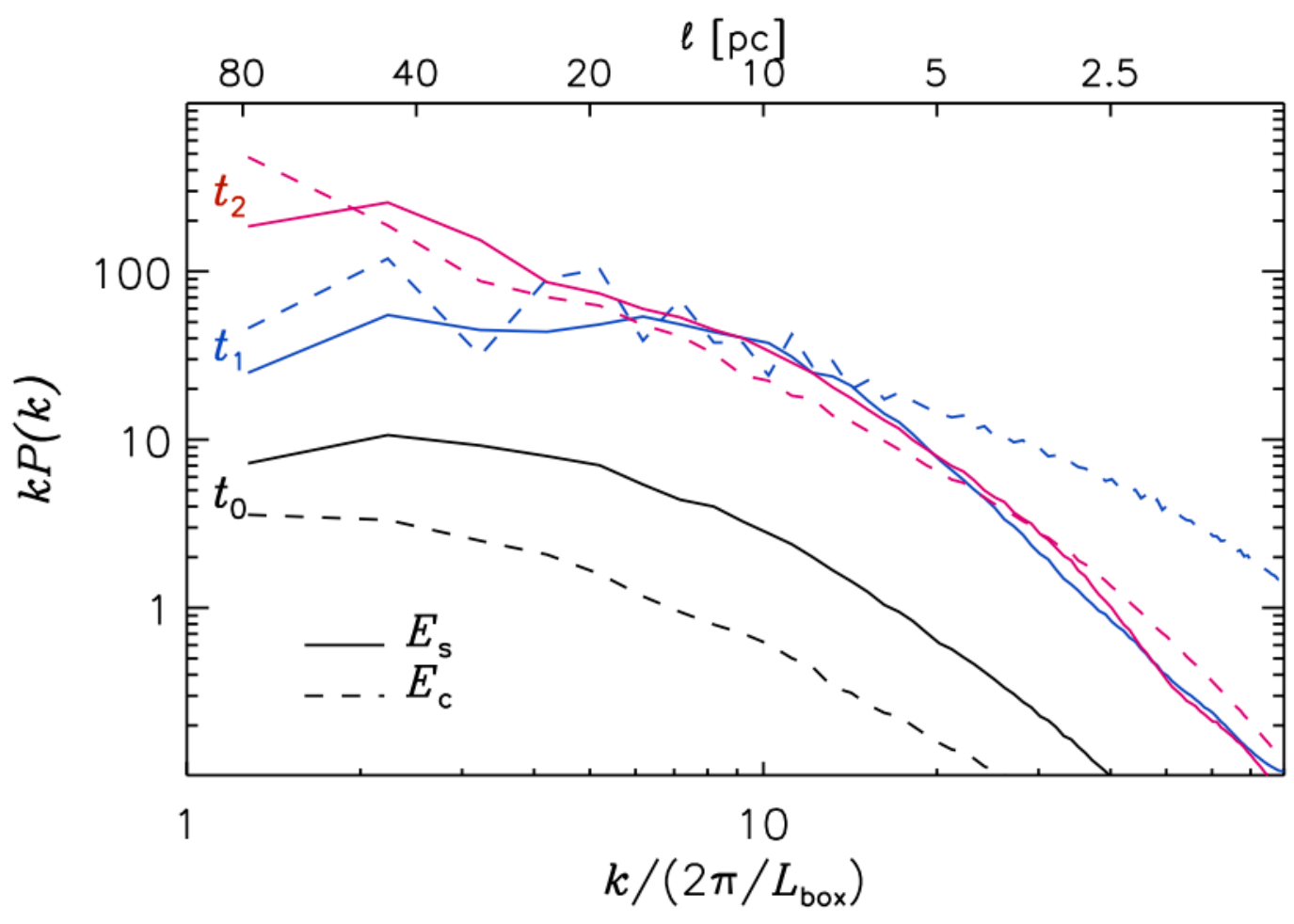}
\includegraphics[width=\columnwidth]{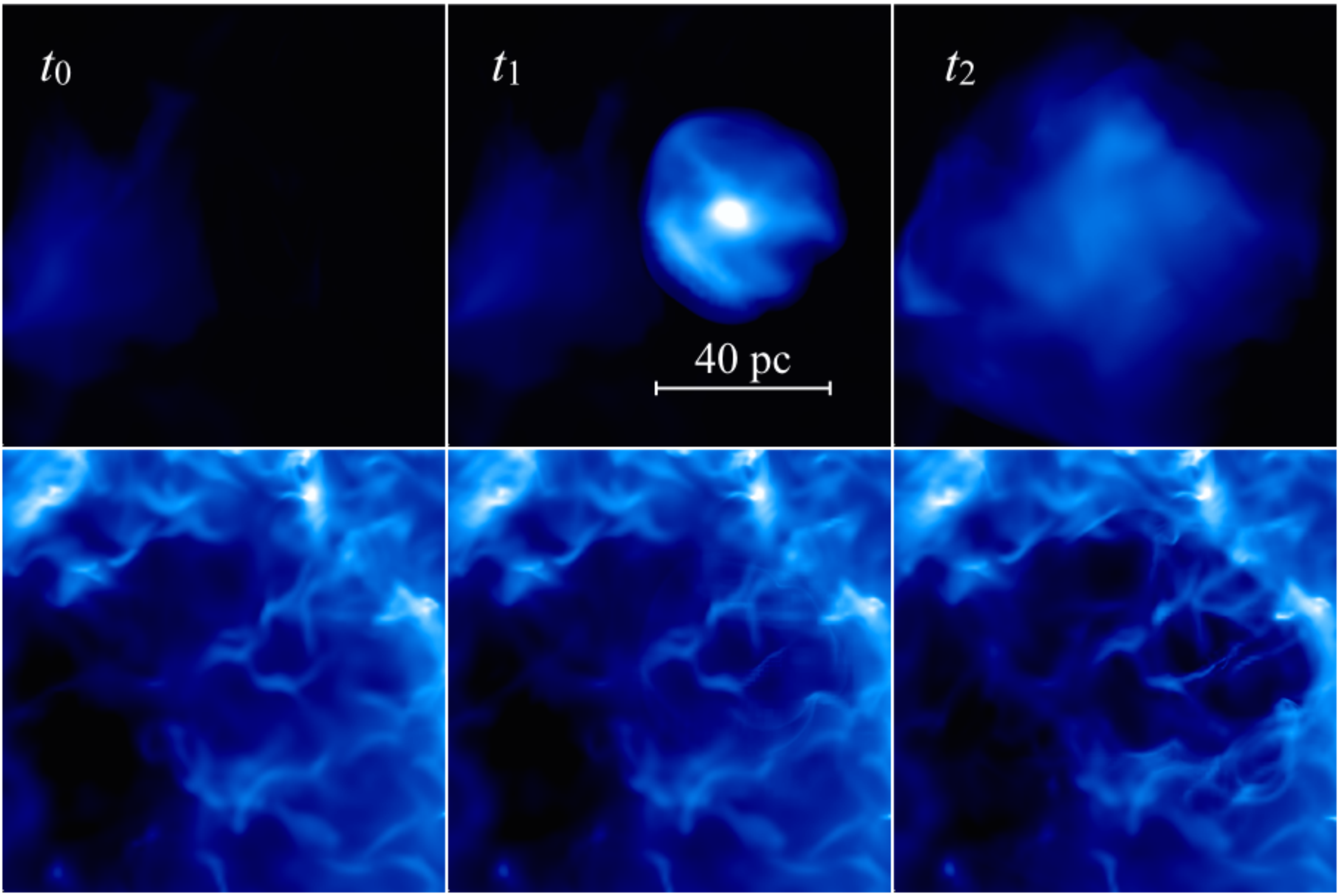}
\caption[]{{\it Left}: Power spectra of solenoidal (solid lines) and compressive (dashed line) velocity components within a (100 pc)$^3$ volume approximately
centered around a SN explosion. The spectra are computed at three different times, $t_0=46.625$ Myr, $t_1=46.750$ Myr and $t_2=46.875$ Myr. The 
time $t_0$ is just before the SN explosion, while the times $t_1$ and $t_2$ are after the explosions, when the diameter of the remnant is approximately 
40 and 80 pc, respectively. Between the times $t_1$ and $t_2$, the expansion of the SN remnant brings the power in both the compressive and solenoidal modes to 
larger scales, as explained in the text. {\it Right}: Squared root of projected temperature (upper row of panels) and logarithm of projected density (lower row of
panels) for the three times at which the power spectra are computed. At time $t_1$ the edge of the remnant is barely visible in the logarithm of projected density,
while at time $t_2$ the gas density within the remnant has significantly decreased in the lowest density regions.}
\label{remnant}
\end{figure*}

SN driving should not be viewed as a purely compressive form of driving. Our simulation indicates that, around the effective
driving scale (see \S 5.2), compressive modes are not dominant over solenoidal ones. We find that the compressive-to-solenoidal
ratio is typically smaller than unity, as shown by the time-averaged power spectra in Figure \ref{powers}.\footnote{The compressive power may exceed
the solenoidal one at snapshots with a SN remnant that expands in a uniform density region, making the baroclinic term negligible as explained below.
This is the case of the specific snapshot of Figure \ref{wavypowers}.}

After a SN explosion, the momentum of the ejecta is gradually transferred to the ISM during the expansion of the SN remnant until the final
momentum-conserving snow-plough phase. As a result, SN-driving covers a wide range of scales and persists for a relatively long time after the
explosion. This complex driving process cannot be primarily compressive because vorticity is readily generated by the baroclinic effect
and amplified by the non-linear advection, as explained below. Even the ISM forcing immediately after the explosion is far from purely compressive,
as hydrodynamical instabilities of the blast wave start already in the interior of the star, generating vorticity and making the ejecta very clumpy and
asymmetric \citep[e.g.][]{Chevalier+Klein78,Herant+Woosley94,Kifonidis+06,Couch+09,Wongwathanarat+15}. In our simulation, the forcing at the
moment of the thermal energy injection is not purely compressive either, because the acceleration from the corresponding pressure force has a
non-zero curl (a baroclinic term) due to the non-uniformity of the medium  (see the argument below).

To examine the ratio of compressive and solenoidal modes, it is helpful to write down the equations of the flow divergence
and vorticity, %
\begin{equation}
\frac{D ({\nabla \cdot \bs u})}{Dt}  + (\partial_i u_j) (\partial_j u_i)  = (\nabla \rho \cdot \nabla p)/\rho^2 - (\nabla^2 p)/\rho,
\label{div}
\end{equation}
and,%
\begin{equation}
\frac{D {\bs \omega}} {Dt}   -  ({\bs \omega} \cdot \nabla)  {\bs u}  +  {\bs \omega}  (\nabla \cdot {\bs u})= (\nabla \rho  \times \nabla p)/\rho^2.
\label{curl}
\end{equation}
The terms on the right-hand sides arise from the pressure term in the Navier-Stokes equation, and $(\nabla \rho  \times \nabla p)/\rho^2$ in
Equation (2) is called the baroclinic term, which contributes to generate vortical motions when the gradients of $\rho$ and $p$ are not
aligned. The  $({\bs \omega} \cdot \nabla) {\bs u}$ term is known as vortex stretching and can amplify the vorticity.

In our simulation, the flow velocity is driven by the pressure term.  If we denote as $p_{\rm s}$ the contribution to the
pressure from the SN thermal energy source, the effective driving acceleration is $-(\nabla p_{\rm s})/\rho$. It follows immediately from
Equations (1) and (2) that the divergence and curl of the effective acceleration are given by $(\nabla \rho \cdot \nabla p_{\rm s})/\rho^2 - (\nabla^2 p_{\rm s})/\rho$
and $(\nabla \rho  \times \nabla p_{\rm s})/\rho^2$, respectively. Because the SN locations are
selected randomly in our simulation and due to the density fluctuations in the ambient gas, the pressure and density gradients ($\nabla \rho$ and $\nabla p_{\rm s}$)
around the boundary of the initial SN sphere are not aligned in general, meaning that, at the instant of the SN energy injection, the
effective driving acceleration at the boundary of the SN sphere is not curl-free. Therefore, the effective driving in our simulation is not purely compressive.

If  $(\nabla \rho \cdot \nabla p_{\rm s})/\rho^2$ dominates over $(\nabla^2 p_{\rm s})/\rho$ (which is supported by the fact that
$|\nabla \ln(\rho) | \cdot |\nabla p| > |\nabla^2 p|$ in a significant fraction of our simulated flow), the ratio of compressive to solenoidal
power generated by the pressure source is determined mainly by the angle, $\theta$, between the directions of $\nabla \rho$ and
$\nabla p_{\rm s}$. Because in our simulation the SN locations are random, one may expect that the angle between the gradients is also random, so that
$\langle (\cos \theta)^2 \rangle \simeq 1/3$ and $\langle (\sin \theta)^2 \rangle \simeq 2/3$. In that case, the effective driving generates the vorticity variance
(corresponding to solenoidal motions) twice faster than the divergence variance. Although of a qualitative nature, the above argument provides an
explanation of why the solenoidal spectrum is typically larger than the compressive one.

By monitoring solenoidal and compressive spectra as a function of time, we can observe that
the expansions of the SN remnants bring both the compressive and solenoidal modes to larger scales, as shown in Figure \ref{remnant}.
The power in compressive modes moves to larger scales according to the simplified 1D model given earlier, which also explains the wavy
features in the spectrum; the solenoidal power is transferred to larger scales by the expansion through the non-linear term in Equation (2),
${\bs \omega} (\nabla \cdot {\bs u})$. Figure \ref{remnant} shows the velocity power spectra in a (100 pc)$^3$ volume centered around a 
SN explosion. The spectra are computed at three different times, one just before the SN explosion, and two after the SN remnant has reached 
an approximate size of 40 and 80 pc. The velocity field is multiplied by a tapered-cosine window function before performing the Helmholtz 
decomposition in Fourier space, to gradually set the velocity to zero at the volume boundaries, making the velocity field periodic. The transfer of
both solenoidal and compressive power to larger scale is clearly seen between the times $t_1$ and $t_2$, as the remnant expands from 40 
to 80 pc.
 
At times or in regions not too close to very young SN remnants, the solenoidal and compressive modes
have a chance to develop cascades toward small scales and to interact with each other, evolving toward a dynamically and
statistically relaxed state. The time-averaged spectra in Figure \ref{powers} indicate that in the relaxed state the solenoidal component of the
velocity is larger than the compressive component at all $k$. The ratio, $\chi(k) \equiv E_{\rm c}(k)/E_{\rm s}(k)$, is $\approx 0.2$ at $k=10$.
At the smallest wave numbers, the modes are almost in
equipartition, $\chi(k) \sim 1$. However, $E_{\rm s}(k)$ is remarkably shallow, while $E_{\rm c}(k) \propto k^{-2}$, so the
compressive-to-solenoidal ratio decreases rapidly towards larger wave numbers, $\chi(k) \propto k^{-0.67}$ in the inertial range.
Although they did not obtain clear power-law scaling and did not identify inertial-range slopes in their SN-driven simulation,
\citet{Balsara+04} also found that $\chi(k)$ decreases with increasing $k$ and $\chi(k) \ll 1$ at large wave numbers.

The different inertial-range slopes of $E_{\rm c}(k)$ and $E_{\rm s}(k)$ and the rapid decrease of $\chi(k)$ with increasing $k$
are in contrast to previous studies, which typically found that $\chi (k)$ is more or less constant in the inertial range.
For example, simulations of weakly compressible turbulence with Mach numbers ${\cal M}_{\rm s} \sim 1$ showed that both
compressive and solenoidal modes have Kolmogorov-like inertial-range spectrum, $E_{\rm c}(k) \propto E_{\rm s}(k) \propto k^{-5/3}$,
and $\chi(k) \sim 0.05$ \citep[e.g.][]{Porter+98,Porter+99,Porter+02}.  At the opposite extreme, simulations with purely compressive
driving find $\chi(k) \sim 1$ in the inertial range, and the same Burgers-like slopes for both modes,
$E_{\rm c}(k) \sim E_{\rm s}(k) \propto k^{-2}$ \citep{Federrath13_4096}. Simulations with solenoidal (or mixed) driving and large
Mach number, ${\cal M}_{\rm s} \gg 1$, yield values $\chi(k) \approx 0.3-0.5$ \citep{Kritsuk+10,Federrath13_4096}, with only
a slight decline towards large wave numbers.

An important difference between these studies and our simulation is that they all adopted a barotropic equation of state, assuming either
adiabaticity or isothermality. The baroclinic effect is thus absent in all the simulations mentioned above. As discussed earlier,
when SNe explode in our simulation, the baroclinic effect from {\it the pressure source} immediately drive solenoidal motions around the
SN spheres at a rate similar to compressible modes.  As the SN remnant and the flow evolve,
the general baroclinic effect in Equation (2) is also likely to play an important role as the flow develops a
cascade and evolves toward relaxation. As compressive motions in the form of shocks or expansions encounter
a dense region in the flow, the baroclinic effect gives rise to shear and vortices around and within the region.
Since the baroclinic term depends on the gradients of the density and pressure, the conversion from
compressive to solenoidal motions is expected to be more efficient at smaller scales. This explains
why the compressive-to-solenoidal ratio decreases rapidly with increasing $k$ in our simulation.
We thus argue that the existence of the baroclinic effect in our simulation is responsible for the different behaviors
of $E_{\rm c}(k)$, $E_{\rm s}(k)$ and $\chi(k)$ compared to previous barotropic simulations.

There has been recent interest in the regime of supersonic turbulence with purely compressive
driving, showing that it affects the probability distribution of gas density \citep{Federrath+08,Molina+12},
the magnetic field amplification by turbulence \citep{Federrath+11PRL}, and the
inertial-range velocity scaling \citep{Schmidt+08prl,Schmidt+09compressive,Federrath13_4096}, with
possible consequences for models of star formation or for turbulence in galaxy clusters
\citep{Porter+15_ICM_compressive_forcing}. However, our result that SN-driving is not primarily
compressive, particularly at MC scales, suggests that features specific to isothermal flows with highly
compressive driving may not apply to ISM turbulence. If SN shock-waves are the main energy source for
MC turbulence, the driving acceleration would consist of a significant fraction of solenoidal modes, which arise
through the baroclinic effect, when the SN shock impacts a cloud. Thus, idealized isothermal simulations
with purely solenoidal driving may better capture MC turbulence than isothermal simulations with purely
compressive driving, and previous results on turbulent fragmentation based on solenoidal driving  may not require
significant correction. A careful study of the full implication for star formation of our result would require the evaluation
of the compressive-to-solenoidal ratio specifically for the clouds formed in the SN-driven turbulence, which we pursue
in a separate work \citep{Pan+15compress}.

\subsection{Energy Injection Scale}

\begin{figure}[t]
\includegraphics[width=\columnwidth]{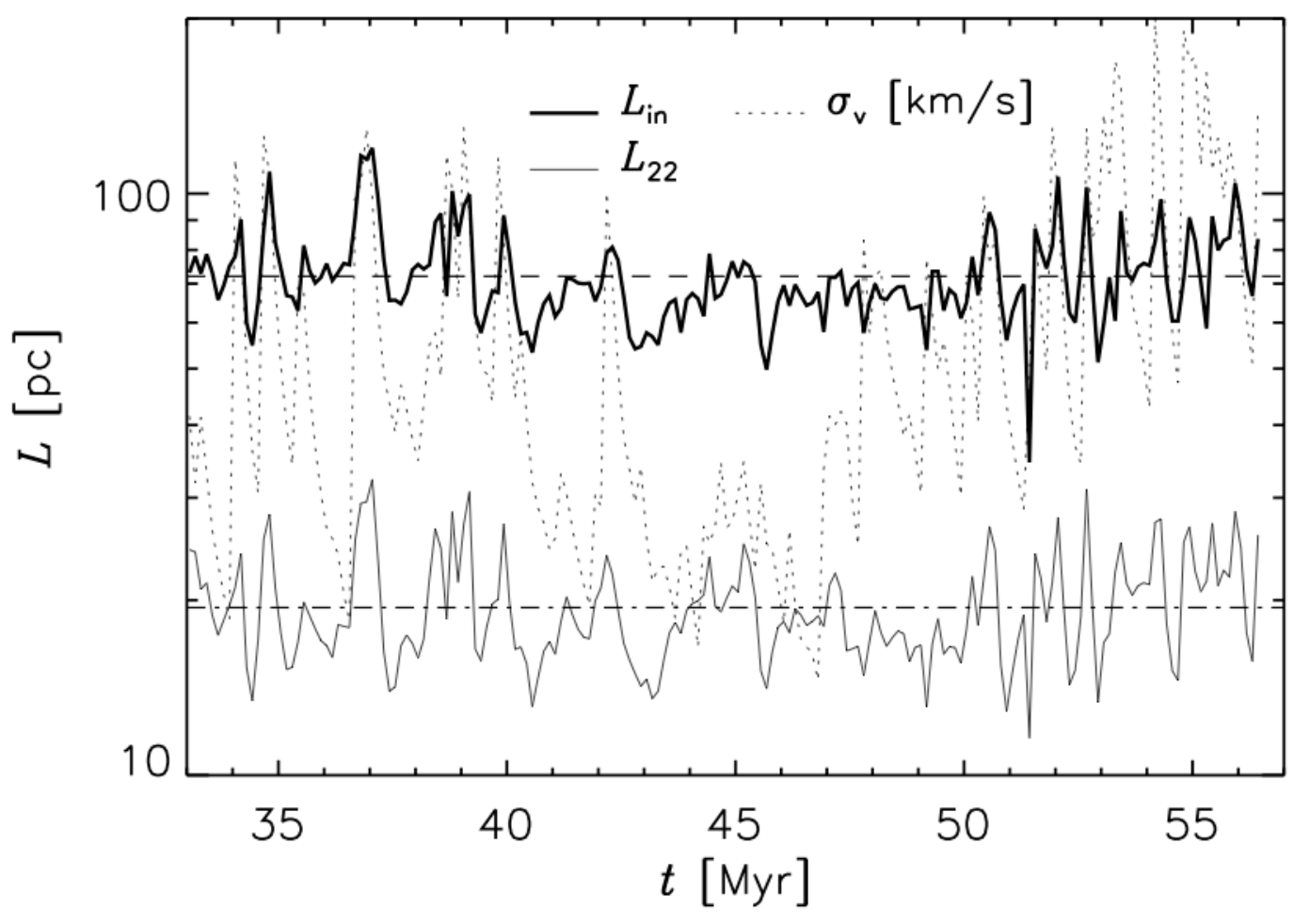}
\caption[]{Time evolution of the energy-injection scale, $L_{\rm in}$ (thick solid line), computed from the integral of the inverse wave number, $1/k$,
weighted by the velocity power spectrum, $E(k)$ (see equation (\ref{L})). The time-averaged value of 70.5 pc is shown by the horizontal dashed line.
The dotted line shows the rms velocity, $\sigma_{\rm v}$ (see the dashed line in Figure \ref{fig_vrms} for the actual values of $\sigma_{\rm v}$).
Almost all peaks in $\sigma_{\rm v}$ correspond to peaks in $L$. The thin solid line shows the time evolution of the transverse integral scale, $L_{22}$
(see text). The average value, $\langle L_{22} \rangle=19.4$ pc, is shown by the horizontal dashed-dotted line.}
\label{integral}
\end{figure}

Using the velocity power spectrum, $E(k)$, we can define a length, $L_{\rm in}$, that corresponds
approximately  to the scale where most of kinetic energy is contained, and can thus be interpreted
as a characteristic scale of energy injection by SN explosions:
\begin{equation}
L_{\rm in} \equiv {2 \pi \int k^{-1} E(k)  dk \over \int E(k)  dk}.
\label{L}
\end{equation}
The time dependence of $L_{\rm in}$ is shown in Figure \ref{integral}, where we have also plotted the rms
velocity, $\sigma_{\rm v}$. The value of $L_{\rm in}$ oscillates between approximately 50 and 100 pc,
with many of the peaks corresponding also to peaks in $\sigma_{\rm v}$. The time average and standard deviations are 70.5
pc and 12.0 pc respectively.

Figure \ref{integral} also shows the transverse integral scale, $L_{22}$. The longitudinal and transverse integral scales,
$L_{11}$ and $L_{22}$, are defined as the integrals of the two-point velocity correlation functions:
%
\begin{align}
\begin{split}
L_{\rm 11} \equiv  \frac{1}{R_{L} (0)}\int_0^{\infty} R_{L} (\ell )d \ell, \\
L_{\rm 22} \equiv  \frac{1}{R_{N} (0)}\int_0^{\infty} R_{N} (\ell )d \ell,
\label{L11}
\end{split}
\end{align}
%
where $R_{L}$ and  $R_{N}$ are the longitudinal and transverse correlation functions,
%
\begin{align}
\begin{split}
R_{L} (\ell ) = \langle u_{\ell }(\bs {x}) u_{\ell }(\bs {x} + \bs {\ell}) \rangle, \\
R_{N} (\ell ) = \langle u_{n}(\bs {x}) u_{n}(\bs {x} + \bs {\ell}) \rangle,
\label{RLL}
\end{split}
\end{align}
%
with $u_{\ell }$ and $u_{n}$ being the velocity component parallel to $\bs{\ell}$ and
one (of the two) transverse component perpendicular to $\bs{\ell}$, respectively.
$L_{11}$ and $L_{22}$ are typically smaller than the injection length, $L_{\rm in}$. In isotropic turbulence,
exact relations exist between $L_{11}$, $L_{22}$ and $L_{\rm in}$. For example, in incompressible turbulence, $L_{11} = 2 L_{\rm 22} = 3 L_{\rm in}/8$ \citep{Monin+Yaglom75}.
Our simulated flow is highly compressible, and we find that the  longitudinal and transverse integral scales are close to each
other with $\langle L_{\rm 11}\rangle = 19.9$ pc, and $\langle L_{\rm 22}\rangle = 19.4$ pc. The near equality of $L_{\rm 11}$
and $L_{\rm 22}$ suggests that kinetic energies contained in solenoidal and compressive modes are comparable.
Figure \ref{powers} shows that at the scale of $L_{\rm in}$ the solenoidal and compressive spectra are
in equipartition, in the sense that the solenoidal spectrum is twice larger than the compressive one due to the extra degree of freedom.
One can demonstrate that in such case $L_{\rm 11}$ and  $L_{\rm 22}$ should be exactly equal,
as found in our simulation. Furthermore, in that case $L_{11}$ and  $L_{\rm 22}$ are expected to be equal to $L_{\rm in}/4$,
which further explains the ratios, $L_{\rm in}/L_{\rm 11}=3.5$ and $L_{\rm in}/L_{\rm 22}=3.6$.

By comparing four galactic-fountain simulations with different SN rates, \citet{Joung+09} found that the energy injection scale
decreases with increasing SN rate. In the model with approximately the same SN rate as in our simulation they obtained
$L_{\rm in }=87$ pc. This value is approximately consistent with the one derived here, if we account for the fact that \citet{Joung+09}
assumed that 60\% of the SN explosions are spatially correlated, as discussed above in \S 2.1, which enhances the formation
of super-bubbles and so should tend to increase the correlation length.

Using their galactic-fountain simulation, \citet{deAvillez07scaling} computed the longitudinal and transverse integral scales, $L_{11}$ and $L_{22}$.
They found that $L_{22}/L_{11} =0.5-0.6$, consistent with
the ratio in incompressible turbulence, implying that the overall compressibility of their simulated  flow was likely
low. Their measured value of $75.2$ pc for $L_{11}$ would indicate an injection length scale of $L_{\rm in} \simeq 200$ pc.
This injection length scale is significantly larger than $L_{\rm in} \simeq 70$ pc found in our simulation. Based on the dependence of the integral scale on the SN
rate from \citet{Joung+09}, this difference may be attributed to the much smaller SN rate in the simulations analyzed by \citet{deAvillez07scaling}.

\section{Velocity Structure}

In order to characterize the turbulence, we compute the velocity structure functions, first for the whole computational volume, and then for
individual clouds. The velocity structure functions of order $p$ are defined as:
\begin{equation}
S_p(\ell)\equiv\langle|u({\bs x}+{\bs \ell})-u({\bs x})|^p\rangle\propto \ell\,^{\zeta(p)}
\label{str}
\end{equation}
where the velocity component $u$ is parallel (longitudinal structure function)
or perpendicular (transversal structure function) to the vector $\bs \ell$
and the spatial average is over all positions $\bs x$.

\citet{Boldyrev2002} proposed an extension to supersonic turbulence of the intermittency model by She and L{\'e}v{\^e}que
\citep{She+Leveque94,Dubrulle94}:

\begin{equation}
\zeta(p)/\zeta(3)=p/9 + 1-\left( 1/3\right) ^{p/3}.
\label{boldyrev}
\end{equation}
This velocity scaling has been found to provide a very accurate prediction for numerical simulations of highly supersonic and
super--Alfv\'{e}nic turbulence \citep{Boldyrev+02,Padoan+04prl,Pan+Scannapieco11SL}.
\citet{Padoan+04prl} showed that, as the rms Mach number of the turbulence increases, the structure function scaling
varies from the She-L{\'e}v{\^e}que scaling of incompressible turbulence to the Boldyrev's scaling, which has been interpreted as
a gradual change of the Hausdorff dimension of the most dissipative structures from 1 (dissipation in filaments)
to 2 (dissipation in sheets). Although the scaling of equation (\ref{boldyrev}) may not apply to flows driven by purely compressive
forces \citep{Schmidt+09compressive}, we have shown in \S 5.1 that the compressive modes are not dominant in SN-driven
turbulence.

Because of the limited extent of the turbulence inertial range in numerical simulations, the structure functions are usually power laws
only over a very limited range of scales, if at all. Thus, the scaling exponents are usually derived by normalizing the structure functions
to the third-order one, which yields power laws that extend well into the (numerical or physical) dissipation range of scales. This useful
property is known as `extended self-similarity' \citep{Benzi+93}.

The third-order structure function always yields a slope larger than unity in supersonic turbulence, while $\zeta(3)=1$ is an exact result
in incompressible turbulence, the so called `4/5 law' first derived by \citet{Kolmogorov41}.
This is because in supersonic turbulence the third-order structure functions should be computed with some density-weighting factor. For example, a
density weight inspired by the assumption of constant energy transfer of a compressible flow, $u \sim (\ell/\rho)^{1/3}$ \citep{Fleck96}, was proposed
by \citet{Kritsuk+07} and further tested by \citet{Koval+Lazarian07} and by \citet{Schmidt+08prl}. In this work, we compute the structure functions
either using the tracer-particle positions and velocities, or using the gas velocity values on a uniform mesh, with a density-weighting method that is
equivalent to the use of tracer particles. This density-weighting method is different from that proposed by \citet{Kritsuk+07}.

\begin{figure}[t]
\includegraphics[width=\columnwidth]{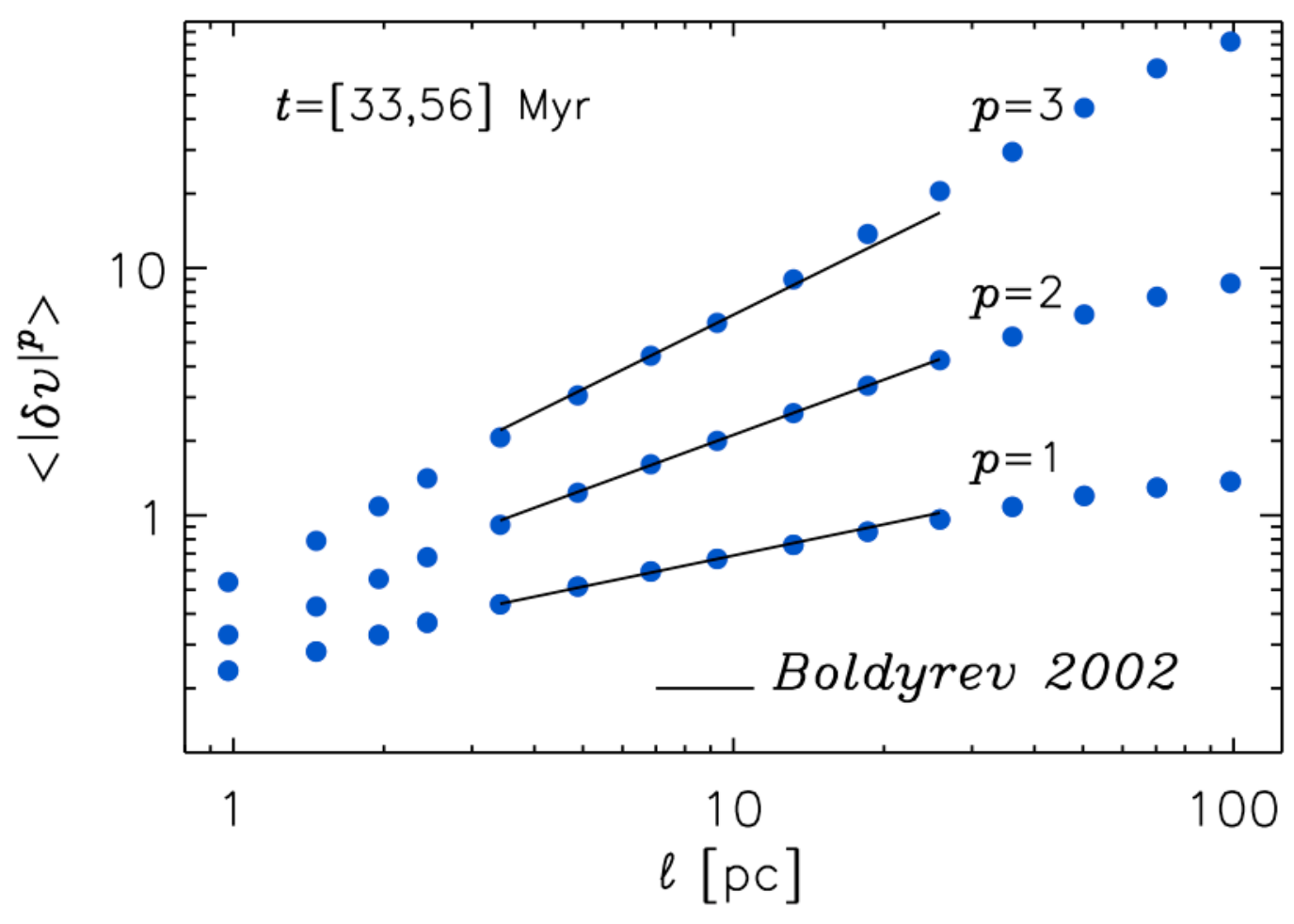}
\caption[]{Density-weighted longitudinal velocity structure functions, $S^{\rm dw}_p(\ell)$ (see equation (\ref{dw})), of orders $p=1$, 2 and 3,
plotted versus the separation $\ell$, computed over the whole computational volume (filled circles).
The plots show the time average of the structure functions from 24 snapshots covering uniformly the time interval
$t=33-56$ Myr (one snapshot per Myr). The solid lines corresponds to the structure function slopes predicted by \citet{Boldyrev2002}
and confirmed in simulations of randomly-driven isothermal turbulence \citep{Boldyrev+02,Padoan+04prl,Pan+Scannapieco11SL},
$\zeta(1)=0.42$, $\zeta(2)=0.74$, $\zeta(3)=1.0$. Although they provide an excellent fit, these solid lines are not obtained by fitting
the data.}
\label{structure_mesh}
\end{figure}

To calculate the structure functions based on tracer particles in a MC, we search particle pairs at given
distances, $\ell$, compute their relative velocities, and average over all the particle pairs to obtain the relative velocity moments
at different orders, $p$. For example, the $p$-th order structure function is computed as
\begin{equation}
S_p^{\rm tr} (\ell)  \equiv \frac{1}{N}\sum_{n=1}^N |{\bs u}_{1n} -{\bs u}_{2n}|^p,
\label{tr}
\end{equation}
where $N$ is the number of pairs separated by a distance of $\ell$, and ${\bs u}_{1n} - {\bs u}_{2n}$ is the relative velocity of the
$n$-th pair of particles. These tracer-based structure functions have a built-in density weighting, because the number of particle
pairs at a given distance depends on the flow densities at the particle positions (the tracer particles are initialized with a number
density proportional to the local gas density). For structure functions over the entire simulation box, we compute grid-based structure
functions with a density weighting defined as
\begin{equation}
S_p^{\rm dw}(\ell) \equiv { \langle \rho({\bs x}) \rho({\bs x}+{\bs \ell}) |{\bs u}({\bs x}+{\bs \ell}) -{\bs u}({\bs x})|^p\rangle \over \langle \rho({\bs x}) \rho({\bs x}+{\bs \ell})\rangle}.
\label{dw}
\end{equation}
It is straightforward to prove that $S_p^{\rm tr} (\ell)$ and $S_p^{\rm dw}(\ell)$ are statistically equivalent to each other, if
the number of tracers is large enough to sufficiently sample the flow density field. To show this, consider two infinitesimal
volumes, $dV_1$ and $dV_2$, around two points, 1 and 2, separated by a distance of $\ell$. When computing the tracer-based
structure functions, these two volumes would contribute to a total number of
$N_{12} = ({\bar n}_p/{\bar \rho})^2  \rho_1 \rho_2 dV_1 dV_2$ particle pairs at a distance of $\ell$, where ${\bar n}_{p}$ and  ${\bar \rho}$
are the mean particle number density and mean flow density, respectively.  In other words, the two points 1 and 2 are essentially
counted  $N_{12} \propto \rho_1 \rho_2$ times in the computation of $S_p^{\rm tr} (\ell)$. This is equivalent to a density-weighting
factor of  $\propto \rho_1 \rho_2$, suggesting that $S_p^{\rm tr} (\ell)$ is equal to the grid-based structure function, $S_p^{\rm dw}(\ell)$.

\begin{figure}[t]
\includegraphics[width=\columnwidth]{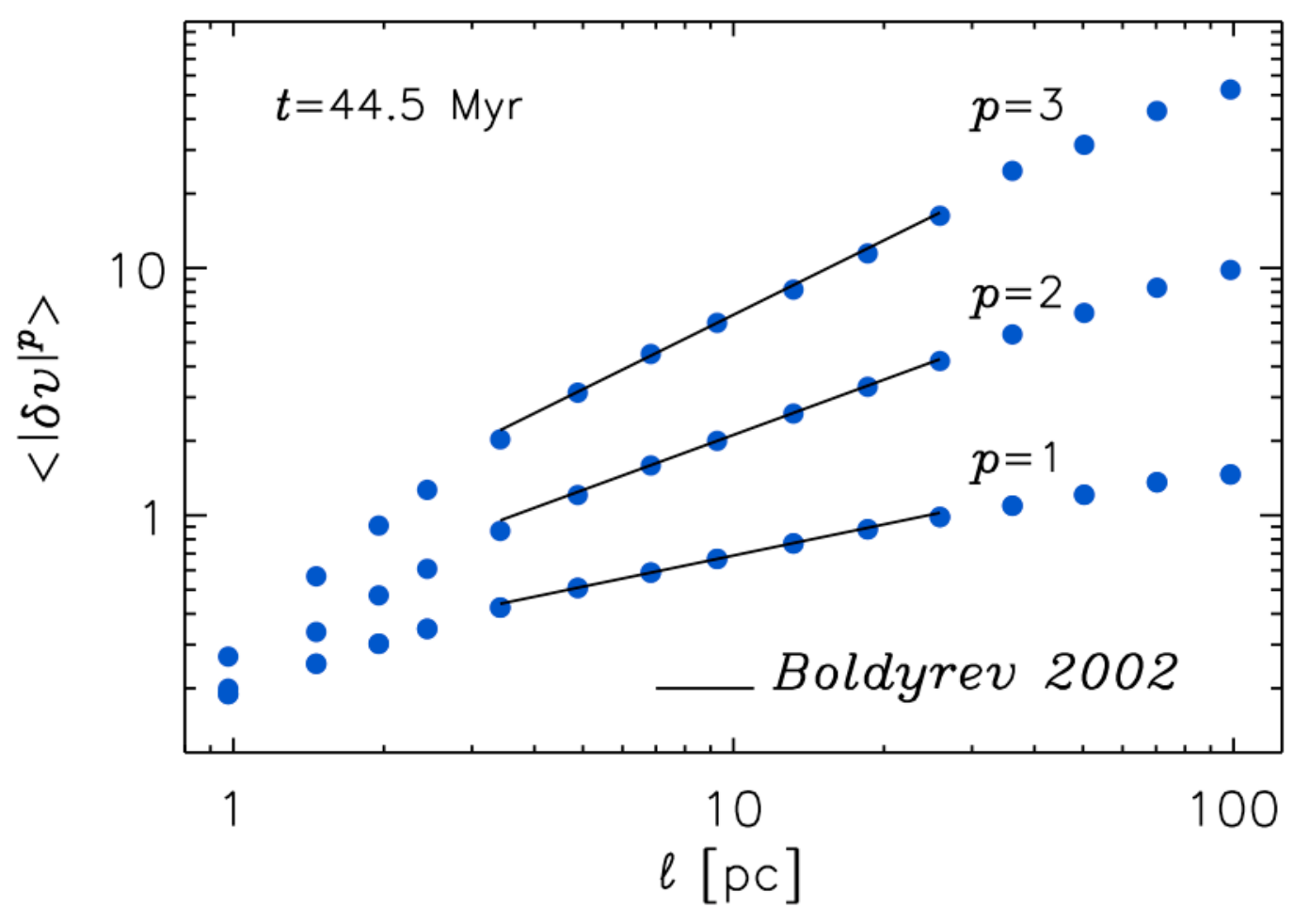}
\caption[]{Same as in Figure \ref{structure_mesh}, but for a single snapshot at $t=44.5$ Myr. }
\label{structure_mesh_2}
\end{figure}

The density-weighting scheme adopted here is of particular interest in the light of a result derived by \citet{Falkovich+10}.
Motivated by Kolmogorov's 4/5 law for incompressible flows, \citet{Falkovich+10} obtained an exact relation for compressible turbulence,
\begin{multline}
\langle \rho({\bs x}) \rho({\bs x}+{\bs \ell}) u_i({\bs x}) u_i({\bs x}+{\bs \ell}) u_j({\bs x}+{\bs \ell})\rangle +  \\
\langle\rho({\bs x}) u_j({\bs x}) p({\bs x}+{\bs \ell}) \rangle \propto \ell_j,
\label{Falkovich}
\end{multline}
where $p$ is the pressure of the flow. In highly supersonic turbulence, the pressure term may be neglected, and we have
\begin{equation}
\langle \rho({\bs x}) \rho({\bs x}+{\bs \ell}) u_i({\bs x}) u_i({\bs x}+{\bs \ell}) u_j({\bs x}+{\bs \ell})\rangle \propto \ell_j.
\label{Falkovich_sup}
\end{equation}
The quantity $\langle \rho({\bs x}) \rho({\bs x}+{\bs \ell}) u_i({\bs x}) u_i({\bs x}+{\bs \ell}) u_j({\bs x}+{\bs \ell})\rangle$ is closely related
(although not exactly equivalent) to the density-weighted third order structure function, $S_3^{\rm dw}$. We therefore expect a
linear scaling for our density-weighted third order structure function, $S_3^{\rm dw} \propto \ell$, which turns out to be confirmed
by our simulation for both the entire flow and individual MCs.

\subsection{Global Velocity Scaling}

\begin{figure}[t]
\includegraphics[width=\columnwidth]{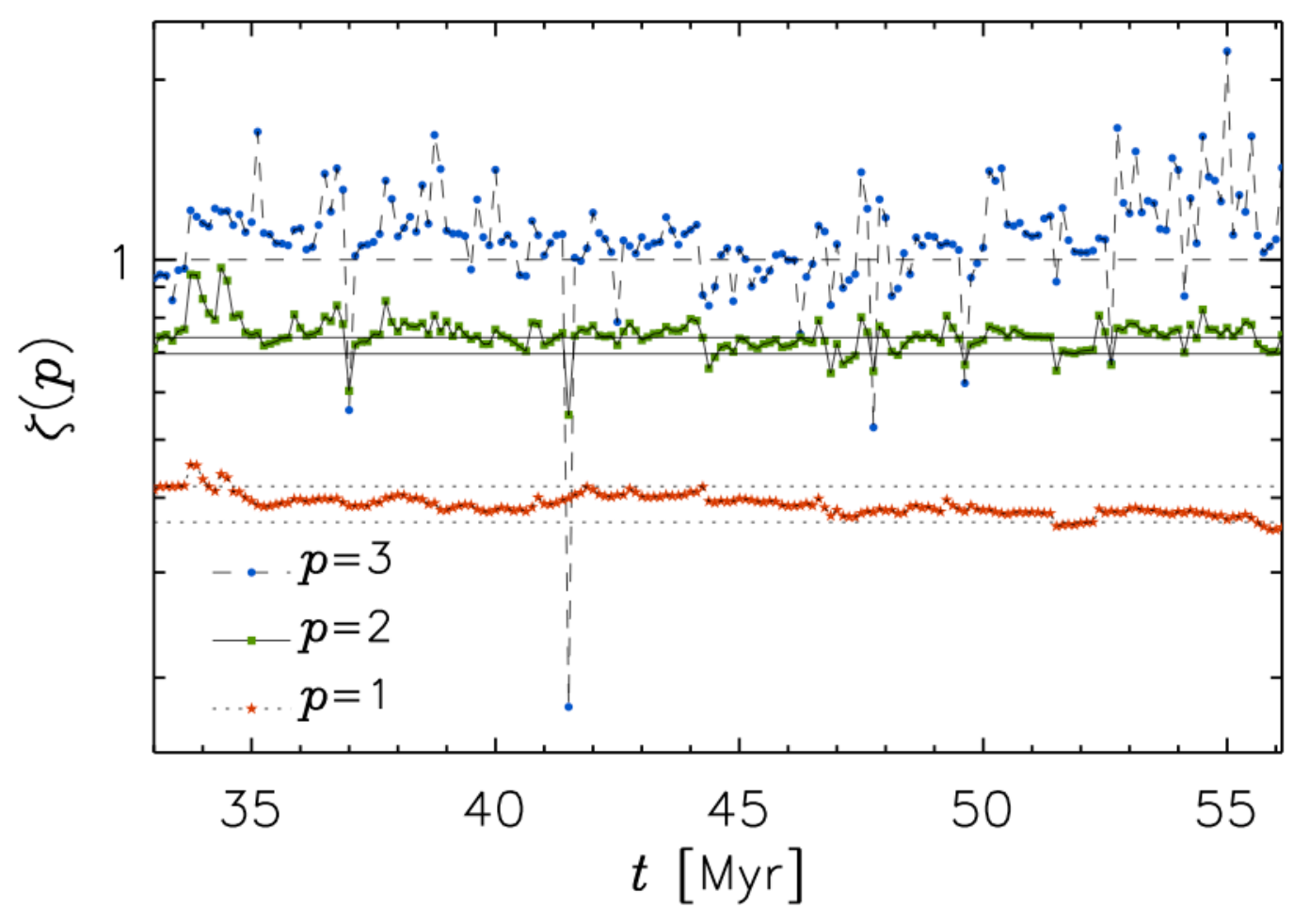}
\caption[]{Structure function slopes versus time, for structure functions computed in each of the 188 snapshots in the time interval $t=33-56$ Myr.
The horizontal dotted lines show the values of the first order slope, $\zeta(1)$, predicted by the She-L{\'e}v{\^e}que intermittency model \citep{She+Leveque94}
(lower line) and the Boldyrev model \citep{Boldyrev2002} (upper line). The horizontal solid lines correspond to the predictions for $\zeta(2)$ by the same
models, and the horizontal dashed line corresponds to $\zeta(3)=1$, an exact mathematical result for incompressible turbulence \cite{Kolmogorov41}.}
\label{structure_slopes}
\end{figure}

We first compute the velocity structure functions averaged over the whole computational volume. The velocity field is first remapped onto a uniform mesh of
512$^3$ cells, and then the structure functions are computed with the density-weighting  method described above. The volume contains large voids of very
low density, with a very low number of tracer particles, so we prefer this mesh-based method instead of the equivalent tracer-particle method (which we
use below for the structure functions inside MCs). To speed up the calculations, we only consider velocity differences along the three main orthogonal
directions, and 16 values of cell distances, $\ell$. Even with this limitation, the sample size is large enough to yield reliable low-order statistics.

Figure \ref{structure_mesh} shows the first, second, and third order longitudinal velocity structure functions, $S^{\rm dw}_p(\ell)$, plotted versus the separation
$\ell$, computed over the whole volume as described above, and time-averaged using 24 snapshots covering uniformly the time interval $t=33-56$ Myr (one
snapshot per Myr). They are well approximated by power laws in the approximate range of scales between 3 and 30 pc. The values of the exponents are
$\zeta(1)=0.39 \pm 0.01$, $\zeta(2)=0.75 \pm 0.01$, $\zeta(3)=1.13 \pm 0.02$.

To compare with values previously found in studies of randomly-driven supersonic isothermal turbulence, in this and following figures we over-plot lines
showing the slopes from eq. \ref{boldyrev}. Notice that eq. \ref{boldyrev} only gives the exponents of the velocity structure functions without density
weighting and normalized to the third-order one. Thus, in this comparison, we make the reasonable assumption that the exponents normalized to the
third order should be the same for the unweighted structure functions as for the density-weighted ones. However, future works should recompute the
velocity structure functions of randomly-driven, supersonic isothermal turbulence simulations using the same density-weighting scheme as in this work.

While the first and second order structure functions in Figure \ref{structure_mesh} have exponents within 3 sigma and 1 sigma respectively from the
values of equation \ref{boldyrev}, $\zeta(3)$ is significantly larger than unity, despite the time averaging. We find that this deviation from unity, and even
from a power law, occurs in snapshots during periods with the highest SN rate, when it is more likely that a snapshot is very close in time to a very recent
SN explosion. During the brief, initial period of the SN bubble expansion, SN driving has a direct effect on a range of scales. On the contrary, in snapshots that are
not too close in time to SN explosions, the SN remnants have already expanded to large scale, the turbulence has had time to relax, and the third order
structure function is found to be a nearly perfect power law with $\zeta(3)=1$. Figure \ref{structure_mesh_2} shows an example of velocity structure functions
from a single snapshot, at $t=44.5$ Myr, that is not too close in time to a SN explosion. The third order structure function is now a power law, with the expected
numerical decay below approximately 3-4 pc, and the third order slope is indistinguishable from unity. The exponents in this single snapshot are $\zeta(1)=0.41 \pm 0.01$,
$\zeta(2)=0.77 \pm 0.02$, $\zeta(3)=1.00 \pm 0.02$, consistent with equation \ref{boldyrev} within one or two sigma.

\begin{figure}[t]
\includegraphics[width=\columnwidth]{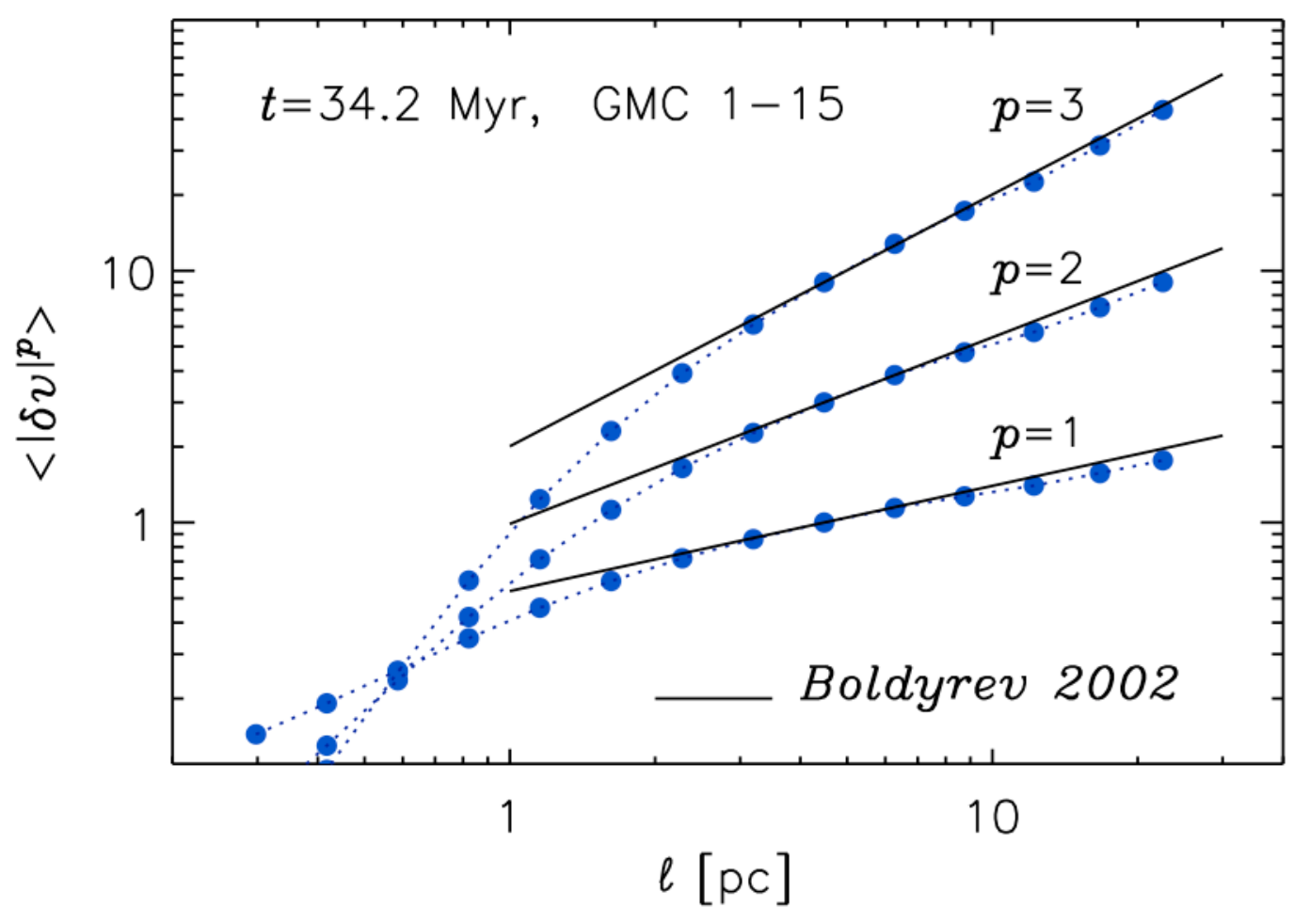}
\caption[]{Velocity structure functions obtained from the average of the structure functions of the 15 most massive clouds selected at time $t=34.2$ Myr.
The structure functions of the clouds are computed from the position and velocity of their tracer particles (see text). As in Figures \ref{structure_mesh}
and \ref{structure_mesh_2}, the solid lines are the model predictions for the structure function slopes, not fits to the data.}
\label{vel_structure_gmcs0.14}
\end{figure}

To better quantify the uncertainty of these exponents, and to illustrate their time dependence, we compute their values for each snapshot in the time interval
$t=33-56$ Myr (8 snapshots per Myr), by fitting the structure functions as a function of $\ell$ in the range between 3 and 30 pc, and plot them versus time in
Figure \ref{structure_slopes}. The two pairs of solid and dotted horizontal lines show the predicted values for incompressible (lower values) and supersonic
(higher values) turbulence (She-L{\'e}v{\^e}que and Bodyrev scaling respectively). One can see
that the values are usually within the predicted ones for the first and second order exponents, while the third order exponent is systematically above unity during
the first and last thirds of that time interval, when the SN rate and the kinetic energy appear to be a bit higher than in the middle third (see Figure \ref{fig_energies1}).
The time average of the single-snapshot values are $\zeta(1)=0.39 \pm 0.04$, $\zeta(2)=0.75 \pm 0.05$, $\zeta(3)=1.1 \pm 0.2$, well within one sigma of the
values from equation \ref{boldyrev}.

The logarithmic fluctuations in Figure \ref{structure_slopes} (the y-axis of the plot is logarithmic) are significantly larger for the third order than for the first and
second orders, meaning that deviations from the expected scaling due to the effect of SN driving is stronger for higher order statistics. Therefore, it is not convenient
to take advantage of extended self-similarity and measure the exponents by plotting the structure functions versus the third order one, instead of versus $\ell$, at least
not for the first and second order case, as their time evolution would then have much larger fluctuations, due to the larger fluctuations of the third order one.

\citet{Joung+MacLow06sn} and \citet{deAvillez07scaling} computed the velocity scaling exponents from galactic-fountain simulations and found a scaling law
consistent with Boldyrev's prediction. However, because of insufficient dynamic range below the driving scale, their simulations did not yield a power-law inertial
range, and so the exponents could be computed only relative to the third order one (taking advantage of the extended self-similarity). Therefore, their result cannot
be compared directly with the ISM velocity scaling derived from observations. Furthermore, their simulations did not reach the necessary spatial resolution to describe the
evolution of individual SN remnants and their interaction with MCs in detail, and to study the velocity scaling within individual clouds, to test if MC turbulence is
consistent with SN driving.

Because our simulation yields power-law velocity structure functions as a function of $\ell$, and the values of the exponents are the same as in previous
numerical studies of MC turbulence based on random driving with a large-scale volume force, we conclude that the use of an artificial force in those
previous studies did not result in incorrect velocity scaling, so no major corrections to IMF and SFR models based on turbulent fragmentation should be
needed. However, because deviations from the average scaling laws are found in snapshots that are very close in time to SN explosions, it may be worthwhile
investigating the possibility of minor effects on star formation resulting from such deviations. \\

\begin{figure}[t]
\includegraphics[width=\columnwidth]{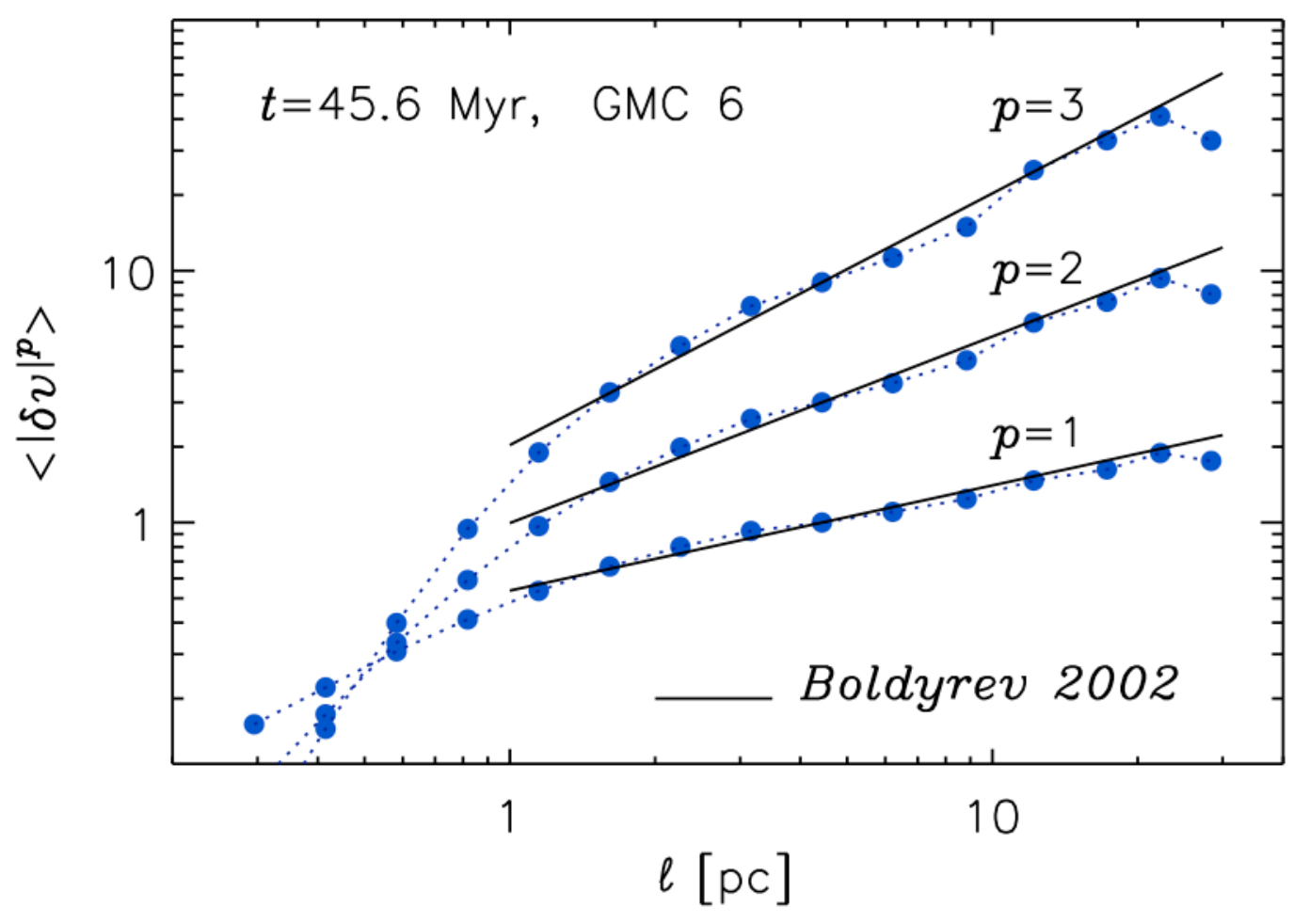}
\caption[]{Same as in Figure \ref{vel_structure_gmcs0.14}, but for a single MC, the seventh most massive one among those selected at time $t=45.6$ Myr
(MC 6 in the top panel of Figure \ref{clouds}).}
\label{vel_structure_gmc6}
\end{figure}

\subsection{Velocity Scaling within MCs}

In order to test if SN driving can generate the same velocity scaling also inside MCs, despite their density contrast, we compute the velocity structure
functions inside MCs selected from our simulation as described in \S 3. To better constrain the scaling exponents, we have computed the velocity
structure functions for the 15 most massive clouds of each snapshot. Because of the very complex cloud shapes, we find it convenient to compute the
structure functions based on the position and velocity of the tracer particles, $S^{\rm tr}_p(\ell)$ (see eq. \ref{tr}). MCs are regions of density enhancement,
so their velocity field is sampled well by the tracer particles. In fact, they contain so many tracer particles that, to speed up the calculation, we randomly
select a number of particle pairs 500 times smaller than all possible pairs in each cloud, resulting in approximately 2 to 200 million pairs per cloud.
As a further simplification, we also compute the differences of each velocity component irrespective of their orientation relative to the separation vector,
$\bs \ell$, so we do not distinguish between transversal and longitudinal structure functions.

\begin{figure}[t]
\includegraphics[width=\columnwidth]{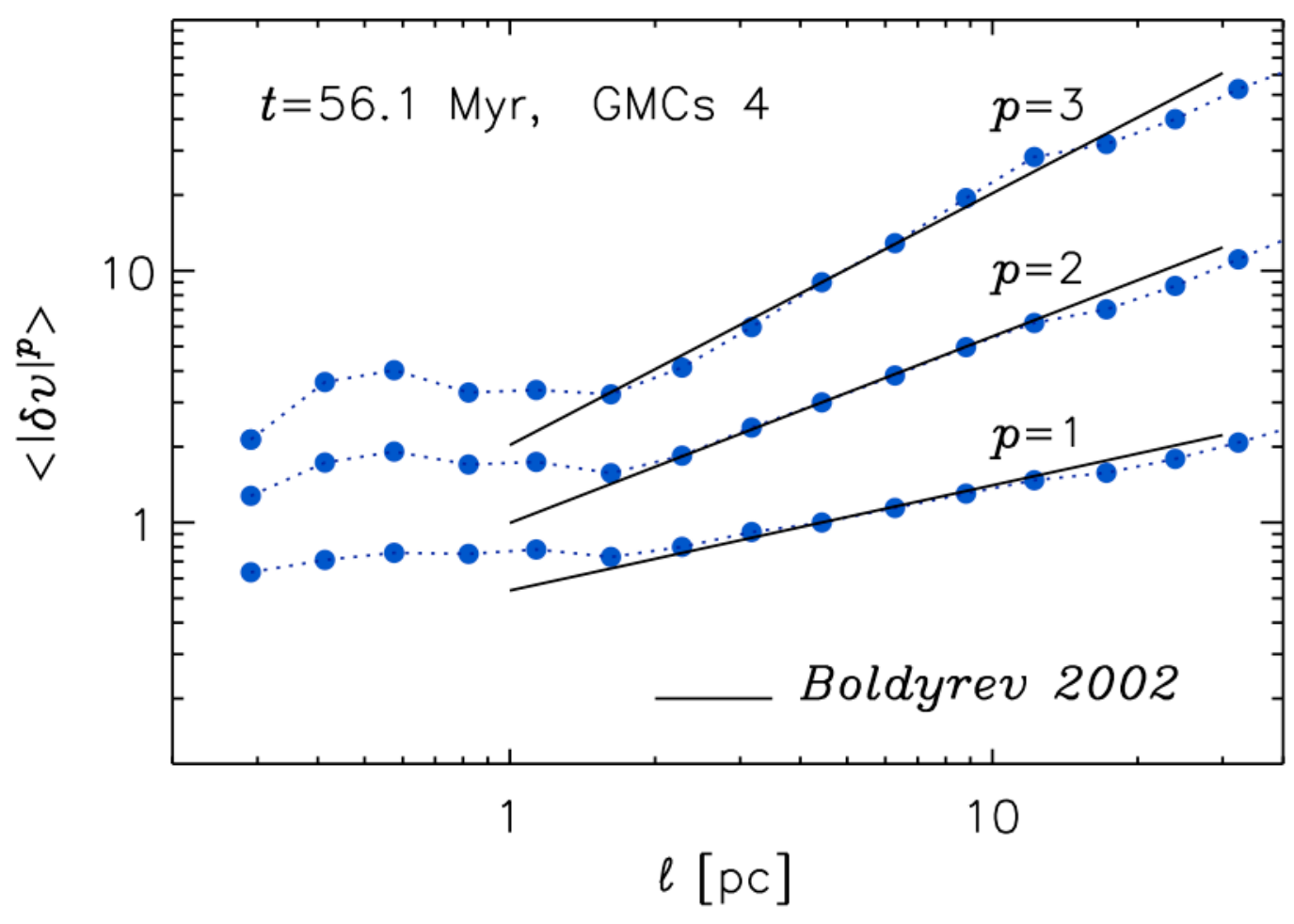}
\caption[]{Same as in Figure \ref{vel_structure_gmc6}, but for the fifth most massive MC among those selected at time $t=56.1$ Myr (MC 4 in the bottom panel of
Figure \ref{clouds}). This is an example of structure functions clearly affected by deviations at small scales due to the effect of nearby and recent SN explosions.
During the brief period of time of the early expansion of a SN bubble, excess energy is found at small or intermediate scales relative to the `undisturbed' structure
functions.}
\label{vel_structure_gmc4}
\end{figure}

We find velocity differences growing with scale up to 10-30 pc, depending on the cloud size. As an example, Figure \ref{vel_structure_gmcs0.14}
shows the average of the structure functions of the 15 most massive clouds selected at time $t=34.2$ Myr. The structure functions inside these
GMCs are consistent with the global ones and are well approximated by power laws down to a scale of approximately $10 dx=2.4$ pc.
At a separation $\ell= 4 dx = 0.96$ pc, the velocity differences are only approximately 30\% below the value extrapolated from the inertial-range
scaling, as shown by the first order structure function in  Figure \ref{vel_structure_gmcs0.14}.

In Figures \ref{vel_structure_gmc6} and \ref{vel_structure_gmc4} we show examples of structure functions of individual MCs, one with no evidence
of the direct effect of SN driving (Figure \ref{vel_structure_gmc6}), and one with significant deviations at $\ell \le 1$ pc, due to a recent nearby SN
explosion (Figure \ref{vel_structure_gmc4}).

These results show that, despite the large contrast between the MC density and the average ambient density, the kinetic energy of SN explosions is
effectively transferred into turbulence within individual clouds, where it establishes the usual scaling laws of supersonic turbulence, all the way to the smallest
scales where the simulation is affected by numerical dissipation. Real MCs also exhibit power-law velocity scaling \citep[e.g.][]{Heyer+Brunt04,Padoan+06perseus}.
The slope of $\zeta(2)=0.8\pm 0.1$ derived for the Perseus region by \citet{Padoan+06perseus} using the method by \citet{Lazarian+Pogosyan2000} is
formally consistent with the scaling laws derived here for the MCs of our simulation. On the other hand, the principle-component-analysis results by
\citet{Heyer+Brunt04} give a very large slope, $\zeta(2)=1.12\pm 0.04$, which is hard to interpret, as it is steeper than the scaling from the Burgers equation
that models an infinitely compressible flow. A careful study of the consistency between the structure function slopes of the SN-driven simulation and of
the observations requires the computation of synthetic observations and is beyond the scope of this work.

All the plots in this section adopt an arbitrary normalization of the structure functions. The actual normalization of MC turbulence, that is the velocity
dispersion of clouds of a given size, is discussed below, comparing the velocity-size Larson relation of our clouds with that from the observations. \\

\section{Virial Parameter and Cloud Structure}

\citet{Larson81} interpreted the MC velocity-size relation he discovered as due to a turbulent cascade in the ISM. Because of the very large Reynolds
number of the observed motions in the cold ISM, the development of a turbulent cascade is unavoidable, and both analytical arguments
and numerical simulations have demonstrated that Larson relations can be viewed as the natural consequence of supersonic turbulence \citep[e.g.][]{Kritsuk+13}.
Nevertheless, the velocity-size relation has also been interpreted as the consequence of the MC self-gravity \citep[e.g.][]{Solomon+87,Heyer+09},
because MC virial masses are often comparable to the masses estimated from the CO luminosity.

\begin{figure}[t]
\includegraphics[width=\columnwidth]{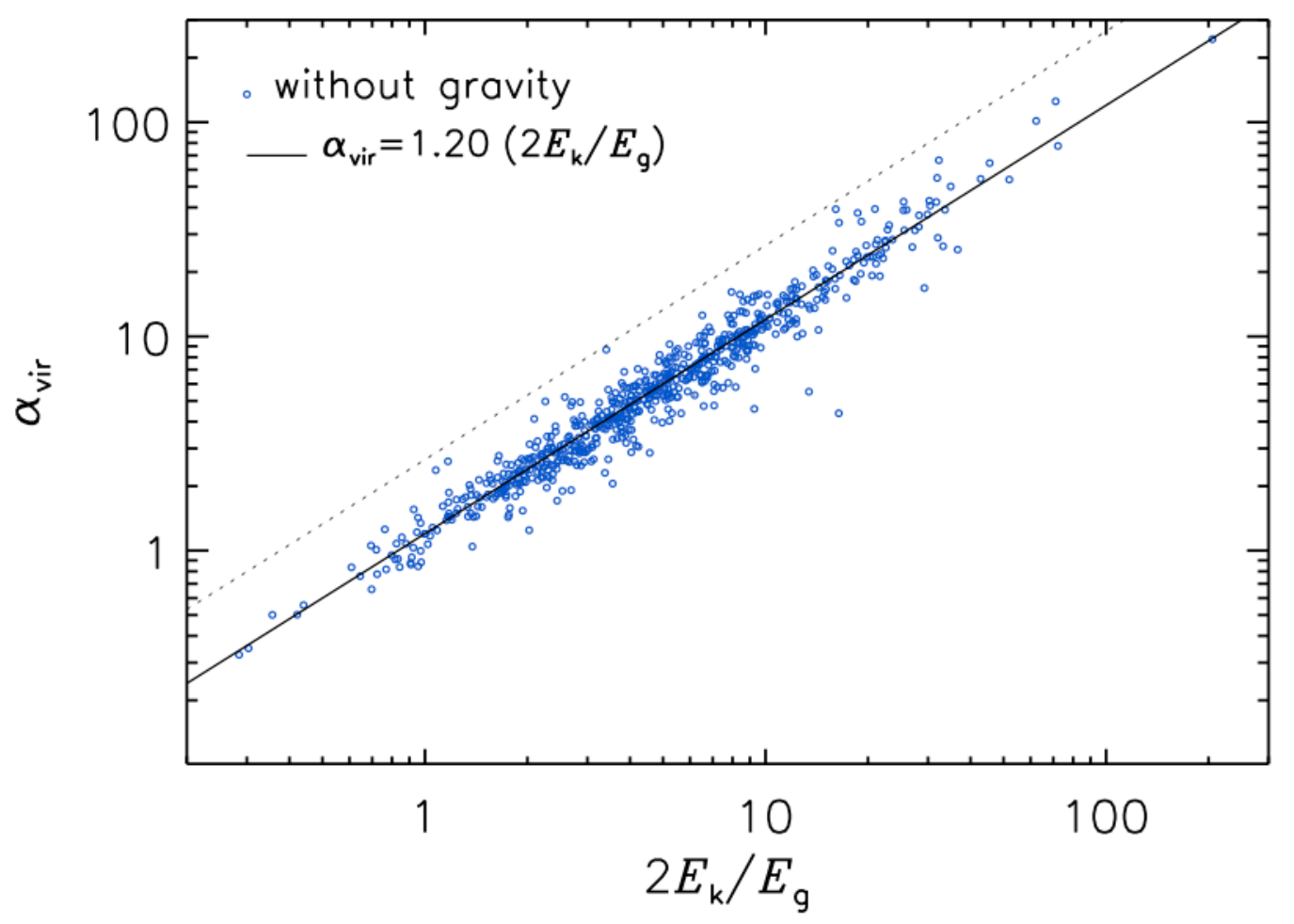}
\caption[]{Virial parameter versus energy ratio for clouds selected from three snapshots of the simulation prior to the inclusion of self-gravity and covering a
time span of 4 Myr. The solid line shows the average ratio of 1.20 between $\alpha_{\rm vir}$ and $2E_{\rm k}/ E_{\rm g}$. The dotted line marks the maximum
value of that ratio.}
\label{alpha_alpha_nog}
\end{figure}

The relative importance of turbulence and self-gravity is measured by the virial parameter, introduced by \citet{Bertoldi+McKee92}:
\begin{equation}
\alpha_{\rm vir} \equiv {5 \sigma^2_{\rm v} R_{\rm cl} \over{G M_{\rm cl}}} = {40\over 3\pi^2}\left({t_{\rm ff}\over t_{\rm dyn}}\right)^2 \sim {2 E_{\rm k}\over{E_{\rm g}}},
\label{alpha}
\end{equation}
where $\sigma_{\rm v}$ is the one-dimensional velocity dispersion, and the dynamical time is defined as:
\begin{equation}
t_{\rm dyn} = R_{\rm cl}/\sigma_{\rm v,3D}.
\label{tdyn_def}
\end{equation}
The last equality in (\ref{alpha}) is exact in the case of an idealized spherical cloud of uniform density. For more realistic cloud mass distributions,
the virial parameter is only an approximation of the ratio of kinetic and gravitational energies.

To estimate the relative importance of turbulence and self-gravity in clouds from our simulation, we have computed the virial parameter and the kinetic and
gravitational energies of clouds selected from six snapshots, three before and three after the inclusion of gravity in the simulation, using a threshold density
$n_{\rm H,min}=100$ cm$^{-3}$. The virial parameter of a cloud is measured using the positions and velocities of the tracer
particles belonging to that cloud, and defining the cloud radius as the rms of the particle positions:
\begin{equation}
R_{\rm cl} \equiv  \left[{1 \over N}\,\sum_ {i=1}^{3} \, \sum_{n=1}^{N} ( x_{i,n} - \bar{x}_i )^2  \right ]^{1/2},
\label{radius}
\end{equation}
where $\bar{x}_i \equiv  \sum_{n=1}^{N} x_{i,n} /N$ are the components of the mean particle position (the cloud's barycenter) and $N$ is the
total number of tracer particles in the cloud.

\begin{figure}[t]
\includegraphics[width=\columnwidth]{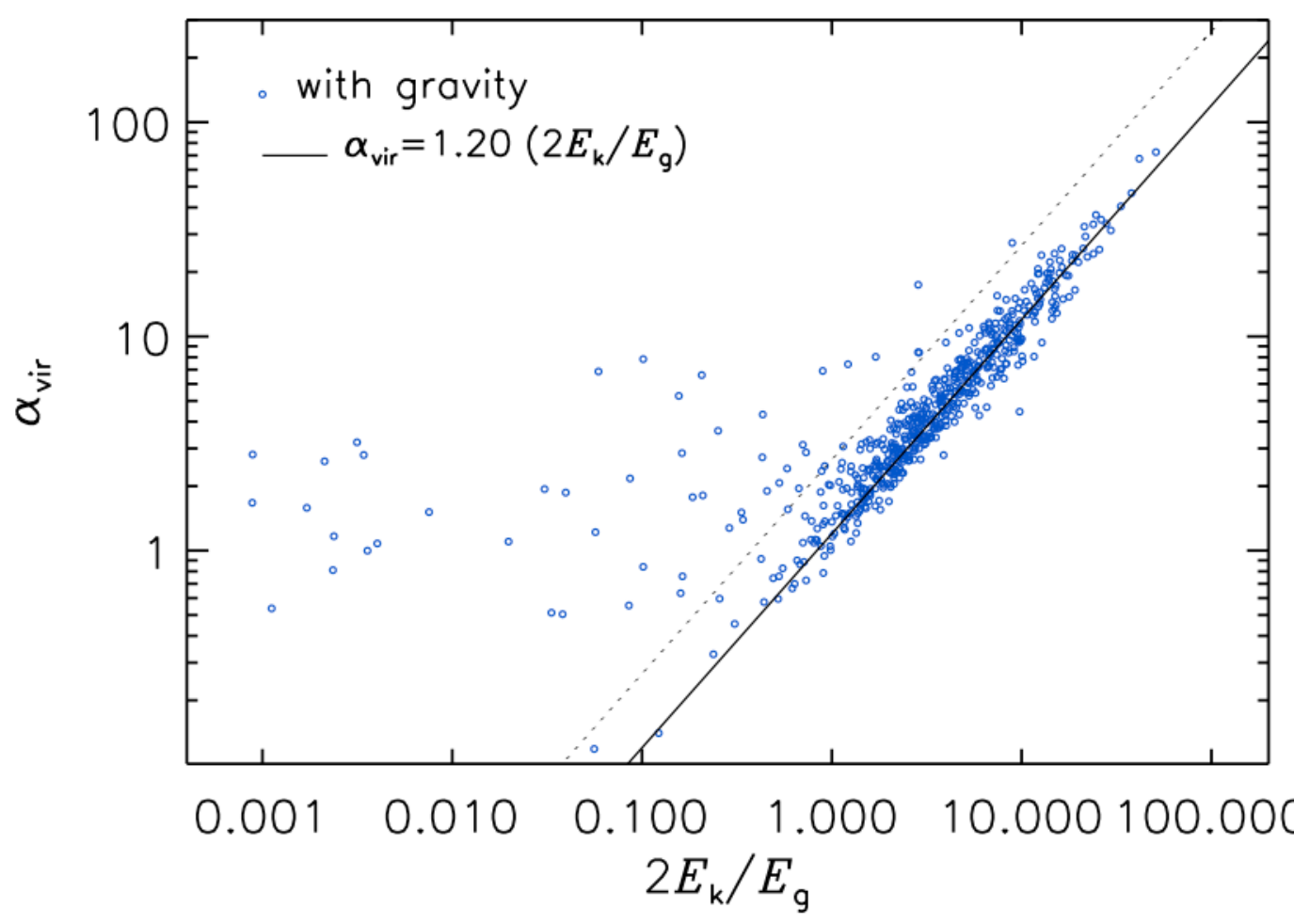}
\caption[]{Virial parameter versus energy ratio for clouds selected from three snapshots after the inclusion of self-gravity and covering the last 4 Myr of the
simulation. The solid and dotted lines are the same as in Figure \ref{alpha_alpha_nog}, showing that the average ratio of $\alpha_{\rm vir}$ and $2E_{\rm k}/ E_{\rm g}$
and its scatter are nearly unchanged for most clouds, relative to their value before the inclusion of self-gravity. Clouds with very low values of $2E_{\rm k}/ E_{\rm g}$
contain collapsing cores whose gravitational energy has been included in the computation of the cloud $E_{\rm g}$. They are star-forming clouds with relatively low
value of $\alpha_{\rm vir}$, but do not show evidence of global collapse.}
\label{alpha_alpha_g}
\end{figure}

While the virial parameter only depends on global MC properties (mass, radius and rms velocity), the ratio of kinetic to gravitational energy is
sensitive to the cloud internal structure (mass distribution and shape) and to correlations between density and velocity \citep[e.g.][]{Federrath+Klessen12}.
Thus, the comparison between $\alpha_{\rm vir}$ and $2E_{\rm k}/ E_{\rm g}$ is a useful tool to probe the internal structure of MCs and its evolution under
the effect of self-gravity. As in the case of the virial parameter, we compute $E_{\rm k}$ and $E_{\rm g}$ of a cloud using the velocities and positions of the
tracer particles in that cloud:
\begin{equation}
E_{\rm k} = {1 \over 2}\,m \sum_{i=1}^{N} u_{i}^2,
\label{Ek}
\end{equation}
\begin{equation}
E_{\rm g} = {1 \over 2}\,G\, m^2 \sum_{i=1}^{N} \sum_{j=1}^{N} {1 \over r_{ij}},
\label{Eg}
\end{equation}
where $N$ is the total number of tracer particles in the cloud, $m$ is the mass associated to a tracer particle, $u_{i}$ is the modulus of the velocity
of the $i$-th particle, and $r_{ij}$ is the distance between the $i$-th and the $j$-th particles. This expression for $E_{\rm g}$ assumes that the cloud
is isolated, as in the definition of the virial parameter. In future work, this expression should be contrasted with a formulation that accounts self-consistently
for the surrounding mass distribution by using the actual gravitational potential from the simulation \citep[e.g.][]{Federrath+Klessen12}.

Figure \ref{alpha_alpha_nog} shows the comparison between $\alpha_{\rm vir}$ and $2E_{\rm k}/ E_{\rm g}$ for clouds selected from three snapshots
covering a time span of 4 Myr before the inclusion of self-gravity. Figure \ref{alpha_alpha_g} shows the same plot, but based on three snapshots after
the inclusion of self-gravity, covering the last 4 Myr of the simulation.
On average, the clouds of Figure \ref{alpha_alpha_g} are selected 9 Myr after the
inclusion of self-gravity, while their average free-fall time, based on the density estimated as $n_{\rm H,cl}=M_{\rm cl}/(4 \pi R_{\rm cl}^3/3)$ (see Figure
\ref{density_hist}), is $4.4\pm2.1$ Myr,
and their average dynamical time, $t_{\rm dyn}\equiv R_{\rm cl}/\sigma_{\rm v,3D}$, is $2.6\pm1.5$ Myr.
Thus, self-gravity has been active for approximately two cloud free-fall times and four cloud dynamical times on average, with significant
changes occurring only in regions that have too small filling factors (e.g. small collapsing cores) to change global statistics
significantly. Indeed, gravity is not expected to be important for clouds with $\alpha_{\rm vir}> 1$. Selecting clouds with
$\alpha_{\rm vir}\le 1.0$, 0.5 and 0.25, we find a mean free-fall time of $2.1\pm1.3$ Myr, $1.5\pm1.3$ Myr and $1.2\pm1.0$ Myr
respectively. Thus, for the clouds where it should be most important, self-gravity has been active for 4 to 8 free-fall times
on average. To establish whether the absence of significant trends during this time interval will continue for the lifetime of the
various structures requires, as discussed in Section 2.1, future simulations that include the supernova feedback from the massive
stars produced by the simulation itself.

\begin{figure}[t]
\includegraphics[width=\columnwidth]{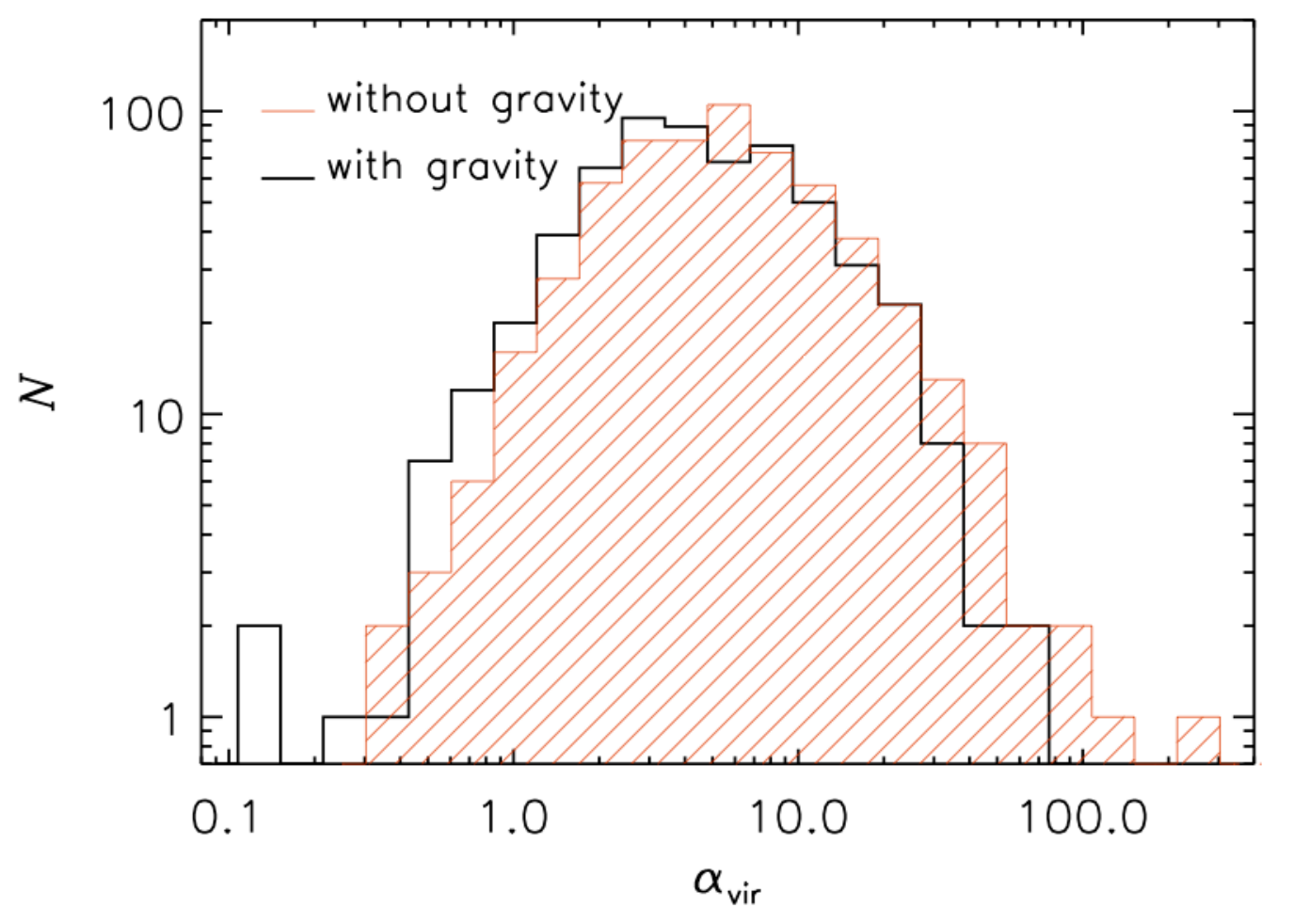}
\caption[]{Probability distributions of the viral parameter for the same three snapshots before self-gravity as in Figure \ref{alpha_alpha_nog} (shaded histogram)
and for the three snapshots after self-gravity as in Figure \ref{alpha_alpha_g} (unshaded histogram).}
\label{alpha_hist}
\end{figure}

Figure \ref{alpha_alpha_nog} shows that $\alpha_{\rm vir}$ provides a remarkably good approximation of the energy ratio, $2E_{\rm k}/ E_{\rm g}$,
despite the complexity of the cloud structure. We find the ratio $\alpha_{\rm vir} / (2E_{\rm k}/ E_{\rm g})=1.20$ on average, constant over three orders of magnitude in
$E_{\rm k}/ E_{\rm g}$ (we have verified that it is constant also over the full range of cloud masses) and with a small scatter (the standard deviation
is approximately 20\% of the mean); it is also nearly unchanged after gravity is included in the simulation. Figure \ref{alpha_alpha_g} shows that
most clouds follow approximately the same relation of $\alpha_{\rm vir}$ versus
$2E_{\rm k}/ E_{\rm g}$ as in the case without self-gravity, except that a few of them have significantly lower values of $2E_{\rm k}/ E_{\rm g}$,
because they contain collapsed cores that have also been included in the computation of the total $E_{\rm g}$. The comparison of the two figures,
without and with gravity, shows the main effect of self-gravity is to cause the collapse of dense cores inside the clouds (hence star formation), while the
cloud virial parameter is not strongly affected.

In Figure \ref{alpha_hist} we show the probability distribution of $\alpha_{\rm vir}$ for the three snapshots without gravity (shaded histogram) and with gravity
(unshaded histogram), where we have included also the star-forming clouds with very low values of $2E_{\rm k}/ E_{\rm g}$. The two distributions are very
similar, with the one including gravity slightly shifted towards lower values; the mean values are 8.5 without gravity and 6.6 with gravity. The small shift between
the two distributions shows that self-gravity causes some amount of cloud contraction, but not a significant change in global cloud structure. This is further
confirmed by the histograms of cloud density shown in Figure \ref{density_hist}, where the cloud density is defined as $n_{\rm H,cl}=M_{\rm cl}/(4 \pi R_{\rm cl}^3/3)$.
The histogram for the clouds with self-gravity is only slightly shifted to higher density, with the mean value changing from $183$ cm$^{-3}$ to $264$ cm$^{-3}$, before
and after the introduction of gravity respectively. This small increase in cloud density shows that self-gravity does not cause a global cloud collapse, even if it drives
star formation through the collapse of dense cores within MCs. Only clouds with $\alpha_{\rm vir} \lesssim 10$ contribute to star formation, according to Figure \ref{alpha_alpha_g}, with most star formation occurring in clouds with $\alpha_{\rm vir} \lesssim 3$, in agreement with recent studies of the SFR in supersonic
turbulence, showing that the SFR is mainly controlled by the virial parameter \citep{Padoan+Nordlund11sfr,Federrath+Klessen12,Padoan+12sfr}.
Future simulations, where self-gravity is active for much longer than the 11 Myr period of our run, and where selfconsistent
supernova feedback is included, will be needed to further test the above results.

\begin{figure}[t]
\includegraphics[width=\columnwidth]{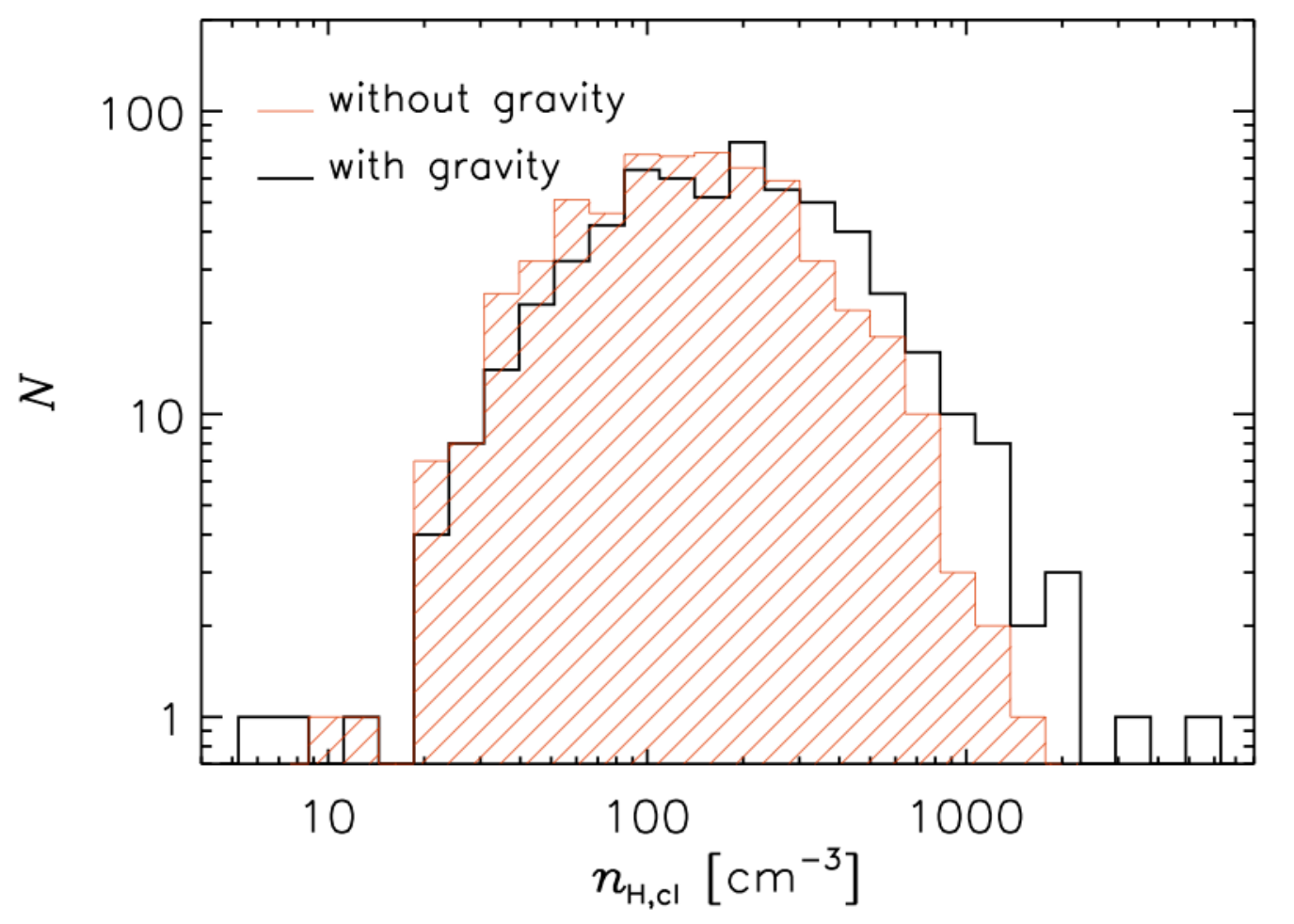}
\caption[]{Probability distributions of mean cloud density for the same three snapshots before self-gravity as in Figure \ref{alpha_alpha_nog} (shaded histogram)
and for the three snapshots after self-gravity as in Figure \ref{alpha_alpha_g} (unshaded histogram).}
\label{density_hist}
\end{figure}

\section{MC Lifetimes}

While the virial parameter estimates the relative importance of turbulence and self-gravity, the comparison of the cloud lifetime with either the dynamical time
or the free-fall time provides a definitive assessment of the actual dynamical state of clouds. For example, short-lived clouds are evidently not gravitationally
bound, while their virial parameter may be of order unity and thus would not allow to draw a definitive conclusion about their dynamical state.

Being so difficult to constrain observationally, MC lifetimes are a valuable outcome of numerical modeling. They can be measured in our simulation thanks to the
introduction of tracer particles, but only up to a maximum age of 23 Myr, the time interval with tracer particles in our simulation (11 Myr if we considered only clouds
during the time interval with self-gravity).
Although this maximum age is comparable to or smaller than the lifetime of the largest clouds, it is at least significantly
longer than the mean free-fall time (4.4 Myr)
and the mean dynamical time (2.6 Myr) of the clouds in the simulation, so it allows us to evaluate the influence of
gravity on the cloud lifetimes for most
clouds in our study.

Dobbs and Pringle (2013) measured MC lifetimes by defining the continuation of a cloud at a later time as that with the largest number of particles in common with
that cloud. The time when the number of particles (and the mass) in common drops to half or less of that in the original cloud marks the end of the cloud lifetime.
Cloud precursors and formation times are defined in the same way, by considering the clouds at earlier times with the largest number of particles in common with
the original cloud. As pointed out by the authors, this method has the drawback that clouds precursors or continuations are often not found with more than half of
the original mass because of changes in density contours, rather than a real cloud dispersion. For example, based on this method, a cloud may seem to have
dispersed after a certain time, but may later reappear.

After verifying in our own simulation that this lifetime definition is indeed very uncertain, we have chosen to estimate cloud lifetimes with a different method that is
independent of the specific density contours of clouds in past and future snapshots. Because the formation of a cloud implies converging flows (even
in the absence of self-gravity), and its dispersion requires diverging flows, we simply define the lifetime of a cloud as the time interval during which the cloud
radius, defined always by all the tracer particles belonging to that cloud, is within a factor of two from the radius at the time the cloud is selected. In other words,
we follow the cloud tracer particles in the past, until their radius has doubled in size, which marks the formation time of the cloud, and in the future also until the
radius has doubled in size, which marks the dispersion time of the cloud.

\begin{figure}[t]
\includegraphics[width=\columnwidth]{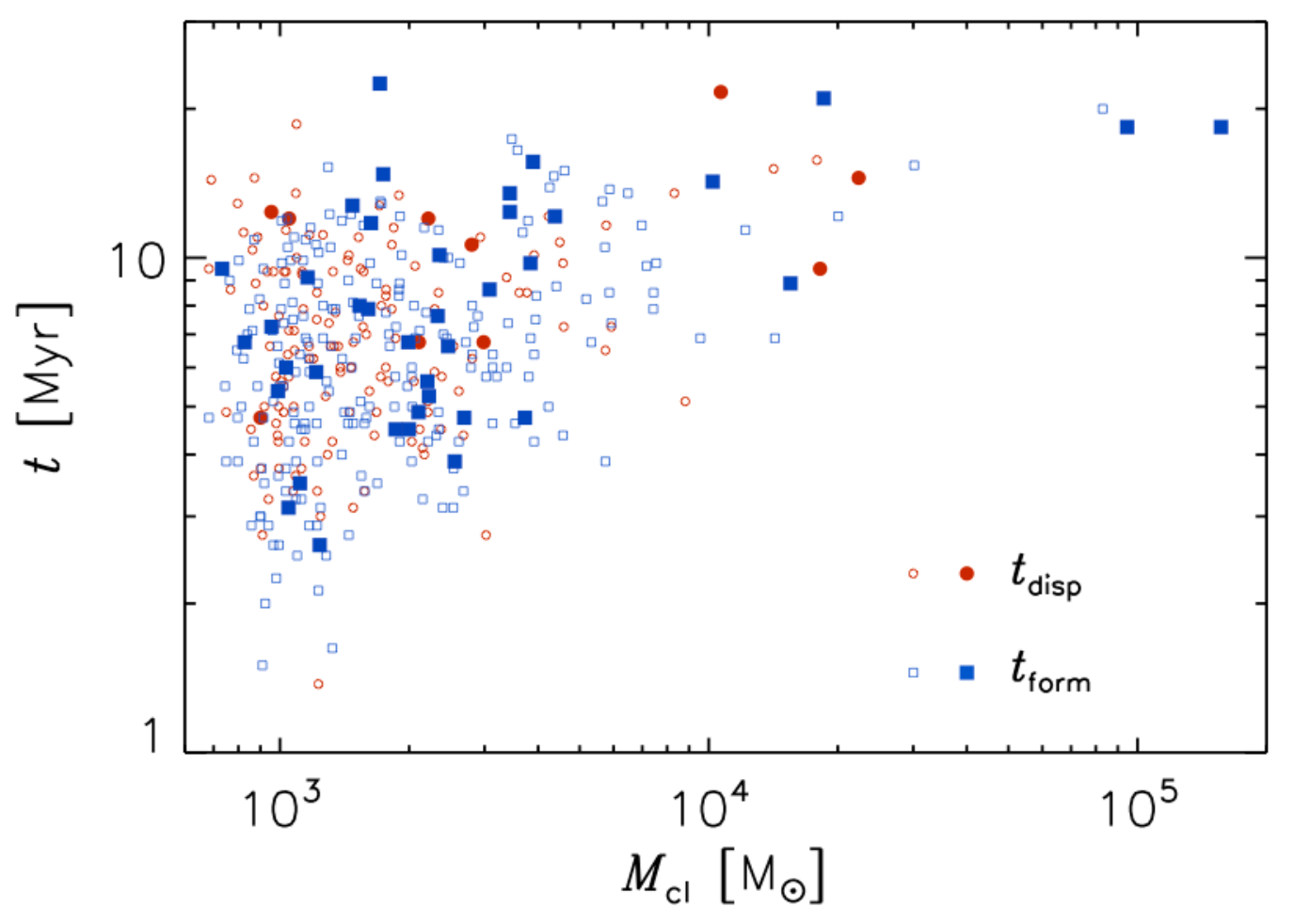}
\caption[]{Dispersion time (empty circles) and formation time (empty squares) for clouds selected from 12 snapshots of our simulation, uniformly distributed at
intervals of 2 Myr, to cover the whole time interval with tracer particles. Filled circles mark the values of dispersion time of clouds selected from the first
snapshot of the series; filled squares mark the values of formation time of clouds from the last snapshot.}
\label{t_form_death}
\end{figure}

Figure \ref{t_form_death} shows the formation time, $t_{\rm form}$ (interval between cloud formation and time of cloud selection), and dispersion time,
$t_{\rm disp}$ (interval between time of cloud selection and cloud dispersion) versus the cloud mass, using all the clouds more massive than approximately
700 M$_{\odot}$ from each of 12 snapshots covering the whole time interval with tracer particles (the time between snapshots is 2 Myr). The sum of
the two times gives the cloud lifetime, $t_{\rm life} = t_{\rm form}+t_{\rm disp}$, but we have plotted the two times separately (empty squares and circles)
because in many cases we can identify the cloud formation time and not its dispersion time (for clouds selected towards the end of the simulation),
or vice-versa (for clouds selected shortly after the introduction of tracer particles). The average values of $t_{\rm form}$ and $t_{\rm disp}$ are biased by
the large number of upper limits (not plotted to avoid confusion). In order to obtain a nearly unbiased estimate of the average values, we
consider only clouds selected from the first and the last snapshots of the series (filled circles and filled squares, respectively, in Figure \ref{t_form_death}).
The first snapshot has 12 clouds more massive than 700 M$_{\odot}$, yielding
10 measured values of $t_{\rm disp}$ and only two upper limits; the last snapshots has 40 clouds more massive than 700 M$_{\odot}$, 39 of which with a measured
value of $t_{\rm form}$ and only one with an upper limit to $t_{\rm form}$. Based on these measurements alone, we find $\langle t_{\rm disp} \rangle=11.1$ Myr
and $\langle t_{\rm form} \rangle=9.2$ Myr. By assuming that clouds are selected at a random moment of their lifetime, we should expect
$\langle t_{\rm life} \rangle  = 2\,\langle t_{\rm form}\rangle = 2\,\langle t_{\rm disp} \rangle$, and so we can average together all the 49 values and obtain
$\langle t_{\rm life}\rangle = 21.4$ Myr.

\begin{figure}[t]
\includegraphics[width=\columnwidth]{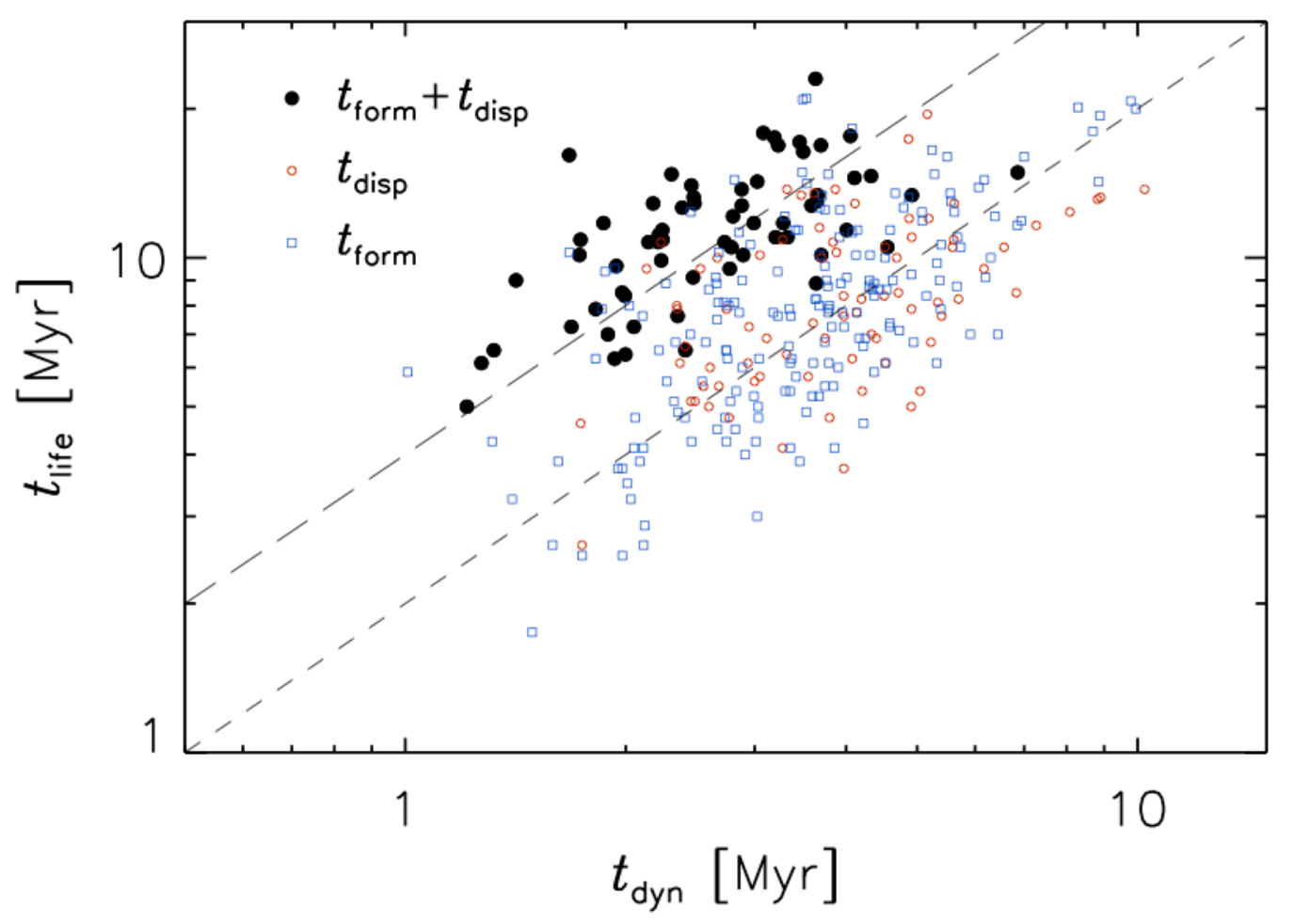}
\caption[]{Cloud lifetime versus cloud dynamical
time for clouds with measured dispersion and formation times (filled circles). Clouds with only dispersion or
formation times are plotted as empty circles and empty squares respectively.
The dashed line corresponds to $ t_{\rm life}=2\, t_{\rm dyn}$, the long-dashed line to
$t_{\rm life}=4\, t_{\rm dyn}$.}
\label{t_cross}
\end{figure}

\begin{figure}[t]
\includegraphics[width=\columnwidth]{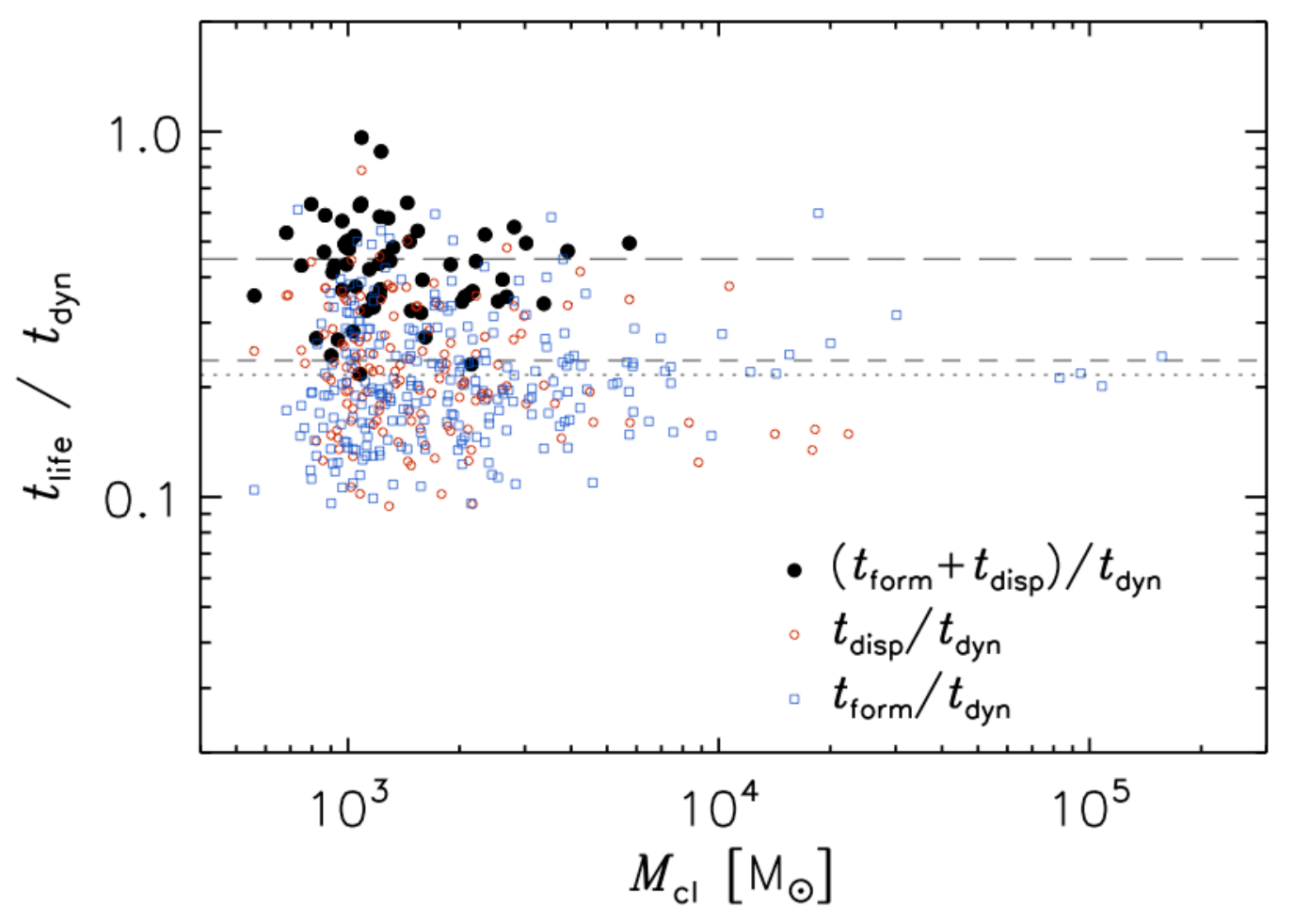}
\caption[]{Ratio of lifetime to dynamical
time versus mass for the same clouds as in Figure \ref{t_cross}. The long-dashed line shows the mean ratio for the clouds
with both dispersion and formation times measured (filled circles), $\langle (t_{\rm form}+t_{\rm disp})/t_{\rm dyn} \rangle = 4.5$.
The dotted and the dashed lines the mean ratio for clouds with only formation or dispersion times respectively,
$\langle t_{\rm form}/t_{\rm dyn} \rangle = 2.2$, $\langle t_{\rm disp}/t_{\rm dyn} \rangle = 2.4$. All these average values are 
consistent with $\langle t_{\rm life}/t_{\rm dyn} \rangle \approx 4$, independent of cloud mass.}
\label{t_life}
\end{figure}

This lifetime is based on clouds covering over two orders of magnitude in mass, and only on 39 measurements.
We can further improve our estimate of MC lifetimes by using all measurements and, at the same time, derive
the mass dependence of the lifetimes, if we normalize the lifetime to the cloud dynamical
time, $t_{\rm dyn}$. Figure \ref{t_cross} plots $t_{\rm life}$ versus $t_{\rm dyn}$ of 64 clouds for which we have measured values of both
$t_{\rm form}$ and $t_{\rm disp}$ (filled circles), besides the formation and dispersion times for all the other clouds where these are measured (empty squares
and circles respectively). The plot in Figure \ref{t_cross} shows an approximately linear correlation between $t_{\rm life}$ and $4\,t_{\rm dyn}$.
The mean and standard deviation of the ratio of the two times are
\begin{equation}
t_{\rm life} / t_{\rm dyn} = 4.5 \pm 1.4.
\label{tlife_over_tdyn}
\end{equation}
As shown by Figure \ref{t_life}, this estimate
is derived from clouds with $M_{\rm cl} \lesssim 6 \times 10^3$ M$_{\odot}$, and thus it should be considered only as an extrapolation when applied to more
massive MCs. Nevertheless, this result for the $t_{\rm life} / t_{\rm dyn}$
ratio is consistent with the measured
formation times of the most massive MCs in the simulation. Figure \ref{t_life} shows that the four most massive GMCs, with masses around
$10^5$ M$_{\odot}$, have $\langle t_{\rm form}/t_{\rm dyn}\rangle$ close to two, which would imply
$\langle t_{\rm life}/t_{\rm dyn} \rangle \approx 4$, in the absence of selection biases due to the limited time interval covered by the tracer particles.

Furthermore, while we cannot use all the values of $t_{\rm form}$ and $t_{\rm disp}$ from Figure \ref{t_form_death} to estimate an unbiased average
lifetime from $\langle t_{\rm life} \rangle \approx 2 \langle t_{\rm form}\rangle \approx 2 \langle t_{\rm disp} \rangle$, we can still use all of the corresponding
ratios, $t_{\rm form}/t_{\rm dyn}$ and $t_{\rm disp}/t_{\rm dyn}$, to estimate an unbiased average ratio of lifetime to dynamical time from
$\langle t_{\rm life}/t_{\rm dyn} \rangle \approx 2 \langle t_{\rm form}/t_{\rm dyn}\rangle \approx 2 \langle t_{\rm disp}/t_{\rm dyn} \rangle$,
if we assume that this ratio is independent of cloud mass. Indeed, the dashed and dotted lines in Figure \ref{t_life} show that
$\langle t_{\rm form}/t_{\rm dyn}\rangle \approx  \langle t_{\rm disp}/t_{\rm dyn} \rangle \approx 2$, using values over the whole mass range,
consistent with the estimate $\langle t_{\rm life}/t_{\rm dyn} \rangle \approx 4$,
based on clouds with $M_{\rm cl} \lesssim 6 \times 10^3$ M$_{\odot}$.

This result shows that both the formation and the dispersion of the MCs in our sample take two dynamical times, on average.
This is an indication that both the formation and the dispersion of the MCs in our sample  is controlled by the turbulence, 
with little influence of self-gravity. Because of the non-negligible scatter in the ratio of cloud lifetime
to dynamical time, one may expect that at least the clouds with the largest ratios may have longer lifetime due to their self-gravity.
This is not the case: we have
verified that there is actually a positive correlation between
$t_{\rm life}/t_{\rm dyn}$
and $\alpha_{\rm vir}$,meaning that larger values of
$t_{\rm life}/t_{\rm dyn}$
are usually due to smaller values of
$t_{\rm dyn}$
because of larger $\sigma_{\rm v}$ (hence larger $\alpha_{\rm vir}$), rather than longer $t_{\rm life}$ as a consequence
of a lower $\alpha_{\rm vir}$. Thus, there is no significant
imprint of self-gravity in the cloud lifetimes, even if more than half of our clouds are selected at a time after
self-gravity has been included in the simulation.
Future simulations, where selfconsistent supernova
feedback allows longer runs with selfgravity, are needed to test if this lack of significant imprint of
selfgravity continues, and if it extends to MCs with longer lifetimes and to higher surface density MCs.

We should also
stress the caveat that, for the most massive MCs of $\sim 10^5$ M$_{\odot}$ we could only measure formation times, and not dispersion times,
due to their long lifetime (and dynamical time) and the limited
duration of the simulation. Thus, we cannot rule out that, at least the most massive MCs, could have dispersion times significantly longer than
two dynamical times.
However, that would imply dispersion times longer than 20 Myr (lifetimes longer than 40 Myr) for such clouds, a timescale over which
the extra energy injection from SN explosions of locally formed massive stars would presumably succeed in dispersing the clouds, even if the general ISM
turbulence could not (see discussion in \S 2.1).

To derive actual values of cloud lifetime as a function of cloud mass, taking advantage of our result
(\ref{tlife_over_tdyn}),
we can use the expression
(\ref{tdyn_Mcl})
for the average cloud dynamical time derived in \S 10.4, which gives an average cloud lifetime of
\begin{equation}
t_{\rm life} = 22.5 \, {\rm Myr}\, (M_{\rm cl}/10^4 {\rm \,M}_{\odot})^{0.25}.
\label{tlife_versus_Mcl}
\end{equation}

\section{Magnetic Field in MCs and MC Formation}

Our simulation adopts a mean magnetic-field strength consistent with the Galactic one (see \S 2), so the magnetic field inside clouds selected
from the simulation should be comparable to that in real MCs. We have already shown in Figures \ref{fig_energies1} and \ref{fig_energies2} that the mean
magnetic energy is not far from equipartition with the mean thermal and kinetic energies averaged over the whole volume,
while the energy ratios are much larger in the dense gas. This clear energy separation in dense gas, with $\langle E_{\rm k,d}/E_{\rm m,d} \rangle=25.1$ and
$\langle E_{\rm m,d}/E_{\rm th,d}\rangle=9.8$, is the necessary consequence of the near equipartition at the largest scales. Being only mildly super-Alfv\'{e}nic,
large-scale compressive motions cannot compress the mean magnetic field by a large factor, so the density enhancement of MCs is largely achieved with
compressions along field lines, resulting in a mean magnetic field strength in the dense gas not much larger than the total mean field. The mean magnetic field
of 4.6 $\mu$G is amplified by the SN-driven turbulence to an rms value of 7.2 $\mu$G, averaged over the whole volume and between $t=33$ and 56 Myr. The
rms field strength in the dense gas is 12.8 $\mu$G, not even a factor of two larger.

\begin{figure}[t]
\includegraphics[width=\columnwidth]{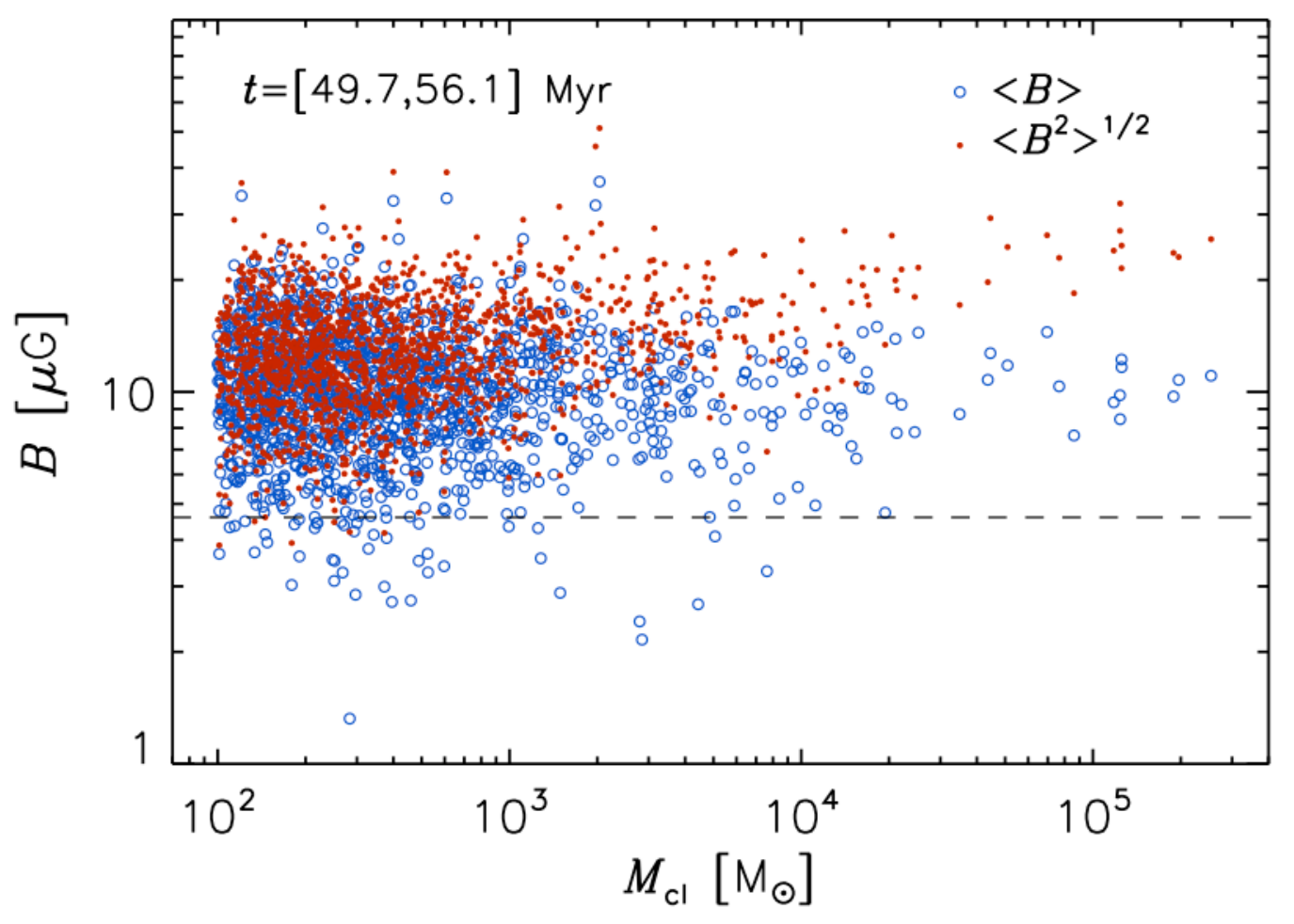}
\caption[]{Magnetic field strength versus cloud mass for the same sample as in the Larson relations of Figures \ref{larson_vel} and \ref{larson_mass}.
The dashed line shows the mean magnetic field averaged over the whole computation volume (also the initial mean field). Empty circles correspond to
the mean value of the magnetic field of all tracer particles in each cloud, while filled circles give the rms value.}
\label{B}
\end{figure}

To investigate the role of the magnetic field in individual MCs, we consider the same catalog of 1,547 clouds as in the comparison with the observations
discussed in the next section. The clouds are selected from 7 snapshots during the final 6 Myr of the simulation, at a resolution of 0.49 pc and with a
density threshold of $n_{\rm H,min}=200$ cm$^{-3}$. We compute both the mean and the rms magnetic field of each cloud using the values sampled by the
tracer particles, $\langle B \rangle = \Sigma_{\rm i}B_{\rm i}/N$ and $\langle B^2 \rangle^{1/2} = (\Sigma_{\rm i}B^2_{\rm i}/N)^{1/2}$, where $B_{\rm i}$ is the magnetic
field strength sampled by the particle i in a given cloud, and N is the total number of particles in that cloud. These magnetic field values are plotted versus cloud mass
in Figure \ref{B}, where the horizontal dashed line represents the mean magnetic field in the computational volume, $B_0=4.6$ $\mu$G. The mean field in the clouds is
approximately 10 $\mu$G on the average, only twice larger than $B_0$, and independent of cloud mass. We have verified that the mean magnetic field strength
of the clouds is also independent of their mean gas density.

The relatively small increase of the cloud mean magnetic field relative to $B_0$ and its independence of gas density are characteristic
of trans-Alfv\'{e}nic supersonic turbulence \citep{Padoan+Nordlund97MHD,Padoan+Nordlund99MHD}, and further illustrates that MCs
must be formed by compressive motions primarily along magnetic field lines, due to the non-negligible magnetic pressure prior to the
compression and cooling of the low-density gas. As the gas is being compressed into a nascent MC by random large-scale motions,
the increasing density and decreasing cooling time cause a drop in both the Alfv\'{e}n and sound speeds \citep{Padoan+10_Como}. As a result,
the turbulence within a MC is super-Alfv\'{e}nic and highly supersonic, while the larger-scale flows responsible for its formation are trans-Alfv\'{e}nic and
mildly supersonic. Because in super-Alfv\'{e}nic  turbulence the magnetic field is amplified by compressions, as shown by a positive
$B-n$ correlation \citep{Padoan+Nordlund97MHD,Padoan+Nordlund99MHD}, dense cores formed by shocks within MCs \citep{Padoan+01cores}
have an enhanced magnetic-field strength on average. Furthermore, cores are topologically the ultimate zero-dimensional destination of
a fluid element undergoing compression, as they can be viewed as the intersection of filaments that are formed by the intersection of
postshock sheets. Much of the flow turbulent energy is dissipated by the time it `stagnates' into a core. Due to this drop in turbulent energy,
together with the increase in magnetic-field strength and density, the turbulence inside dense cores is trans-Alfv\'{e}nic and trans-sonic.
In summary, super-Alfv\'{e}nic and supersonic MC turbulence is the natural consequence of large-scale trans-Alfv\'{e}nic trans-sonic
turbulence and also the natural origin of small-scale trans-Alfv\'{e}nic trans-sonic turbulence in prestellar cores.

\begin{figure}[t]
\includegraphics[width=\columnwidth]{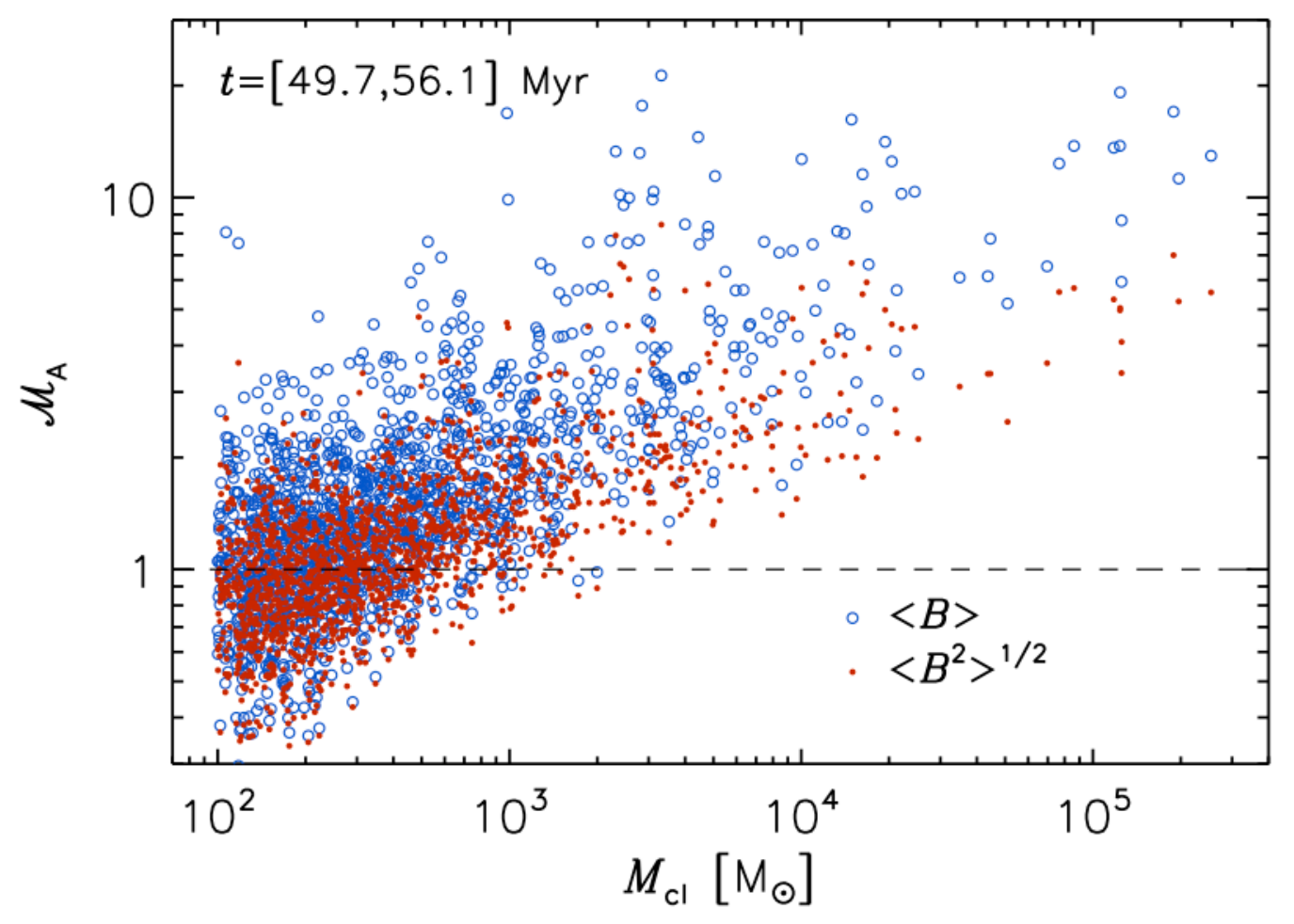}
\caption[]{Alfv\'{e}nic rms Mach number versus cloud mass for the same clouds as in Figure \ref{B}, computed with the cloud mean magnetic field (empty circles),
or the cloud rms magnetic field (filled cirlces).}
\label{Mach}
\end{figure}

The small-scale enhancement of the magnetic field within MCs is partly illustrated in Figure \ref{B} by the values of the rms field in the clouds
that is approximately a factor of two larger than the mean field in the most massive clouds. As a more direct demonstration of the
super-Alfv\'{e}nic nature of MC turbulence, Figure \ref{Mach} shows the cloud rms Alfv\'{e}nic Mach number versus the cloud mass.
The Mach number is computed as the ratio of the cloud rms velocity and the cloud Alfv\'{e}n velocity, where the latter is computed either
with the mean magnetic field (empty circles) or with the rms magnetic field (filled circles), and using the mean density sampled by the tracer
particles. Nearly all clouds with mass larger than $10^3$ M$_{\odot}$ are super-Alfv\'{e}nic, even considering their amplified field strength.
For the 41 GMCs with masses larger than $10^4$ M$_{\odot}$, the average Alfv\'{e}nic Mach number is 8.3 with respect to the mean field,
and 3.9 with respect to the rms field.

The super-Alfv\'{e}nic nature of the turbulence in the clouds from our simulation is consistent with the observational evidence. Based on the comparison
between simulations of MHD turbulence and MC observations, \citet{Padoan+Nordlund97MHD,Padoan+Nordlund99MHD} suggested that MC turbulence
was better characterized by supersonic turbulent flows with ${\cal M}_{A}\gg 1$ than flows with  ${\cal M}_{A}\approx 1$. This result was later confirmed with the
aid of synthetic observations \citep{Padoan+04power} and synthetic Zeeman splitting measurements \citep{Lunttila+2008,Lunttila+2009}. Taking advantage
of the anisotropy of MHD turbulence, \citet{Heyer+Brunt2012} demonstrated that the densest regions of the Taurus MC complex are characterized by
super-Alfv\'{e}nic turbulence, while in low density regions the motions are sub or trans-Alfv\'{e}nic, also consistent with the picture from our simulation, where
MCs are formed by large-scale trans-Alfv\'{e}nic turbulence, and thus fed preferentially by motions along magnetic field lines, as discussed above
\citep{Nordlund+Padoan03,Padoan+10_Como}.

\begin{figure}[t]
\includegraphics[width=\columnwidth]{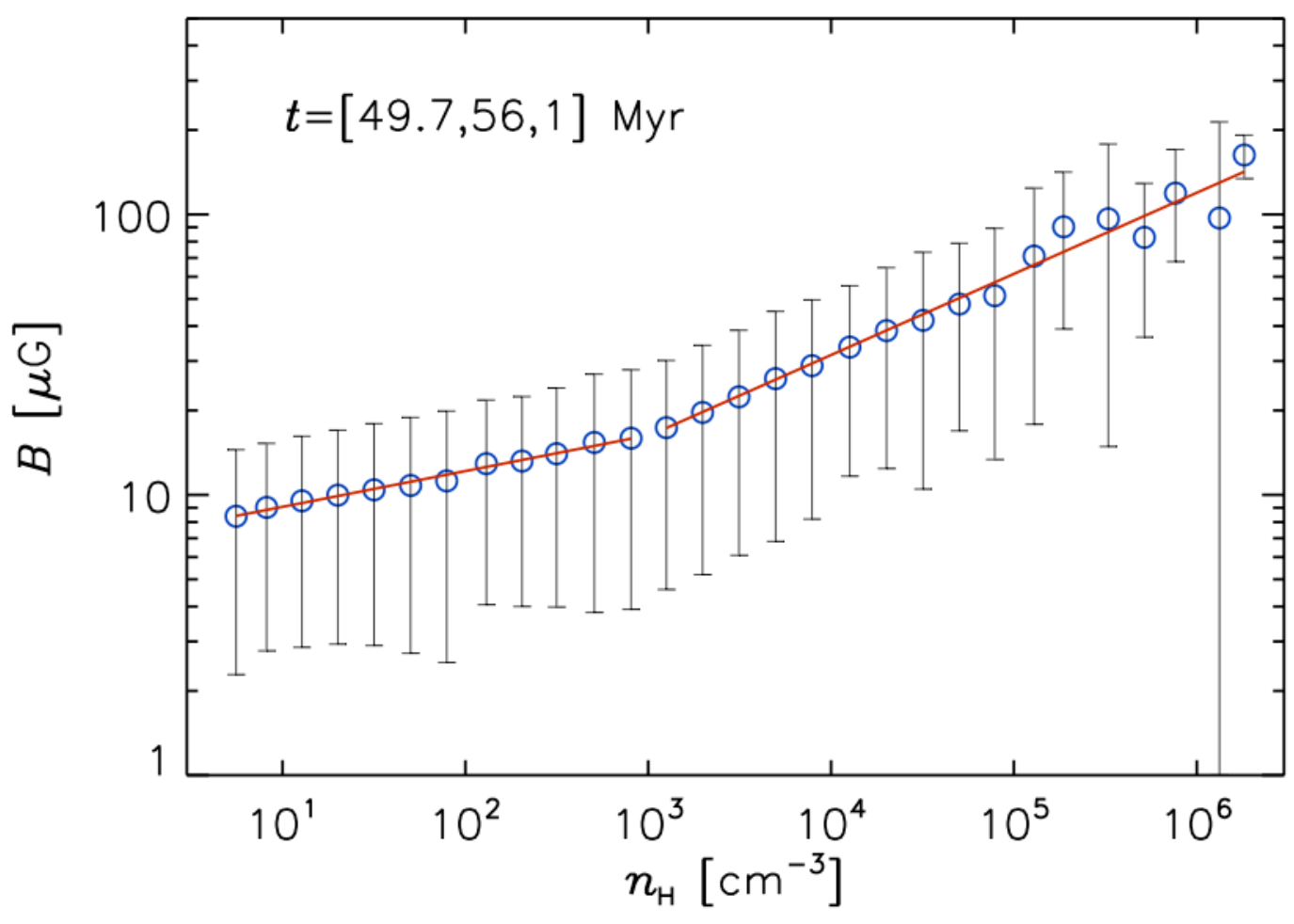}
\caption[]{$B-n$ relation obtained from averaging the $B-n$ relations of all 1563 clouds in our sample. The circles correspond to the mean value of
$n_{\rm H}$ and $B$ in each density bin, while the error bars mark the values two standard deviations above and below the mean, to illustrate the
scatter in the $B-n$ relation. The $B-n$ relation of an individual cloud is computed using the values of $B$ and $n_{\rm H}$ of all the tracer particles in
that cloud. Because the clouds are selected as regions with density $n_{\rm H}>100$ cm$^{-3}$, no tracer particles belonging to a cloud has density lower
than that. To extend the relation to densities $n_{\rm H}<100$ cm$^{-3}$, we define a cloud volume delimited by the smallest and largest coordinates
of the tracer particles of that cloud, and include the values of $B$ and $n_{\rm H}$ of each cell of that volume to compute the relation. The two solid lines
are least-square fits for densities $n_{\rm H} < 10^3$ cm$^{-3}$ and  $n_{\rm H} > 10^3$ cm$^{-3}$, giving $B \sim n_{\rm H}^{0.13}$ and
$B \sim n_{\rm H}^{0.29}$, respectively.}
\label{bn}
\end{figure}

To further characterize the cloud turbulence, we have computed the $B-n$ relation inside all the clouds of our sample, using again the values
of $B$ and $n$ of the tracer particles inside the clouds. We divide the density in logarithmic bins, and compute the mean magnetic field strength
and its standard deviation in each density bin. We then average these values among all the clouds, using weights proportional to the number
of tracer particles in the density bins of each cloud. Figure \ref{bn} shows this mean $B-n$ relation. We also illustrate the large scatter by
plotting error bars that correspond to twice the standard deviation above and below the mean values. The two solid lines are least-square
fits for $n_{\rm H} < 10^3$ cm$^{-3}$, $B \sim n_{\rm H}^{0.13}$, and for $n_{\rm H} > 10^3$ cm$^{-3}$, $B \sim n_{\rm H}^{0.29}$.

The stronger dependence of the magnetic-field strength on density at $n_{\rm H} > 10^3$ cm$^{-3}$ than at lower density is qualitatively
consistent with the observations \citep{Crutcher+10}. The slope we derive is much smaller than that derived by \citet{Crutcher+10} at high densities,
$B \sim n_{\rm H}^{0.65}$. However, their slope does not refer to the mean magnetic field at a given density, but to its maximum value. Considering the
large number of measured upper limits well below such maximum values, the dependence of the mean $B$ on density could be significantly shallower
than the estimated slope of the upper envelope of the $B-n$ relation. Furthermore, the Bayesian analysis by \citet{Crutcher+10} assumes a uniform
distribution of the magnetic field strength, while this distribution is exponential  in super-Alfv\'{e}nic turbulence \citep{Padoan+Nordlund99MHD}
(we have verified it is exponential also in our clouds). Finally, and most importantly, the $B-n$ relation in Figure \ref{bn} is not computed for a selection
of dense cores, as in the observations, but using every single tracer particle in the cloud, so it should not be compared quantitatively to the observational
$B-n$ relation. Such a comparison would require synthetic Zeeman observations of a selection of dense cores, as in \citet{Lunttila+2008,Lunttila+2009}.
It would also require higher spatial resolution, because most of the observed cores with the largest detected magnetic-field strengths, at densities of order
$10^5-10^7$ cm$^{-3}$, have sizes substantially smaller than the spatial resolution of our simulation. The higher resolution would also allow to better resolve
the dynamo amplification in dense cores \citep{Federrath+11}, which would tend to increase the slope of the $B-n$ relation.


\section{Comparison with Outer-Galaxy MCs}

To further test our results, we carry out a comparison of the properties of our MCs with those of observed MCs.
This is a preliminary approach based on the derivation of projected quantities, such as column density, equivalent
radius and line-of-sight velocity dispersion. Follow-up studies with synthetic observations taking into consideration chemistry and
radiative transfer are also needed.

Our observational sample of choice for this comparison is the MC catalog by \citet{Heyer+01}, extracted from a decomposition of the $^{12}$CO
FCRAO Outer Galaxy Survey \citep{Heyer+98}. Besides the large dynamic range of the survey, its main advantage is that for the Outer Galaxy there is
no blending of emission from separate MCs along the line of sight, or at least the problem is strongly mitigated compared with the Inner Galaxy. As a result,
a very large number of clouds can be reliably selected over a broad range of cloud masses and sizes. The catalog contains a total number of 10,156 objects,
up to a mass of approximately $8\times 10^5$ M$_{\odot}$ and a size of 45 pc. It is estimated to be complete down to a mass of approximately 600 M$_{\odot}$
and a cloud size of 3 pc.

Inner-Galaxy MC catalogs are far less reliable and complete because of velocity blending, so they are not suitable for the comparison we
pursue here. For example, the recent catalog of Inner-Galaxy MCs \citep{Rathborne+09,Roman-Duval+09} extracted from the UB--FCRAO
Galactic Ring Survey \citep{Jackson+06} contains objects between 1 and $10^6$ M$_{\odot}$, but is estimated to be complete only above
$4\times 10^4$ M$_{\odot}$  \citep{Roman-Duval+09}. We suspect a more realistic completeness limit may be $2\times 10^5$ M$_{\odot}$, because the
mass distribution is a power law only above that mass \citep{Roman-Duval+09}, which corresponds to a size limit of approximately 20 pc, judging from
the mass-size relation and from the lack of a power law in the size distribution below 20 pc. This severe incompleteness suggests that the cloud
surface density may be overestimated by a large factor. The MC mass distribution is expected to be a power law down to small masses, so the abrupt
departure from a power law below $2\times 10^5$ M$_{\odot}$ indicates that much of the missing mass from smaller clouds is incorrectly assigned to
larger ones due to velocity blending. The failure to select individual three-dimensional clouds is also demonstrated by the absence of a velocity-size
correlation and, possibly, by the extremely low values of the cloud virial parameters (the distribution peaks at $\alpha_{\rm vir}\approx0.2$), which would
imply a larger star formation rate and a stronger signature of global collapse than observed. In summary, the differences between Galactic-Ring and
Outer-Galaxy MCs may not be as large as often assumed.
Thus, although
our comparison is primarily with Outer-Galaxy MCs, the results may be applicable to
Galactic clouds in general.

\begin{figure}[t]
\includegraphics[width=\columnwidth]{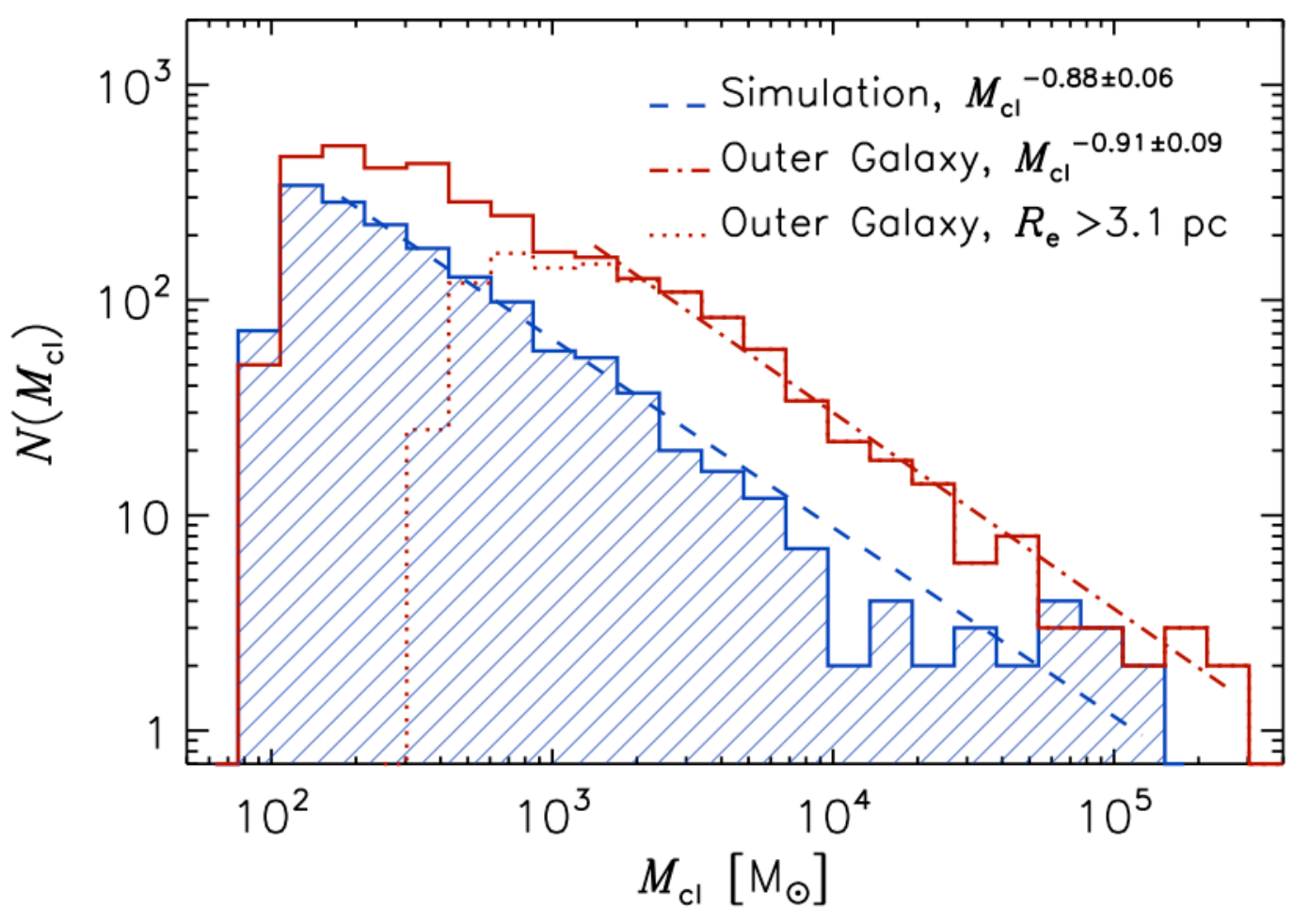}
\caption[]{Probability distribution of cloud mass from the sample of 1,547 clouds of our highest-resolution catalog ($512^3$ cells, or 0.49 pc) and with the
threshold density that best matches the observed mass-size relation, $n_{\rm H,min}=200$ cm$^{-3}$ (shaded histogram).
The dashed line is the result of a least-square fit yielding a slope of $-0.88\pm 0.06$. The unshaded, solid-line histogram
shows the mass distribution from the observational sample of 3,228 Outer-Galaxy MCs (see text), selected from the larger sample in \citet{Heyer+01}. The dotted-line
histogram is the mass distribution of the observational sample, excluding clouds with radius $R_{\rm e}\le 3.1$ pc, the size-completeness limit of the Outer-Galaxy survey.
It shows a contribution to the mass distribution by clouds below this completeness limit up to a mass of approximately 1,000 M$_{\odot}$. The least-square-fit to the
observational mass distribution above $1.5\times 10^3$ M$_{\odot}$ has a slope of $-0.99\pm 0.09$ and is shown by the dashed-dotted line.}
\label{fig_mass_dist}
\end{figure}

As in \citet{Heyer+01}, we consider only the subset of 3,901 clouds with circular velocities $v_{\rm c} < -20$ km s$^{-1}$, because of kinematic distance accuracy.
We further select clouds with mass $M_{\rm cl} > 100$ M$_{\odot}$, as that is the mass limit for our numerical cloud catalogs, resulting in a total sample of
3,228 Outer-Galaxy MCs. Given the distances to the clouds and the angular resolution of the survey, the spatial resolution varies between 0.4 pc and 3.8 pc.
Therefore, the cloud extraction of our highest resolution catalog with $dx = 0.48$ pc matches well the highest resolution in the observations.
The main limitation of the survey is the
velocity resolution, only slightly better than 1 km s$^{-1}$, which, combined with the measurement of line width based on the equivalent width instead of the
antenna-temperature-averaged velocity dispersion, results in a minimum velocity dispersion of clouds of approximately 0.5 km s$^{-1}$. However, we
show that the data can be used to test both the slope and the normalization of the Larson velocity-size relation from the simulation despite this low
velocity resolution.

A explained in \S 3, we illustrate this comparison using clouds selected from 7 snapshots from the last 6 Myr of the simulation. However, we have verified
that all the observational MC properties discussed in this section are essentially the same when derived from a catalog of clouds selected from the last 6 Myr
prior to the inclusion of gravity. This confirms that global MC properties are primarily the result of SN-driven turbulence, with little modification due to self-gravity,
apart from the slight increase in mean cloud density shown in \S 7, and an increase in the mass of the largest cloud.

Of the 12 cloud catalogs
described in \S 3, we choose the one with the highest-resolution ($512^3$ cells, or 0.49 pc) and with the threshold density that best matches the observed
mass-size relation, $n_{\rm H,min}=200$ cm$^{-3}$, for all the plots in this section, and we compute velocity dispersions using the tracer particles, thus taking
advantage of the highest resolution of the simulation, $dx=0.24$ pc, due to the large number density of tracer particles within dense clouds. This catalog
contains 1,547 objects.

\subsection{Mass Distribution}

\begin{figure}[t]
\includegraphics[width=\columnwidth]{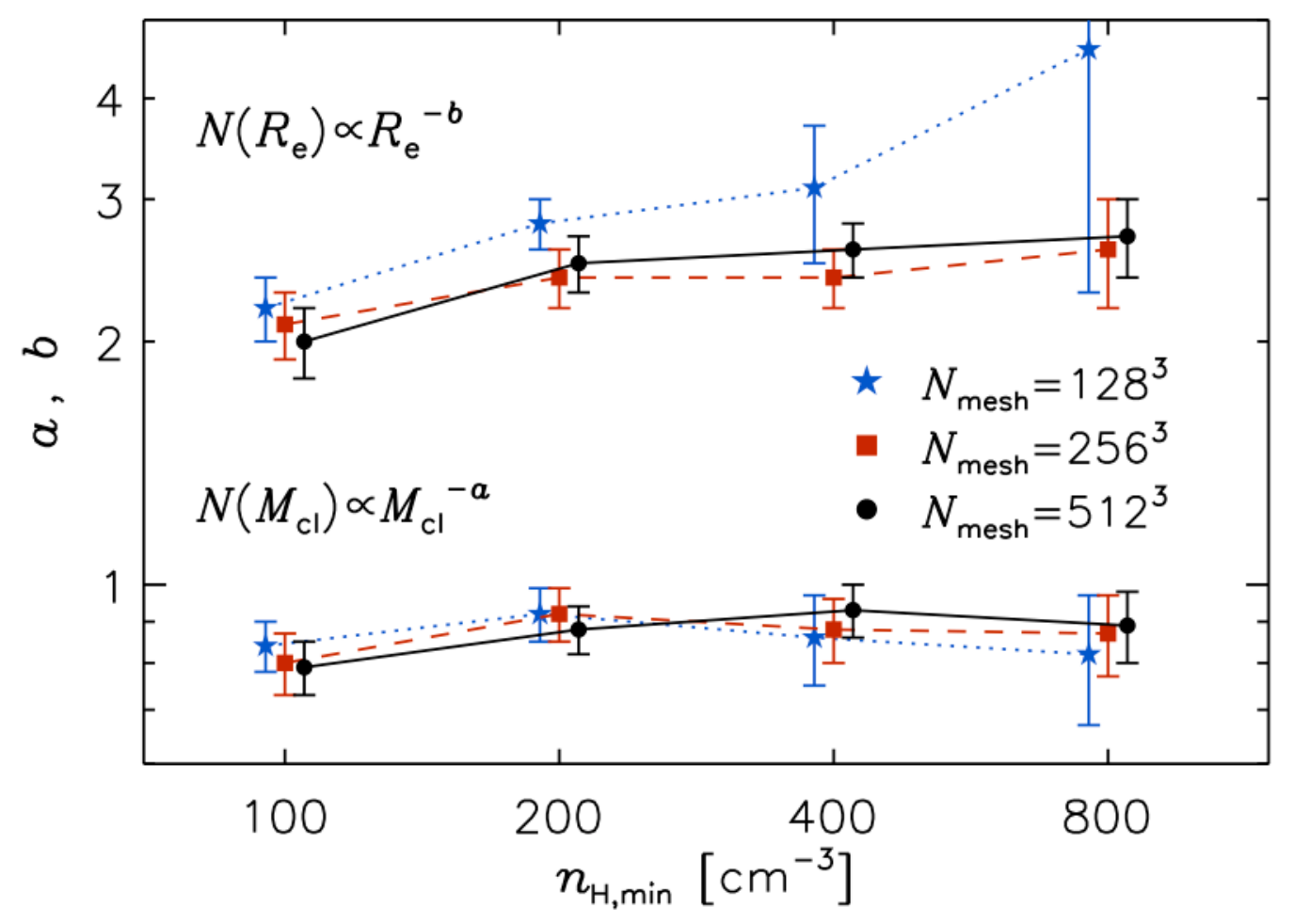}
\caption[]{Exponents of the power-law fits to the probability distributions of cloud masses (lower plots) and cloud sizes (upper plots) for all twelve
catalogs of MCs selected from the simulation. The exponents are plotted as a function of threshold density, $n_{\rm H,min}$, and for three different
values of spatial resolution of cloud selection.}
\label{exponents_size}
\end{figure}

Figure \ref{fig_mass_dist} shows the mass distribution of our clouds (shaded histogram). The histogram is well approximated by a power law,
with a slope of $-0.88\pm0.06$ in the approximate mass range $200 - 10^5$ M$_{\odot}$ (dashed line). The figure also shows the
mass distribution of the MCs from the observational sample that is also nicely fit by a power law, with a slope of $-0.91\pm0.09$
in the approximate mass range $1.5\times 10^3 - 2\times 10^5$ M$_{\odot}$ (dashed-dotted line). \citet{Heyer+01} derived a slope of
$-0.80 \pm 0.03$ by including all clouds down to the completeness limit of 600 M$_{\odot}$. Our slope is a bit steeper because we are
a bit more conservative on the completeness limit. The dotted-line histogram in Figure \ref{fig_mass_dist} shows the mass distribution for
a sub-sample where we include only clouds above the size completeness limit of 3.1 pc, the value derived by \citet{Heyer+01}. The comparison
with the histogram of the full sample shows that some MCs with sizes below the size completeness limit are found with masses up to approximately
1,000 M$_{\odot}$, so we consider the catalog to be complete only above that mass value.

The mass distribution of our clouds is consistent with that of the MCs from the Outer Galaxy Survey. Furthermore, this is true for the mass
distribution from all numerical cloud catalogs we have compiled, independent of the numerical resolution and threshold density of cloud selection.
As shown in Figure \ref{exponents_size}, the exponent of the power-law fit to the mass distribution has a weak dependence on resolution and
threshold density and, within the 1-$\sigma$ error bars shown in the plot, it is consistent with the observational exponent in all cases.

The largest cloud in our $n_{\rm H,min}=200$ cm$^{-3}$ catalog has a mass of $1.3\times 10^5$ M$_{\odot}$, a few times smaller than the
largest MC in the observational sample. However, Figure \ref{fig_mass_dist} shows that this maximum mass is consistent with the largest mass
expected from our sample size and the slope of the mass distribution. If we simulated a region larger than 250 pc, and thus collected a much larger cloud sample, we
would likely derive a power-law mass distribution extended to larger masses. The observations are consistent with a power-law mass distribution
for clouds more massive than those in the Outer Galaxy Survey. For example, in their analysis of the MC sample by \citet{Solomon+87},
\citet{Williams+McKee97} found a comparable slope of $-0.81\pm 0.14$ in the mass range $3\times10^5 - 5.6 \times10^6$ M$_{\odot}$.
Although more uncertain, the slope they obtained from the analysis of the sample by \citet{Scoville+87}, $-0.67\pm0.25$, is also consistent with
the slope of the Outer-Galaxy MCs at lower masses. More recently, \citet{Roman-Duval+10} found a slope of $-0.64\pm0.25$
in the mass range $4\times10^4$--$10^6$ M$_{\odot}$ from the analysis of the Galactic Ring Survey (a more conservative completeness
limit of $8\times10^4$ M$_{\odot}$ gives a slope of $-0.86\pm0.25$).

\subsection{Size Distribution}

\begin{figure}[t]
\includegraphics[width=\columnwidth]{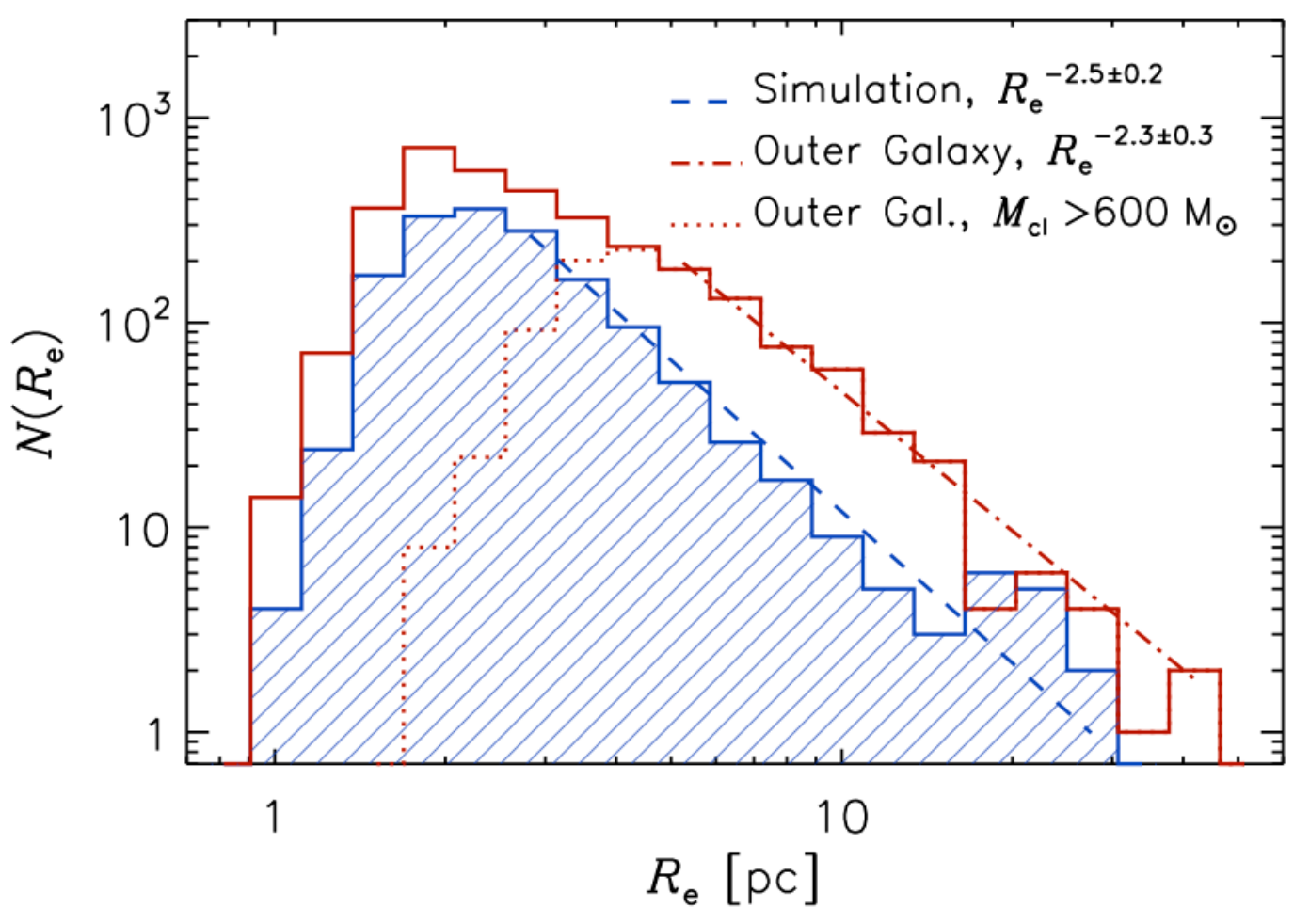}
\caption[]{Probability distribution of cloud size for the same cloud catalog from the simulation (shaded histogram) and the same observational sample
(unshaded, solid-line histogram) as in Figure \ref{fig_mass_dist}. The dotted-line histogram is the size distribution of the observational sample, excluding
clouds with mass $M_{\rm cl} < 600$ M$_{\odot}$, the mass-completeness limit of the Outer-Galaxy survey. The dashed line is a fit to the tail of
the histogram from the simulation, with slope  $-2.5\pm0.2$, and the dashed-dotted line a fit to the observational size distribution, with slope
 $-2.3\pm0.3$.}
\label{fig_size_dist}
\end{figure}

Following \citet{Heyer+01} and most observational works, as a measure of a cloud's size we adopt the equivalent radius, $R_{\rm e}\equiv(A_{\rm cl}/\pi)^{0.5}$,
where $A_{\rm cl}$ is the cloud projected area. The probability distribution of $R_{\rm e}$ for the clouds from the simulation is shown in Figure \ref{fig_size_dist}
(shaded histogram). It is well approximated by a power law with a slope of $-2.5\pm0.2$ in the approximate range of $2-15$ pc. Within the uncertainty, this is
consistent with the slope of $-2.3 \pm 0.3$ of the observational sample in the approximate equivalent-radius range of $5-50$ pc. Furthermore, Figure \ref{exponents_size}
shows that the size distributions of clouds selected from the simulation with different threshold density and resolution are also consistent with the observations,
within the 1-$\sigma$ uncertainty (except for the catalog with $n_{\rm H,min}=200$ cm$^{-3}$ and the lowest-resolution). As in the case of the mass distribution,
we have been slightly more conservative in the estimation of the size completeness limit, based on evidence that around the value of 3.1 pc, the completeness
limit estimated by \citet{Heyer+01}, we still find some contribution from clouds with mass below the mass completeness limit of 600 M$_{\odot}$, as illustrated
by the dotted histogram in Figure \ref{fig_size_dist}. As a result, we find the same slope as in \citet{Heyer+01}, but with a three times larger uncertainty. The same
power law seems to apply to even larger clouds. For example, \citet{Sanders+85} find a slope of $-2.3\pm0.2$ for a sample of 80 clouds in the approximate size
range of $20-80$ pc.

\begin{figure}[t]
\includegraphics[width=\columnwidth]{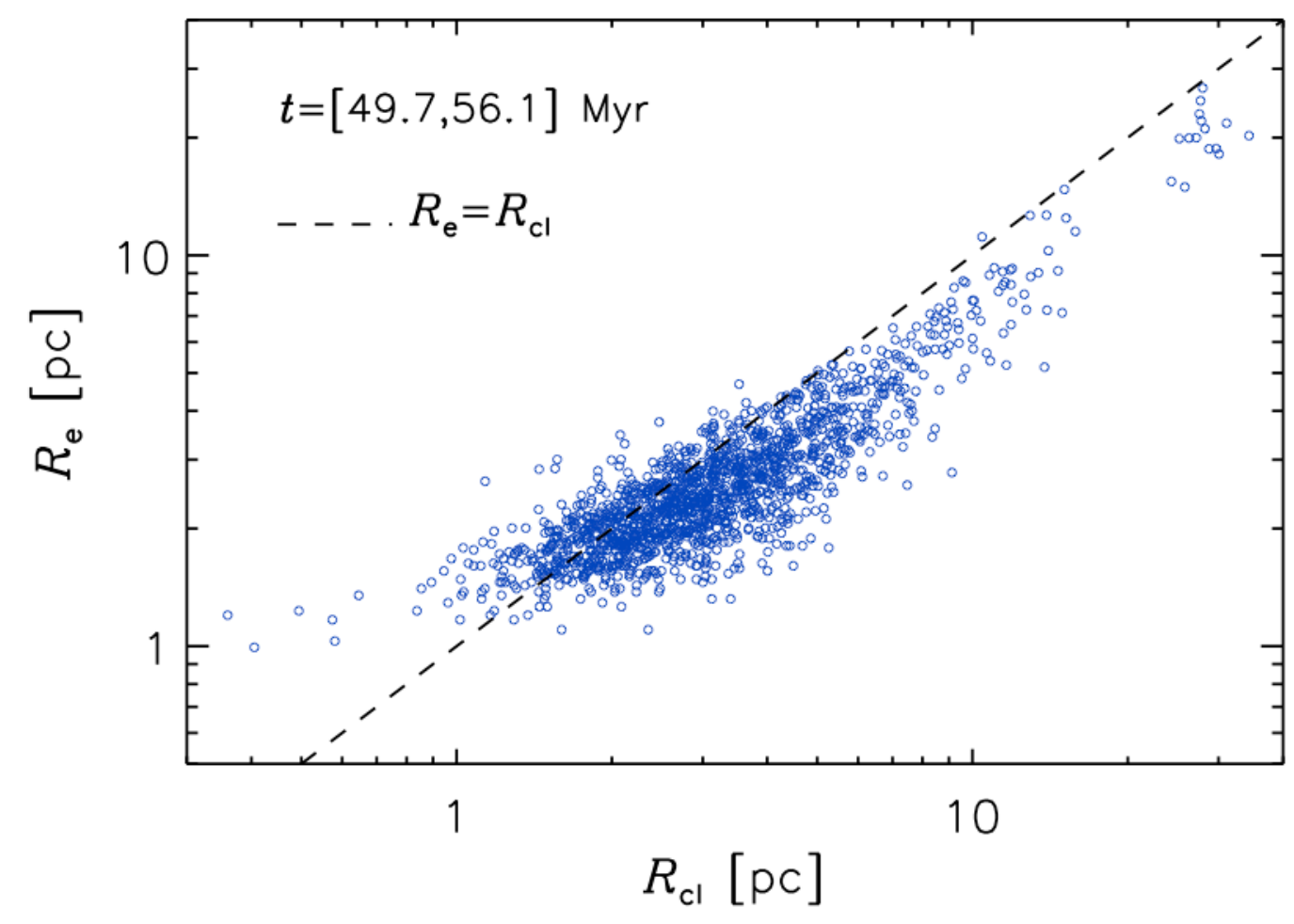}
\caption[]{Equivalent radius versus the three-dimensional cloud radius for all the clouds in the same simulation sample as in Figures \ref{fig_mass_dist} and \ref{fig_size_dist}. }
\label{reff}
\end{figure}

Because we have previously computed a three dimensional cloud radius, $R_{\rm cl}$, from the simulation, we can test the relation between the observable
radius, $R_{\rm e}$, and the three-dimensional one. The comparison is shown in Figure \ref{reff}, where the dashed line corresponds to $R_{\rm e}=R_{\rm cl}$.
$R_{\rm e}$ is smaller than $R_{\rm cl}$ for most clouds with $R_{\rm cl} \gtrsim 2$ pc, and larger than $R_{\rm cl}$ for most clouds with $R_{\rm cl} \lesssim 2$ pc.
The average ratio is $R_{\rm e}/R_{\rm cl}=0.87$ and increases towards smaller radii. As a consequence, the probability distribution of $R_{\rm cl}$ is slightly
shallower than that of $R_{\rm e}$.

\subsection{Velocity-Size Relation}

The velocity-size relation of MCs determines the normalization of the velocity scaling inside individual clouds. In \S 6.2, we have shown that
the energy from SN explosions sets a turbulent cascade inside individual MCs that follows the usual velocity scaling of supersonic turbulence,
but we have not discussed the normalization of the velocity structure functions of individual clouds. To address the velocity normalization, we compute
the internal rms velocity of our clouds based on the velocity of their tracer particles. Being derived from tracer particles, this rms velocity is mass-weighted,
which is a reasonable approximation when comparing it with the antenna-temperature-weighted rms velocity from MC observations. As in the observations,
we compute the one-dimensional (line-of-sight) rms velocity, in the direction perpendicular to the plane (plane of the sky) where we measure the equivalent radius.

\begin{figure}[t]
\includegraphics[width=\columnwidth]{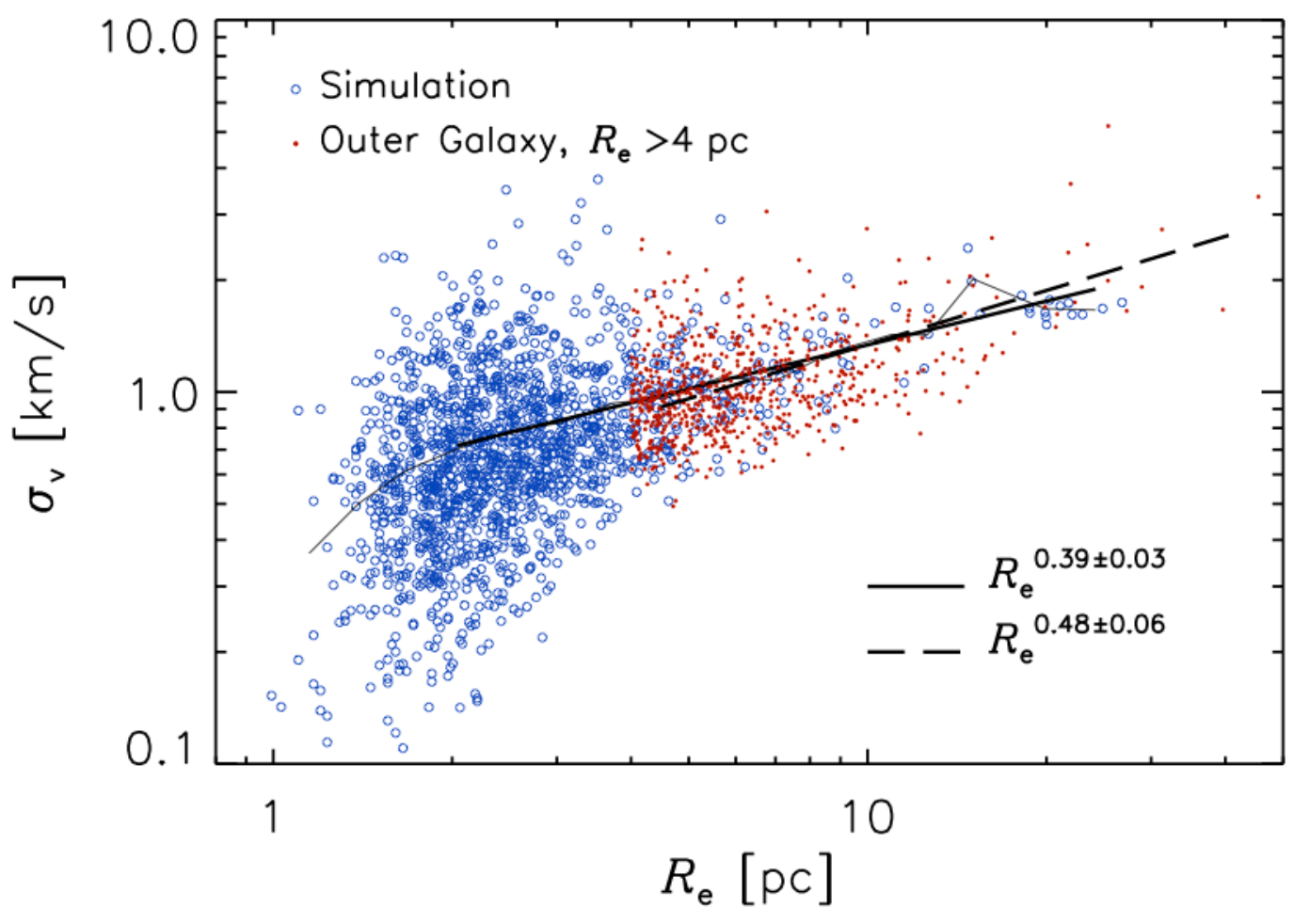}
\caption[]{One-dimensional rms velocity versus size of clouds selected from the simulation (empty circles) and from the Outer Galaxy Survey (filled circles).
The cloud samples are the same as in Figures \ref{fig_mass_dist} and \ref{fig_size_dist}, but the observations are shown only for $R_{\rm e}> 4$ pc, because
for smaller cloud sizes the lower envelope of the velocity-size relation is not resolved by the observation (the minimum value of $\sigma_{\rm v}$ that can be
detected is $\sim 0.5$ km s$^{-1}$). The thin solid line shows the mean values of $\sigma_{\rm v}$ in logarithmic bins of $R_{\rm e}$, and the thick solid line
is a fit to those values, giving a slope of $0.39 \pm 0.07$. The dashed line is the fit to the binned data from the observations, giving a slope of $0.48 \pm 0.06$,
and exactly the same normalization as the simulation at $R_{\rm e}\approx 10$ pc.}
\label{larson_vel}
\end{figure}

Figure \ref{larson_vel} shows the velocity-size relation of the clouds from the simulation (empty circles) and from the observational sample (filled circles).
We have excluded all the observed clouds with $R_{\rm e} < 4$ pc, because the observations cannot detect velocity dispersions smaller than approximately
0.5 km s$^{-1}$, due to the low velocity resolution. The lower envelope of the velocity-size relation of both the simulation and the observations decreases with
decreasing cloud size, and reaches the value of 0.5 km s$^{-1}$ at approximately 4 pc. Thus, the velocity dispersion of a fraction of the clouds smaller than 4 pc
may be significantly overestimated, with that fraction growing towards smaller cloud radii, causing an artificial flattening of the velocity-size relation. As shown
in Figure 6 of \citet{Heyer+01}, the velocity-size relation for the full sample is essentially flat below 4 pc.

Values of $\sigma_{\rm v}$ from the simulation are not to be trusted for the smallest cloud sizes, $R_{\rm e} < 2$ pc, because of the increasing effect of
numerical dissipation towards smaller scales. The thin solid line in Figure \ref{larson_vel} shows the average values of $\sigma_{\rm v}$ in logarithmic
bins of $R_{\rm e}$. While it is nicely fit by a power law for $R_{\rm e} >2$ pc, it clearly drops at smaller cloud sizes. This is consistent with the cloud
structure functions shown in Figure \ref{vel_structure_gmcs0.14}, where the numerical dissipation starts to become important below approximately 2
pc as well.

Despite these limitations imposed by the low velocity resolution of the observations and the numerical dissipation in the simulation, we still have a sufficient
range in $R_{\rm e}$ where the simulation and the observations can be compared. Both the upper and the lower envelopes of the velocity-size relation are
very similar in the two cases. Furthermore, the velocity normalization is nearly identical. The thick solid and dashed lines in Figure \ref{larson_vel} show the
least square fits of the average values of $\sigma_{\rm v}$ in logarithmic bins of $R_{\rm e}$ of the simulation (for $R_{\rm e} >2$ pc) and of the observations
(for $R_{\rm e} > 4$ pc), respectively. From the simulation we get
\begin{equation}
\sigma_{\rm v} = (1.34 \pm 0.04) {\rm \,km \,s}^{-1} (R_{\rm e}/10 {\rm \,pc})^{0.39\pm0.03},
\label{larson_sim}
\end{equation}
and from the observations:
\begin{equation}
\sigma_{\rm v} = (1.34 \pm 0.06) {\rm \, km \,s}^{-1} (R_{\rm e}/10 {\rm \, pc})^{0.48\pm0.06},
\label{larson_obs}
\end{equation}
so the velocity normalization at $R_{\rm e}=10$ pc is indistinguishable in the two cases. This agreement between the simulation and the observations
in the slope, total scatter and normalization of the velocity-size relation is strong evidence that SN driving alone can be responsible for the turbulence
observed in MCs.

\begin{figure}[t]
\includegraphics[width=\columnwidth]{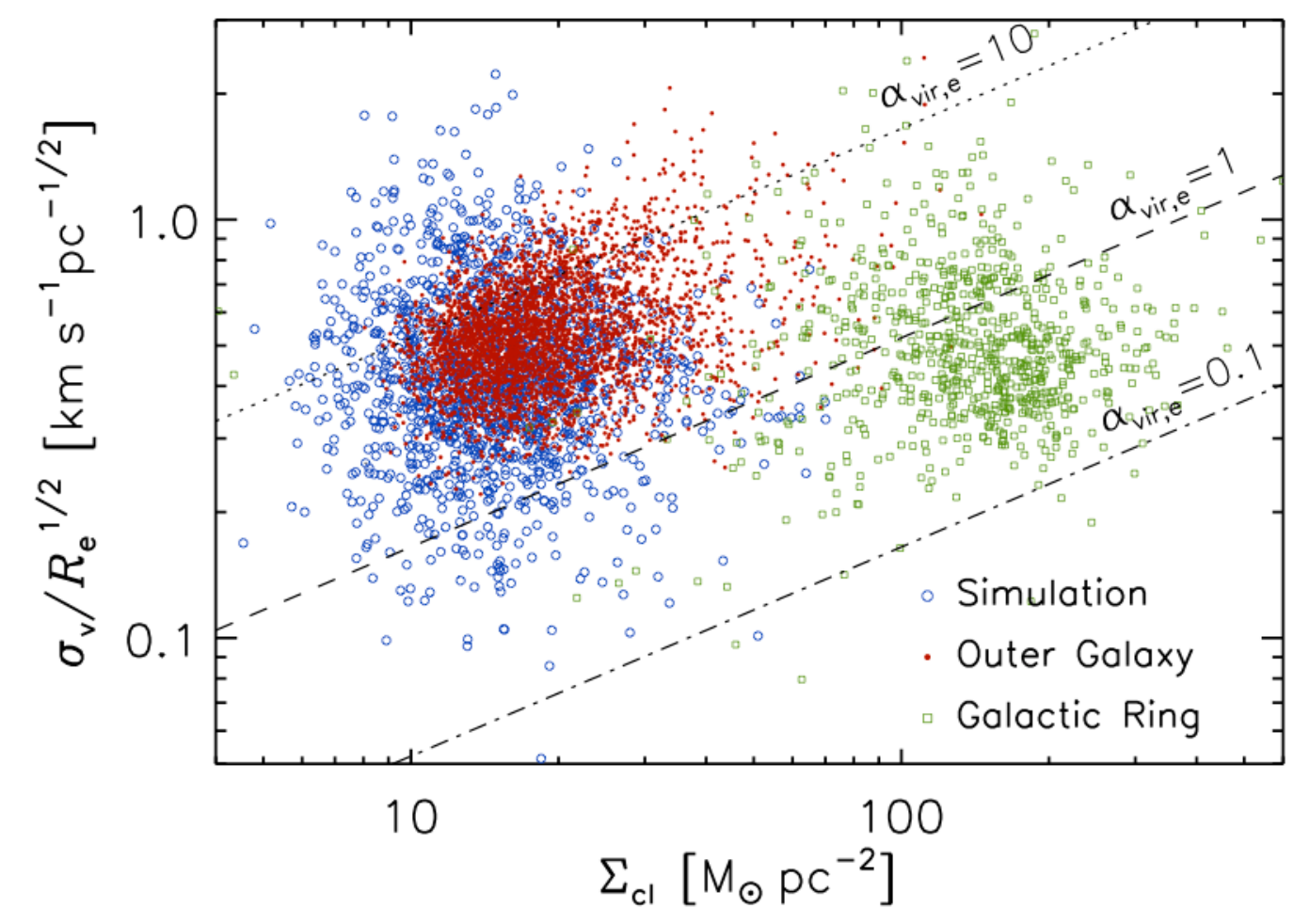}
\caption[]{Normalization of the velocity-size relation versus column density. We normalize the velocity dispersion with $R_{\rm e}^{0.5}$, instead of the shallower
slopes derived in Figure \ref{larson_vel}, to reproduce the plot in Figure 7 of \citet{Heyer+09}. Besides the data points from Figure \ref{larson_vel} (this time including
Outer-Galaxy MCs with $R_{\rm e} < 4$ pc), we also show a sample of MCs from the Galactic Ring Survey \citep{Roman-Duval+10}. The column density of the
Galactic-Ring MCs is on the average 10 times larger than that of the Outer-Galaxy MCs, yet the velocity normalization is essentially the same.}
\label{larson_norm}
\end{figure}

The universality of the MC velocity normalization has been questioned by \citet{Heyer+09}, claiming that it depends on column density, and
thus that the velocity-size relation is controlled by gravity rather than being a natural consequence of the ISM turbulence. To further confirm the
agreement between the simulation and the observations, we show the velocity normalization as a function of column density in Figure \ref{larson_norm}
(we plot  $\sigma_{\rm v}/R_{\rm e}^{0.5}$ as in \citet{Heyer+09}, even if the slope of the velocity-size relation is actually smaller than 0.5).
Because the range of cloud column densities is similar in the two cases, we should not expect a different normalization even if it depended
on surface density. Figure \ref{larson_norm} shows a good overlap between our clouds and the observations. We also plot the values for the
Galactic-Ring clouds in \citet{Roman-Duval+10} that have an average surface density an order of magnitude larger than the Outer-Galaxy clouds
(notice that the difference in surface density is only a factor of five for equal cloud mass or size, and could have been overestimated by a factor of two or
three, as explained in the opening of \S 10). The figure shows that there is no difference in the velocity normalization of Galactic-Ring and Outer-Galaxy MCs,
despite the difference in surface density. Thus, we conclude that the normalization of the velocity-size relation of the MCs in our sample is consistent with being 
controlled by SN-driven turbulence, rather than by the clouds self-gravity. This result is in contradiction to the claim that the velocity normalization of MCs scales with
surface density \citep{Heyer+09}, based on clouds from the sample by \citet{Solomon+87}, analyzed within rectangular maps of different sizes, rather than
a fixed antenna-temperature threshold.

\begin{figure}[t]
\includegraphics[width=\columnwidth]{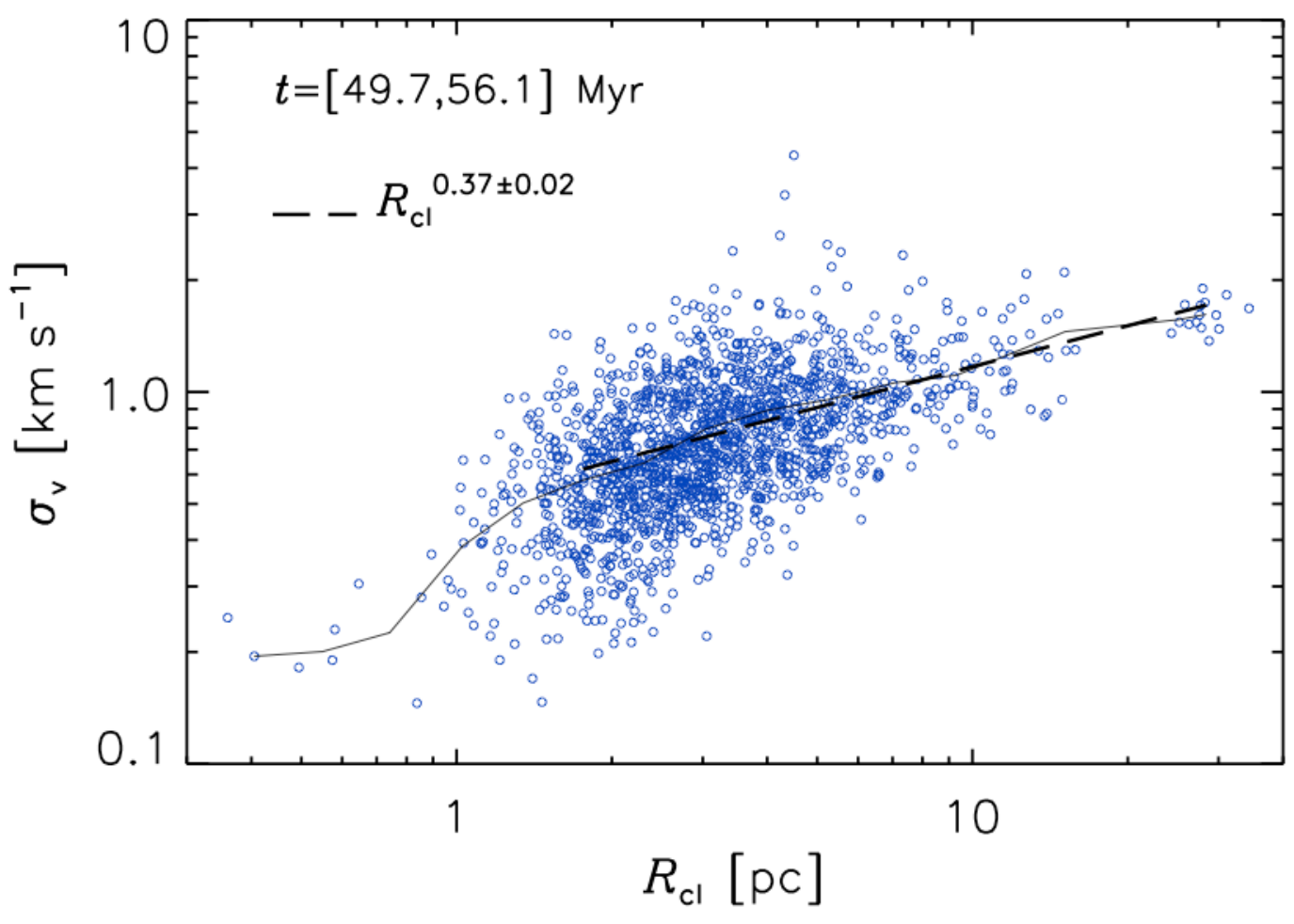}
\caption[]{Cloud rms velocity versus cloud size for the same numerical cloud sample as in the previous figures, but with both rms velocity and cloud size measured
in three-dimensional space, $\sigma_{\rm v}=\sigma_{\rm v,3D}/\sqrt{3}$ and $R_{\rm cl}$. The thin solid line shows the average $\sigma_{\rm v}$
in logarithmic intervals of $R_{\rm cl}$, and the dashed line is the fit to the binned values, with a slope $0.37 \pm 0.02$ for cloud sizes $R_{\rm cl}>1.6$ pc.}
\label{larson_vel_3d}
\end{figure}

Besides being useful to derive the velocity normalization, the observed velocity-size relation may also provide a rather accurate estimate of the slope
of the second order velocity structure function. In Figure \ref{larson_vel_3d} we show the relation between the velocity dispersion derived from the
three-dimensional one, $\sigma_{\rm v,3D}/\sqrt{3}$, and the three-dimensional cloud radius, $R_{\rm cl}$. The thick solid line is the least square
fit to the average values of $\sigma_{\rm v}$ in logarithmic bins of $R_{\rm cl}$ shown by the thin solid line. The fits to the three-dimensional velocity-size
relation for $R_{\rm cl} > 1.6$ pc (where the relation is well approximated by a power law) gives
\begin{equation}
\sigma_{\rm v} = (1.19 \pm 0.04) {\rm \, km \,s}^{-1} (R_{\rm cl}/10 {\rm \, pc})^{0.37\pm0.02}.
\label{larson_sim_3d}
\end{equation}
This relation is consistent with its two-dimensional counterpart (apart from the lower normalization due to the fact that $R_{\rm cl} > R_{\rm e}$ on average),
so the observable relation can be considered as a good estimate of the intrinsic three-dimensional one.
Furthermore, the slope of the relation (\ref{larson_sim_3d}) is also consistent with the slope of the second order structure function, $\zeta(3)/2\approx 0.37$,
of the clouds from the simulation. Thus, we conclude that the observed velocity-size relation provides an estimate of the velocity scaling of MC
turbulence, as long as it is based on MCs with reliable distance measurements and sufficient velocity resolution to detect the lower envelope of the relation.

The velocity-size relation (\ref{larson_sim_3d}) implies the following expression for the dynamical time of MCs as a function of the cloud three-dimensional radius,
using the definition (\ref{tdyn_def}) of dynamical time adopted in \S 8:
\begin{equation}
t_{\rm dyn} = 4.8 \, {\rm Myr}\, (R_{\rm cl}/10 {\rm \,pc})^{0.63}.
\label{tdyn_Re}
\end{equation}
Using the result of \S 8 that $\langle t_{\rm life}/t_{\rm dyn} \rangle \approx 4$, our velocity-size relation implies that our largest MCs with sizes in the
range $R_{\rm cl}\sim10-30$ pc have lifetimes in the range $t_{\rm life}\sim 19-38$ Myr.

\begin{figure}[t]
\includegraphics[width=\columnwidth]{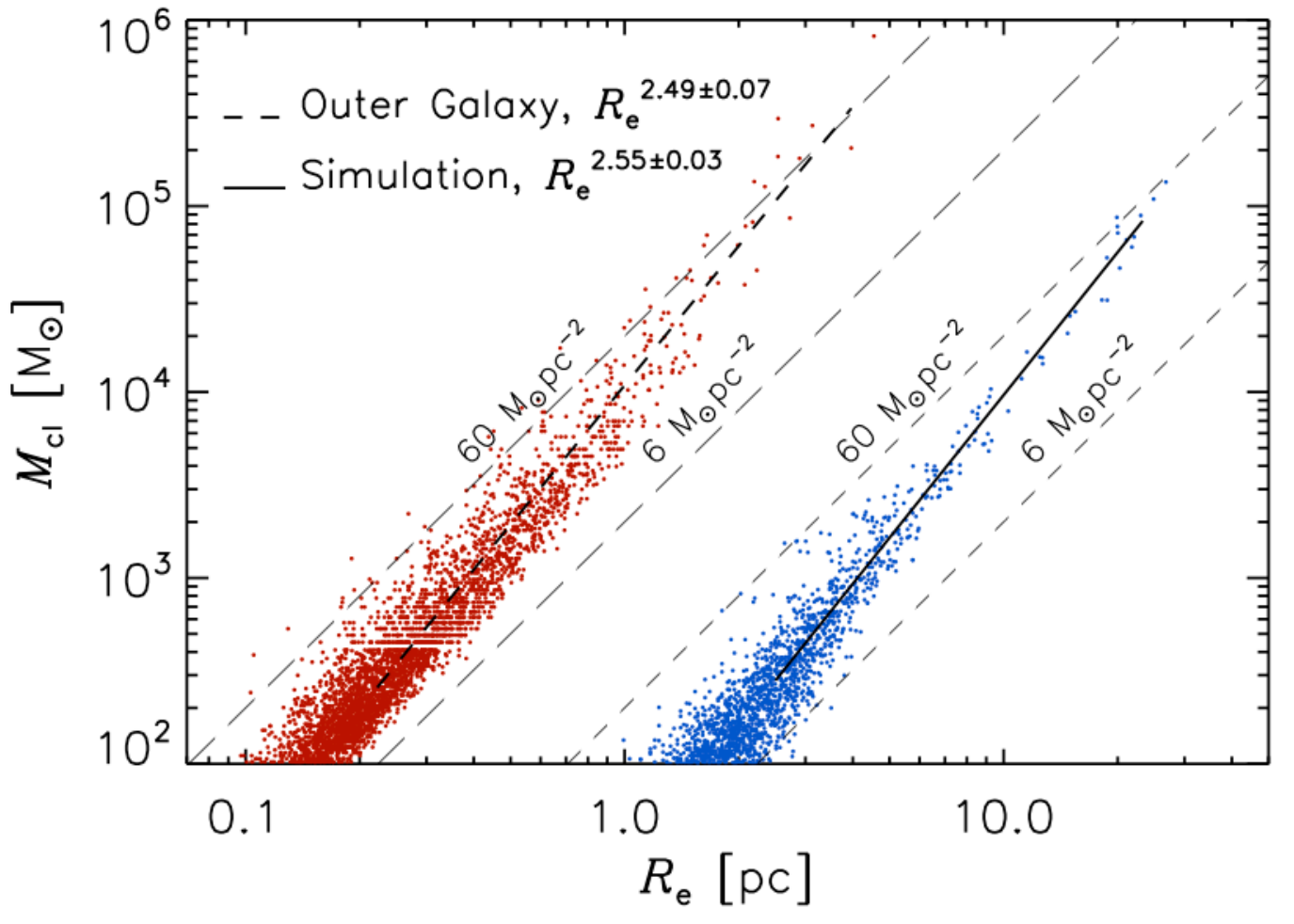}
\caption[]{Mass versus size for the same numerical and observational cloud samples as in Figure \ref{larson_vel}, but including all observed MCs with
mass larger than 100 M$_{\odot}$. The observational points have been
shifted to the left by dividing the observed values of $R_{\rm e}$ by 10, to avoid the overlap with the numerical data points. The thick solid and dashed
lines are fit to the binned data of the simulation and the observations respectively, for $R_{\rm e}>2$ pc. The thin, long-dashed lines show two values of
constant surface density.}
\label{larson_mass}
\end{figure}

\subsection{Mass-Size Relation}

The mass-size relation is plotted in Figure \ref{larson_mass}, this time including all observed clouds above 100 M$_{\odot}$. The values of $R_{\rm e}$
of the observed MCs have been divided by 10, to avoid overlap with the clouds from the simulation. Because of the imposed limit on the minimum cloud
mass, the data is binned and fit only for $R_{\rm e} > 2$ pc for both the simulation and the observations. The resulting fits are
\begin{equation}
M_{\rm cl} = (9.6 \pm 0.3)\times 10^3 {\rm M}_{\odot} \,(R_{\rm e}/10 {\rm \,pc})^{2.55\pm0.03},
\label{mass_size_sim}
\end{equation}
for the simulation, and
\begin{equation}
M_{\rm cl} = (10.9 \pm 0.7)\times 10^3 {\rm M}_{\odot}\, (R_{\rm e}/10 {\rm \,pc})^{2.49\pm0.07},
\label{mass_size_obs}
\end{equation}
for the observations, so the slope of the mass-size relation from the simulation is consistent with the observations. The normalization of the mass--size
relation depends on the threshold antenna temperature of the observational sample and the threshold density of the numerical sample. Of our MC catalogs
described in \S 3, the ones with $n_{\rm H,min}=200$ cm$^{-3}$ have the mass-size normalization closest to that of the Outer-Galaxy MC sample
by \citet{Heyer+01}. This is the reason why all plots in this section are based on the highest-resolution catalog with that value of threshold density.

The total scatter in the relation is also similar between the observations and the simulation, if we account for the facts that the observational sample size is
approximately twice as large as the numerical one, and that we have not added any observational uncertainty to the masses and sizes of the clouds from our
simulation. Because of the similarity in both the slope and the total scatter, we can conclude that the mass-size relation resulting from SN-driven turbulence
is consistent with that of real MCs from the Outer Galaxy Survey. A similar mass-size relation (though with a five times larger normalization) was also
derived from the Galactic Ring Survey \citep{Roman-Duval+10} and is implied by previous estimates of cloud fractal dimensions from various observational
surveys \citep[e.g.][]{Elmegreen+Falgarone96,Sanchez+07} and from simulations of randomly driven turbulence \citep[e.g.][]{Kritsuk+07,Federrath+09}.

Combining the mass-size relation (\ref{mass_size_sim}) with the dynamical time expression in equation (\ref{tdyn_Re}), and adopting the average value 
derived in \S 10.2 for the ratio between the two definitions of cloud radius, $R_{\rm e}=0.87 R_{\rm cl}$, we obtain the following expression for the average
dynamical time of MCs as a function of cloud mass,
\begin{equation}
t_{\rm dyn} = 5.0 \, {\rm Myr}\, (M_{\rm cl}/10^4 {\rm \,M}_{\odot})^{0.25},
\label{tdyn_Mcl}
\end{equation}
hence the expression (\ref{tlife_versus_Mcl}) for the average cloud lifetime as a function of cloud mass anticipated in \S 8.

\subsection{Virial Parameter}

In \S 7, the virial parameter is computed with the three-dimensional radius, $R_{\rm cl}$. Here, as in most MC studies including \citet{Heyer+09},
we define an observable version of the virial parameter using the equivalent radius, $R_{\rm e}$, and thus refer to it as $\alpha_{\rm vir,e}$.
The dependence of $\alpha_{\rm vir,e}$ on $M_{\rm cl}$ is fully determined by the velocity-size and mass-size relations presented above.
Nevertheless, it is presented here as an alternative view of the comparison of the simulation with the observations, and to suggest an empirical
calibration of the virial parameter of MCs.

\begin{figure}[t]
\includegraphics[width=\columnwidth]{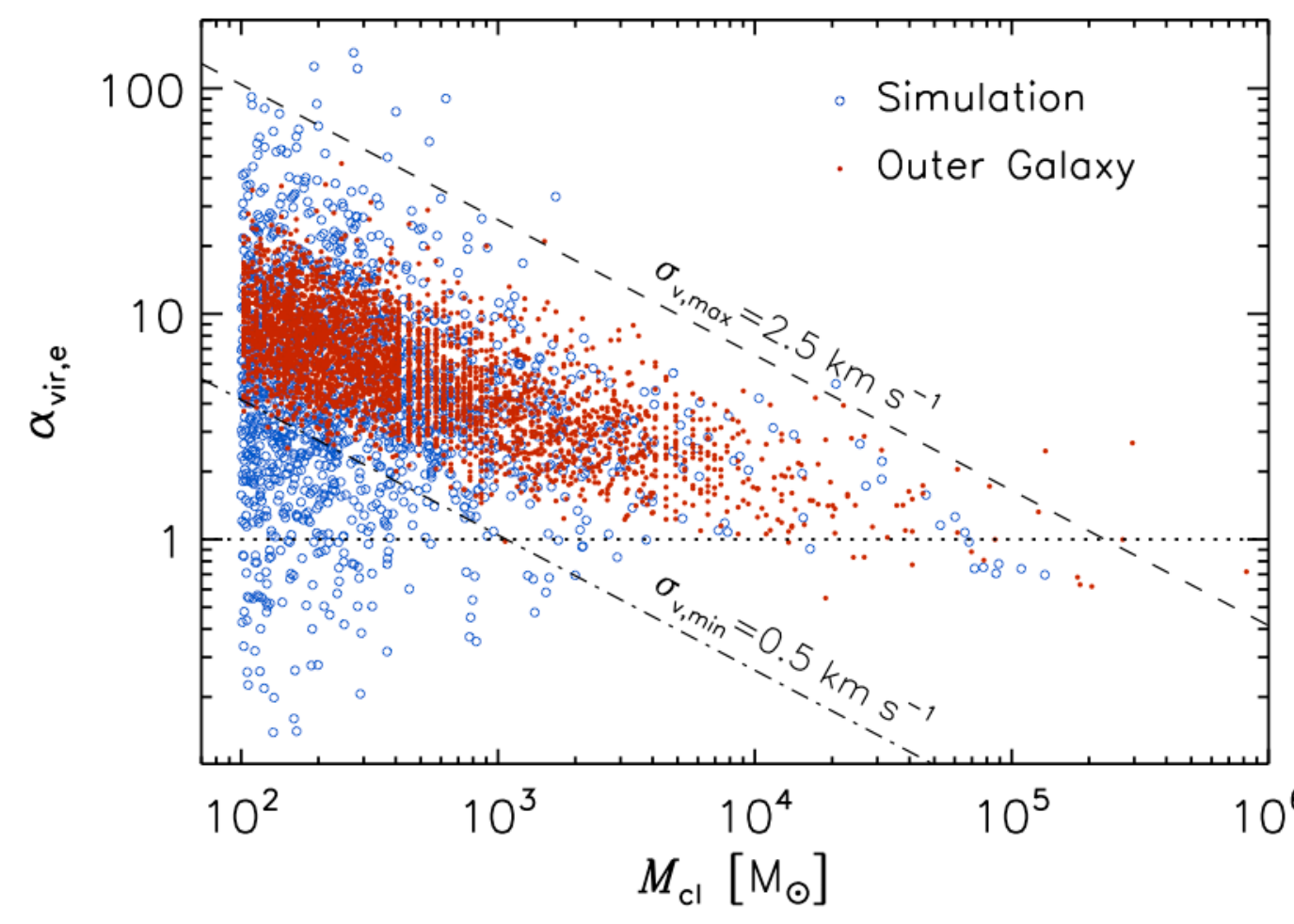}
\caption[]{Virial parameter versus mass for the same clouds as in Figures \ref{larson_mass}.
The dashed line shows a predicted upper envelope assuming that the largest velocity dispersion is $\sim 2.5$ km s$^{-1}$ (independent of cloud mass)
and adopting the fit to the mass-size relation from Figures \ref{larson_mass}. The dashed-dotted line shows the predicted lower envelope, also based
on the mass-size relation and assuming a minimum velocity of $\sim 0.5$ km s$^{-1}$, as in the observations.}
\label{fig_alpha}
\end{figure}

The values of $\alpha_{\rm vir,e}$ are plotted versus $M_{\rm cl}$ in Figure \ref{fig_alpha} for both the simulation (empty circles) and the observations
(filled circles). The most striking feature of this plot is the very large scatter in the values of  $\alpha_{\rm vir,e}$ at a fixed cloud mass, growing with
decreasing cloud mass, which is a direct consequence of the large scatter in the velocity-size relation. As in that relation, the lower envelope for the
observational data is limited by the smallest value of $\sigma_{\rm v}$ to which the observations are sensitive, $\sigma_{\rm v}\approx 0.5$ km s$^{-1}$.
The dashed-dotted line in Figure \ref{fig_alpha} shows the minimum value of $\alpha_{\rm vir,e}$ as a function of $M_{\rm cl}$, computed by setting
$R_{\rm e}$ as a function of $M_{\rm cl}$ using the average mass-size relation from the previous section and setting $\sigma_{\rm v}= 0.5$ km s$^{-1}$.

The virial parameter would be independent of mass if the velocity size relation were $\sigma_{\rm v} \sim R_{\rm e}^{1/2}$ and the mass-size relation
were $M_{\rm cl} \sim R_{\rm e}^2$, the often assumed form of the Larson relations. However, the mass-size relation is such that the cloud surface
density grows with mass, as shown in the previous section, causing the decrease of $\alpha_{\rm vir,e}$ with increasing $M_{\rm cl}$ seen in
Figure \ref{fig_alpha}, only partly mitigated by the exponent of the velocity-size relation being a bit smaller than 0.5. The upper envelope in
Figure \ref{fig_alpha} is even steeper than the average decrease of virial parameter with mass, as a consequence of the nearly flat upper envelope
of the velocity-size relation. The dashed line in Figure \ref{fig_alpha} shows the expected upper envelope based on the mass-size relation and a maximum
rms velocity of 2.5 km s$^{-1}$ that is representative of some of the largest values in both the simulation and the observations.

Once we account for the minimum velocity dispersion in the Outer-Galaxy Survey, the relation of $\alpha_{\rm vir,e}$ with $M_{\rm cl}$
and its scatter for the clouds from the simulation are consistent with those for the observed Outer-Galaxy MCs, as expected from the
agreement found in the velocity-size and mass-size relations. This agreement suggests the possibility of an empirical calibration of the
observed values of the virial parameter based on the results of our simulation. We have shown in \S 7 that the virial parameter
computed with the radius $R_{\rm cl}$ is $\alpha_{\rm vir} \approx 1.2 \,(2 E_{\rm k}/E_{\rm g})$. We have also shown in this section
that, on the average, $R_{\rm cl} \approx R_{\rm e}/0.87$, thus we derive this useful result for the relation between the observed virial
parameter based on the equivalent cloud radius and the energy ratio:
\begin{equation}
{2 E_{\rm k}\over{E_{\rm g}}} \approx 0.96 \,\alpha_{\rm vir,e}
\label{virial_e}
\end{equation}
(assuming negligible saturation of the observed lines, such that the rms velocity can be assumed to be approximately mass weighted).
\citet{Bertoldi+McKee92} modeled extensively the coefficient of the virial parameter of clumps, depending on the mass profile and
the ellipticity of the clumps. Given their complex structure, MCs are not described by radial density profiles and ellipticity parameters
as easily as compact clumps, so an empirical calibration based on simulations as proposed here is needed.

\section{Overview of Results and Conclusions}

The main goal of this work is to test if ISM turbulence driven only by SN explosions can explain the turbulence observed within MCs.
We have addressed this question with an AMR simulation representing an ISM volume of (250 pc)$^3$ and reaching a maximum
resolution of 0.24 pc, with refinement based on density, density gradients and pressure gradients to resolve individual
SN remnants and their interaction with the dense gas. We have studied the SN-driven turbulence over the whole volume and within
individual clouds. We have compiled 12 different catalogs of MCs selected from the simulation using four different values of threshold density
and three different spatial resolution. The properties of these clouds have been studied using tracer particles, hence
taking advantage of the full resolution of the simulation. First we have presented cloud properties based on particle position, velocity
and magnetic field values measured in three-dimensional space; then we have carried out a comparison with real MCs from the
Outer Galaxy Survey by measuring projected quantities, such as the line-of-sight velocity dispersion and the equivalent radius.
Our results are summarized in the following:
\begin{enumerate}

\item Near equipartition of total energies in the whole volume results in a distinct energy separation in the dense gas. While the overall
ISM turbulence is trans-Alfv\'{e}nic and mildly supersonic, the turbulence in the dense gas is highly supersonic and super-Alfv\'{e}nic.

\item Approximately 11\% of the total kinetic energy is transferred to the dense gas, even if most SN explosions occur at low densities.
The dense gas has an average velocity dispersion of 8.5 km s$^{-1}$.

\item During the rapid expansion of a SN remnant, the velocity power spectrum may briefly show strong features at different scales.
During most of the time the power spectrum is statistically relaxed and develops a power-law inertial range that scales with
wavenumber as $E(k)\propto k^{-1.46}$, between approximately 20 and 60 pc.

\item Unlike previous studies of compressible turbulence, the power spectrum of the compressive component of the velocity,
$E_{\rm c}(k)\propto k^{-1.98}$, is much steeper than that of the solenoidal component, $E_{\rm s}(k)\propto k^{-1.31}$.  The
baroclinic effect is the best candidate to explain this result, as previous simulations of supersonic turbulence adopted equations
of state where such effect was absent.

\item SN driving is not purely compressive. The curl of the forcing is the baroclinic term that is comparable to or larger than the divergence
of the forcing (except in the unrealistic cases of a uniform-density medium or a barotropic equation of state). As a result, the power in solenoidal
modes exceeds that in compressive modes almost at any time and any wavenumbers (at the scale of 20 pc, $E_{\rm c}/E_{\rm s} \sim 0.2$).
Thus, isothermal simulations of turbulent fragmentation based on random solenoidal driving are a much better approximation of MC
turbulence than isothermal simulations with purely compressive driving.

\item The time-averaged energy-injection scale of SN-driven turbulence is approximately 70 pc with our SN rate (may be larger with a smaller
SN rate, or vice-versa), and oscillates in time between 50 and 100 pc.

\item The scaling exponents of the first and second order structure functions of velocity, relative to the third order one, in SN-driven turbulence are
consistent with those found in supersonic turbulence driven by an idealized, random volume force, which supports the validity of turbulent
fragmentation and star formation studies where the ISM turbulent cascade within MCs was modeled with such an idealized driving.

\item Based on a new scheme to compute density-weighted velocity structure functions, we obtain a third order exponent close to unity, as in
the incompressible Kolmogorov's `4/5 law'. Thanks to the AMR method and to this density-weighting scheme, the structure functions probe
the inertial range of MC turbulence down to a scale of 2-3 pc, while previous studies of SN-driven turbulence did not resolve an inertial range
and only addressed the relative scaling.

\item The scaling of the velocity structure functions within individual MCs selected from the simulation is generally consistent with the scaling
derived from MC observations. However, deviations are found for MCs directly affected by recent SN remnants.

\item The ratio of cloud virial parameter and kinetic to gravitational energy ratio is $\alpha_{\rm vir} / (2E_{\rm k}/ E_{\rm g})=1.2$, independent
of energy ratio and mass (see point 15 for the observable virial parameter). This structural property of MCs is not significantly affected by self-gravity
during the duration of our simulation, where self-gravity is included in the final 11 Myr, corresponding to about two average cloud free-fall times. 
The ratio becomes very large only in a small fraction of clouds with $\alpha_{\rm vir} \lesssim 10$, where self-gravity causes the collapse of dense cores. 
Even in these star-forming clouds, there is no evidence of global cloud collapse. Self-gravity only causes a slight shift towards larger densities of the 
probability distribution of cloud densities.

\item The formation and dispersion times of MCs are of the order of two cloud dynamical times. Equivalently, the cloud lifetime, defined as the sum of
formation and dispersion times, is approximately four cloud dynamical times. This is evidence indicating that SN-driven turbulence is responsible for 
cloud formation and dispersion, with little influence from self-gravity visible during the duration of our run. Future work, with longer runs,
is needed to determine to what extent this remains true for longer periods of time.

\item The clouds have a mean magnetic field enhanced only by a factor of two relative to the mean magnetic field in the simulated volume The
turbulence is super-Alfv\'{e}nic for all clouds more massive than approximately $10^3$ M$_{\odot}$.

\item The comparison with the MC sample from the Outer Galaxy Survey shows that clouds selected from the simulation have properties consistent
with the observations, such as the mass and size distributions, the velocity-size and mass-size relations and the dependence of virial parameter on
cloud mass. In our run, these properties, including the normalization of the velocity-size relation, are essentially the same for clouds selected either 
before or after the inclusion of gravity in the simulation; they are primarily the result of SN-driven turbulence, with only a minor contribution from
self-gravity.

\item The normalization of the velocity-size relation does not depend on surface density in the simulation, nor in the observations. It is the same for
MCs from the Outer Galaxy Survey as for those in the Galactic Ring Survey whose surface density is significantly higher.

\item The simulation provides a calibration of the observable virial parameter, $\alpha_{\rm vir,e}$, based on the equivalent radius, which allows
a derivation of the energy ratio from observational quantities, ${2 E_{\rm k}/{E_{\rm g}}} \approx 0.96 \,\alpha_{\rm vir,e}$.

\end{enumerate}

Based on these results from the simulation and given its successful comparison with the Outer Galaxy Survey,
we conclude that the SN-driven turbulence in our simulation is consistent with the observed MC turbulence during the duration of our experiment.
Although other energy sources are present in galaxies, and local radiative and
mechanical feedbacks also play a role in the dispersion of star-forming MCs, the
origin, evolution and internal dynamics of MCs in our run are primarily the consequence of
SN-driving, which is able to sustain turbulence at observed levels without help from those extra energy sources.

\acknowledgements

We are grateful to Christoph Federrath and the anonymous referee for providing useful comments and corrections.
Computing resources for this work were provided by the NASA High-End Computing (HEC) Program through
the NASA Advanced Supercomputing (NAS) Division at Ames Research Center, by PRACE through a Tier-0 award providing
us access to the computing resource SuperMUC based in Germany at the Leibniz Supercomputing Center, and by the Port
d'Informaci\'{o} Cient\'{i}fica (PIC), Spain, maintained by a collaboration of the Institut de F\'{i}sica d'Altes Energies (IFAE)
and the Centro de Investigaciones Energ\'{e}ticas, Medioambientales y Tecnol\'{o}gicas (CIEMAT). PP acknowledges support
by the ERC FP7-PEOPLE- 2010- RG grant PIRG07-GA-2010-261359 and by the Spanish MINECO under project AYA2014-57134-P.
TH is supported by a Sapere Aude Starting Grant from The Danish Council for Independent Research.
Research at Centre for Star and Planet Formation was funded by the Danish National Research Foundation and the University
of Copenhagen's programme of excellence.


\begin{thebibliography}{104}
\expandafter\ifx\csname natexlab\endcsname\relax\def\natexlab#1{#1}\fi

\bibitem[{{Balsara} {et~al.}(2004){Balsara}, {Kim}, {Mac Low}, \&
  {Mathews}}]{Balsara+04}
{Balsara}, D.~S., {Kim}, J., {Mac Low}, M.-M., \& {Mathews}, G.~J. 2004, \apj,
  617, 339

\bibitem[{{Benzi} {et~al.}(1993){Benzi}, {Ciliberto}, {Tripiccione}, {Baudet},
  {Massaioli}, \& {Succi}}]{Benzi+93}
{Benzi}, R., {Ciliberto}, S., {Tripiccione}, R., {Baudet}, C., {Massaioli}, F.,
  \& {Succi}, S. 1993, \pre, 48, 29

\bibitem[{{Bertoldi} \& {McKee}(1992)}]{Bertoldi+McKee92}
{Bertoldi}, F., \& {McKee}, C.~F. 1992, \apj, 395, 140

\bibitem[{{Boldyrev}(2002)}]{Boldyrev2002}
{Boldyrev}, S. 2002, ApJ, 569, 841

\bibitem[{{Boldyrev} {et~al.}(2002){Boldyrev}, {Nordlund}, \&
  {Padoan}}]{Boldyrev+02}
{Boldyrev}, S., {Nordlund}, {\AA}., \& {Padoan}, P. 2002, \apj, 573, 678

\bibitem[{{Bournaud} {et~al.}(2010){Bournaud}, {Elmegreen}, {Teyssier},
  {Block}, \& {Puerari}}]{Bournaud+10}
{Bournaud}, F., {Elmegreen}, B.~G., {Teyssier}, R., {Block}, D.~L., \&
  {Puerari}, I. 2010, \mnras, 409, 1088

\bibitem[{{Chevalier} \& {Klein}(1978)}]{Chevalier+Klein78}
{Chevalier}, R.~A., \& {Klein}, R.~I. 1978, \apj, 219, 994

\bibitem[{{Couch} {et~al.}(2009){Couch}, {Wheeler}, \&
  {Milosavljevi{\'c}}}]{Couch+09}
{Couch}, S.~M., {Wheeler}, J.~C., \& {Milosavljevi{\'c}}, M. 2009, \apj, 696,
  953

\bibitem[{{Crutcher} {et~al.}(2010){Crutcher}, {Wandelt}, {Heiles},
  {Falgarone}, \& {Troland}}]{Crutcher+10}
{Crutcher}, R.~M., {Wandelt}, B., {Heiles}, C., {Falgarone}, E., \& {Troland},
  T.~H. 2010, \apj, 725, 466

\bibitem[{{de Avillez} \& {Breitschwerdt}(2005)}]{deAvillez05}
{de Avillez}, M.~A., \& {Breitschwerdt}, D. 2005, \aap, 436, 585

\bibitem[{{de Avillez} \& {Breitschwerdt}(2007)}]{deAvillez07scaling}
---. 2007, \apjl, 665, L35

\bibitem[{{Dobbs}(2015)}]{Dobbs15}
{Dobbs}, C.~L. 2015, \mnras, 447, 3390

\bibitem[{{Dobbs} \& {Pringle}(2013)}]{Dobbs+Pringle13}
{Dobbs}, C.~L., \& {Pringle}, J.~E. 2013, \mnras, 432, 653

\bibitem[{Dubrulle(1994)}]{Dubrulle94}
Dubrulle, B. 1994, Phys. Rev. Lett., 73, 959

\bibitem[{{Elmegreen} {et~al.}(2003){Elmegreen}, {Elmegreen}, \&
  {Leitner}}]{Elmegreen+2003}
{Elmegreen}, B.~G., {Elmegreen}, D.~M., \& {Leitner}, S.~N. 2003, \apj, 590,
  271

\bibitem[{Elmegreen \& Falgarone(1996)}]{Elmegreen+Falgarone96}
Elmegreen, B.~G., \& Falgarone, E. 1996, ApJ, 471, 816

\bibitem[{{Falkovich} {et~al.}(2010){Falkovich}, {Fouxon}, \&
  {Oz}}]{Falkovich+10}
{Falkovich}, G., {Fouxon}, I., \& {Oz}, Y. 2010, Journal of Fluid Mechanics,
  644, 465

\bibitem[{{Faucher-Gigu{\`e}re} {et~al.}(2013){Faucher-Gigu{\`e}re},
  {Quataert}, \& {Hopkins}}]{Faucher-Giguere+13}
{Faucher-Gigu{\`e}re}, C.-A., {Quataert}, E., \& {Hopkins}, P.~F. 2013, \mnras,
  433, 1970

\bibitem[{{Federrath}(2013)}]{Federrath13_4096}
{Federrath}, C. 2013, \mnras, 436, 1245

\bibitem[{{Federrath} {et~al.}(2011{\natexlab{a}}){Federrath}, {Chabrier},
  {Schober}, {Banerjee}, {Klessen}, \& {Schleicher}}]{Federrath+11PRL}
{Federrath}, C., {Chabrier}, G., {Schober}, J., {Banerjee}, R., {Klessen},
  R.~S., \& {Schleicher}, D.~R.~G. 2011{\natexlab{a}}, Physical Review Letters,
  107, 114504

\bibitem[{{Federrath} \& {Klessen}(2012)}]{Federrath+Klessen12}
{Federrath}, C., \& {Klessen}, R.~S. 2012, \apj, 761, 156

\bibitem[{{Federrath} \& {Klessen}(2013)}]{Federrath+Klessen13}
---. 2013, \apj, 763, 51

\bibitem[{{Federrath} {et~al.}(2008){Federrath}, {Klessen}, \&
  {Schmidt}}]{Federrath+08}
{Federrath}, C., {Klessen}, R.~S., \& {Schmidt}, W. 2008, \apjl, 688, L79

\bibitem[{{Federrath} {et~al.}(2009){Federrath}, {Klessen}, \&
  {Schmidt}}]{Federrath+09}
---. 2009, \apj, 692, 364

\bibitem[{{Federrath} {et~al.}(2011{\natexlab{b}}){Federrath}, {Sur},
  {Schleicher}, {Banerjee}, \& {Klessen}}]{Federrath+11}
{Federrath}, C., {Sur}, S., {Schleicher}, D.~R.~G., {Banerjee}, R., \&
  {Klessen}, R.~S. 2011{\natexlab{b}}, \apj, 731, 62

\bibitem[Ferland et al.(1998)]{Ferland+98} Ferland, G.~J.,
Korista, K.~T., Verner, D.~A., et al.\ 1998, \pasp, 110, 761

\bibitem[{{Fleck}(1996)}]{Fleck96}
{Fleck}, Jr., R.~C. 1996, \apj, 458, 739

\bibitem[Franco \& Cox(1986)]{Franco+Cox86} Franco, J., \& Cox, D.~P.\ 1986, \pasp, 98, 1076

\bibitem[{{Gatto} {et~al.}(2015){Gatto}, {Walch}, {Low}, {Naab}, {Girichidis},
  {Glover}, {W{\"u}nsch}, {Klessen}, {Clark}, {Baczynski}, {Peters},
  {Ostriker}, {Ib{\'a}{\~n}ez-Mej{\'{\i}}a}, \& {Haid}}]{Gatto+15}
{Gatto}, A., {Walch}, S., {Low}, M.-M.~M., {Naab}, T., {Girichidis}, P.,
  {Glover}, S.~C.~O., {W{\"u}nsch}, R., {Klessen}, R.~S., {Clark}, P.~C.,
  {Baczynski}, C., {Peters}, T., {Ostriker}, J.~P.,
  {Ib{\'a}{\~n}ez-Mej{\'{\i}}a}, J.~C., \& {Haid}, S. 2015, \mnras, 449, 1057

\bibitem[{{Glover} \& {Clark}(2012)}]{Glover+12}
{Glover}, S.~C.~O., \& {Clark}, P.~C. 2012, \mnras, 421, 116

\bibitem[{{Gnedin} \& {Hollon}(2012)}]{Gnedin+Hollon12}
{Gnedin}, N.~Y., \& {Hollon}, N. 2012, \apjs, 202, 13

\bibitem[{{Habing}(1968)}]{Habing68}
{Habing}, H.~J. 1968, \bain, 19, 421

\bibitem[{{Heiles} \& {Troland}(2005)}]{Heiles+Troland05_B}
{Heiles}, C., \& {Troland}, T.~H. 2005, \apj, 624, 773

\bibitem[{{Hennebelle} \& {Chabrier}(2008)}]{Hennebelle+Chabrier08}
{Hennebelle}, P., \& {Chabrier}, G. 2008, \apj, 684, 395

\bibitem[{{Hennebelle} \& {Chabrier}(2011)}]{Hennebelle+Chabrier11sfr}
---. 2011, \apjl, 743, L29

\bibitem[{{Herant} \& {Woosley}(1994)}]{Herant+Woosley94}
{Herant}, M., \& {Woosley}, S.~E. 1994, \apj, 425, 814

\bibitem[{{Heyer} {et~al.}(2009){Heyer}, {Krawczyk}, {Duval}, \&
  {Jackson}}]{Heyer+09}
{Heyer}, M., {Krawczyk}, C., {Duval}, J., \& {Jackson}, J.~M. 2009, \apj, 699,
  1092

\bibitem[{{Heyer} {et~al.}(1998){Heyer}, {Brunt}, {Snell}, {Howe}, {Schloerb},
  \& {Carpenter}}]{Heyer+98}
{Heyer}, M.~H., {Brunt}, C., {Snell}, R.~L., {Howe}, J.~E., {Schloerb}, F.~P.,
  \& {Carpenter}, J.~M. 1998, \apjs, 115, 241

\bibitem[{{Heyer} \& {Brunt}(2004)}]{Heyer+Brunt04}
{Heyer}, M.~H., \& {Brunt}, C.~M. 2004, \apjl, 615, L45

\bibitem[{{Heyer} \& {Brunt}(2012)}]{Heyer+Brunt2012}
---. 2012, \mnras, 420, 1562

\bibitem[{{Heyer} {et~al.}(2001){Heyer}, {Carpenter}, \& {Snell}}]{Heyer+01}
{Heyer}, M.~H., {Carpenter}, J.~M., \& {Snell}, R.~L. 2001, \apj, 551, 852

\bibitem[{{Heyer} \& {Terebey}(1998)}]{Heyer+Terebey98}
{Heyer}, M.~H., \& {Terebey}, S. 1998, \apj, 502, 265

\bibitem[{{Hopkins}(2012)}]{Hopkins12imf}
{Hopkins}, P.~F. 2012, \mnras, 423, 2037

\bibitem[{{Ianjamasimanana} {et~al.}(2015){Ianjamasimanana}, {de Blok},
  {Walter}, {Heald}, {Cald{\'u}-Primo}, \& {Jarrett}}]{Ianjamasimanana+15_HI}
{Ianjamasimanana}, R., {de Blok}, W.~J.~G., {Walter}, F., {Heald}, G.~H.,
  {Cald{\'u}-Primo}, A., \& {Jarrett}, T.~H. 2015, \aj, 150, 47

\bibitem[{{Iffrig} \& {Hennebelle}(2015)}]{Iffrig+Hennebelle15SN}
{Iffrig}, O., \& {Hennebelle}, P. 2015, \aap, 576, A95

\bibitem[{{Jackson} {et~al.}(2006){Jackson}, {Rathborne}, {Shah}, {Simon},
  {Bania}, {Clemens}, {Chambers}, {Johnson}, {Dormody}, {Lavoie}, \&
  {Heyer}}]{Jackson+06}
{Jackson}, J.~M., {Rathborne}, J.~M., {Shah}, R.~Y., {Simon}, R., {Bania},
  T.~M., {Clemens}, D.~P., {Chambers}, E.~T., {Johnson}, A.~M., {Dormody}, M.,
  {Lavoie}, R., \& {Heyer}, M.~H. 2006, \apjs, 163, 145

\bibitem[{{Joung} \& {Mac Low}(2006)}]{Joung+MacLow06sn}
{Joung}, M.~K.~R., \& {Mac Low}, M.-M. 2006, \apj, 653, 1266

\bibitem[{{Joung} {et~al.}(2009){Joung}, {Mac Low}, \& {Bryan}}]{Joung+09}
{Joung}, M.~R., {Mac Low}, M.-M., \& {Bryan}, G.~L. 2009, \apj, 704, 137

\bibitem[{{Kifonidis} {et~al.}(2006){Kifonidis}, {Plewa}, {Scheck}, {Janka}, \&
  {M{\"u}ller}}]{Kifonidis+06}
{Kifonidis}, K., {Plewa}, T., {Scheck}, L., {Janka}, H.-T., \& {M{\"u}ller}, E.
  2006, \aap, 453, 661

\bibitem[{{Kim} \& {Ostriker}(2015)}]{Kim+Ostriker15SN}
{Kim}, C.-G., \& {Ostriker}, E.~C. 2015, \apj, 802, 99

\bibitem[{Kolmogorov(1941)}]{Kolmogorov41}
Kolmogorov, A.~N. 1941, Dokl. Akad. Nauk. SSSR, 30, 301

\bibitem[{{Kowal} \& {Lazarian}(2007)}]{Koval+Lazarian07}
{Kowal}, G., \& {Lazarian}, A. 2007, \apjl, 666, L69

\bibitem[{{Kritsuk} {et~al.}(2013){Kritsuk}, {Lee}, \& {Norman}}]{Kritsuk+13}
{Kritsuk}, A.~G., {Lee}, C.~T., \& {Norman}, M.~L. 2013, \mnras, 436, 3247

\bibitem[{{Kritsuk} {et~al.}(2007){Kritsuk}, {Norman}, {Padoan}, \&
  {Wagner}}]{Kritsuk+07}
{Kritsuk}, A.~G., {Norman}, M.~L., {Padoan}, P., \& {Wagner}, R. 2007, \apj,
  665, 416

\bibitem[{{Kritsuk} {et~al.}(2010){Kritsuk}, {Ustyugov}, {Norman}, \&
  {Padoan}}]{Kritsuk+10}
{Kritsuk}, A.~G., {Ustyugov}, S.~D., {Norman}, M.~L., \& {Padoan}, P. 2010, in
  Astronomical Society of the Pacific Conference Series, Vol. 429, Numerical
  Modeling of Space Plasma Flows, Astronum-2009, ed. N.~V. {Pogorelov},
  E.~{Audit}, \& G.~P. {Zank}, 15

\bibitem[{{Krumholz} \& {McKee}(2005)}]{Krumholz+McKee05sfr}
{Krumholz}, M.~R., \& {McKee}, C.~F. 2005, \apj, 630, 250

\bibitem[{{Larson}(1981)}]{Larson81}
{Larson}, R.~B. 1981, MNRAS, 194, 809

\bibitem[{{Lazarian} \& {Pogosyan}(2000)}]{Lazarian+Pogosyan2000}
{Lazarian}, A., \& {Pogosyan}, D. 2000, \apj, 537, 720

\bibitem[{{Lehnert} {et~al.}(2013){Lehnert}, {Le Tiran}, {Nesvadba}, {van
  Driel}, {Boulanger}, \& {Di Matteo}}]{Lehnert+13}
{Lehnert}, M.~D., {Le Tiran}, L., {Nesvadba}, N.~P.~H., {van Driel}, W.,
  {Boulanger}, F., \& {Di Matteo}, P. 2013, \aap, 555, A72

\bibitem[{{Lunttila} {et~al.}(2008){Lunttila}, {Padoan}, {Juvela}, \&
  {Nordlund}}]{Lunttila+2008}
{Lunttila}, T., {Padoan}, P., {Juvela}, M., \& {Nordlund}, {\AA}. 2008, \apjl,
  686, L91

\bibitem[{{Lunttila} {et~al.}(2009){Lunttila}, {Padoan}, {Juvela}, \&
  {Nordlund}}]{Lunttila+2009}
---. 2009, \apjl, 702, L37

\bibitem[{{Martizzi} {et~al.}(2015){Martizzi}, {Faucher-Gigu{\`e}re}, \&
  {Quataert}}]{Martizzi+15SN}
{Martizzi}, D., {Faucher-Gigu{\`e}re}, C.-A., \& {Quataert}, E. 2015, \mnras,
  450, 504

\bibitem[{{Molina} {et~al.}(2012){Molina}, {Glover}, {Federrath}, \&
  {Klessen}}]{Molina+12}
{Molina}, F.~Z., {Glover}, S.~C.~O., {Federrath}, C., \& {Klessen}, R.~S. 2012,
  \mnras, 423, 2680

\bibitem[{{Monin} \& {Iaglom}(1975)}]{Monin+Yaglom75}
{Monin}, A.~S., \& {Iaglom}, A.~M. 1975, {Statistical fluid mechanics:
  Mechanics of turbulence. Volume 2 /revised and enlarged edition/}

\bibitem[Neufeld et al.(1995)]{Neufeld+95} Neufeld, D.~A., Lepp,
S., \& Melnick, G.~J.\ 1995, \apjs, 100, 132

\bibitem[{{Nordlund} \& {Padoan}(2003)}]{Nordlund+Padoan03}
{Nordlund}, {\AA}., \& {Padoan}, P. 2003, in Lecture Notes in Physics, Berlin
  Springer Verlag, Vol. 614, Turbulence and Magnetic Fields in Astrophysics,
  ed. E.~{Falgarone} \& T.~{Passot}, 271--298

\bibitem[{{Ostriker} {et~al.}(2010){Ostriker}, {McKee}, \&
  {Leroy}}]{Ostriker+10sfr}
{Ostriker}, E.~C., {McKee}, C.~F., \& {Leroy}, A.~K. 2010, \apj, 721, 975

\bibitem[{{Ostriker} \& {Shetty}(2011)}]{Ostriker+Shetty11}
{Ostriker}, E.~C., \& {Shetty}, R. 2011, \apj, 731, 41

\bibitem[{{Padoan} {et~al.}(2012){Padoan}, {Haugb{\o}lle}, \&
  {Nordlund}}]{Padoan+12sfr}
{Padoan}, P., {Haugb{\o}lle}, T., \& {Nordlund}, {\AA}. 2012, \apjl, 759, L27

\bibitem[{{Padoan} {et~al.}(2014){Padoan}, {Haugb{\o}lle}, \&
  {Nordlund}}]{Padoan+14accr}
---. 2014, \apj, 797, 32

\bibitem[{{Padoan} {et~al.}(2004{\natexlab{a}}){Padoan}, {Jimenez}, {Juvela},
  \& {Nordlund}}]{Padoan+04power}
{Padoan}, P., {Jimenez}, R., {Juvela}, M., \& {Nordlund}, {\AA}.
  2004{\natexlab{a}}, \apjl, 604, L49

\bibitem[{{Padoan} {et~al.}(2004{\natexlab{b}}){Padoan}, {Jimenez}, {Nordlund},
  \& {Boldyrev}}]{Padoan+04prl}
{Padoan}, P., {Jimenez}, R., {Nordlund}, {\AA}., \& {Boldyrev}, S.
  2004{\natexlab{b}}, Physical Review Letters, 92, 191102

\bibitem[{{Padoan} {et~al.}(2001){Padoan}, {Juvela}, {Goodman}, \&
  {Nordlund}}]{Padoan+01cores}
{Padoan}, P., {Juvela}, M., {Goodman}, A.~A., \& {Nordlund}, {\AA}. 2001, \apj,
  553, 227

\bibitem[{{Padoan} {et~al.}(2006){Padoan}, {Juvela}, {Kritsuk}, \&
  {Norman}}]{Padoan+06perseus}
{Padoan}, P., {Juvela}, M., {Kritsuk}, A., \& {Norman}, M.~L. 2006, \apjl, 653,
  L125

\bibitem[{{Padoan} {et~al.}(2010){Padoan}, {Kritsuk}, {Lunttila}, {Juvela},
  {Nordlund}, {Norman}, \& {Ustyugov}}]{Padoan+10_Como}
{Padoan}, P., {Kritsuk}, A.~G., {Lunttila}, T., {Juvela}, M., {Nordlund}, A.,
  {Norman}, M.~L., \& {Ustyugov}, S.~D. 2010, in American Institute of Physics
  Conference Series, Vol. 1242, American Institute of Physics Conference
  Series, ed. G.~{Bertin}, F.~{de Luca}, G.~{Lodato}, R.~{Pozzoli}, \&
  M.~{Rom{\'e}}, 219--230

\bibitem[{Padoan \& Nordlund(1997)}]{Padoan+Nordlund97MHD}
Padoan, P., \& Nordlund, {\AA}. 1997, astro-ph/9706176

\bibitem[{Padoan \& Nordlund(1999)}]{Padoan+Nordlund99MHD}
---. 1999, ApJ, 526, 279

\bibitem[{{Padoan} \& {Nordlund}(2002)}]{Padoan+Nordlund02imf}
{Padoan}, P., \& {Nordlund}, {\AA}. 2002, \apj, 576, 870

\bibitem[{{Padoan} \& {Nordlund}(2011)}]{Padoan+Nordlund11sfr}
---. 2011, \apj, 730, 40

\bibitem[{{Padoan} {et~al.}(1997){Padoan}, {Nordlund}, \&
  {Jones}}]{Padoan+Nordlund97imf}
{Padoan}, P., {Nordlund}, A., \& {Jones}, B.~J.~T. 1997, \mnras, 288, 145

\bibitem[{{Pan} {et~al.}(2015){Pan}, {Padoan}, {Haugbolle}, \&
  {Nordlund}}]{Pan+15compress}
{Pan}, L., {Padoan}, P., {Haugbolle}, T., \& {Nordlund}, A. 2015, ArXiv
  e-prints

\bibitem[{{Pan} \& {Scannapieco}(2011)}]{Pan+Scannapieco11SL}
{Pan}, L., \& {Scannapieco}, E. 2011, \pre, 83, 045302

\bibitem[{{Porter} {et~al.}(1999){Porter}, {Pouquet}, {Sytine}, \&
  {Woodward}}]{Porter+99}
{Porter}, D., {Pouquet}, A., {Sytine}, I., \& {Woodward}, P. 1999, Physica A
  Statistical Mechanics and its Applications, 263, 263

\bibitem[{{Porter} {et~al.}(2002){Porter}, {Pouquet}, \&
  {Woodward}}]{Porter+02}
{Porter}, D., {Pouquet}, A., \& {Woodward}, P. 2002, \pre, 66, 026301

\bibitem[{{Porter} {et~al.}(2015){Porter}, {Jones}, \&
  {Ryu}}]{Porter+15_ICM_compressive_forcing}
{Porter}, D.~H., {Jones}, T.~W., \& {Ryu}, D. 2015, ArXiv e-prints

\bibitem[{{Porter} {et~al.}(1998){Porter}, {Woodward}, \&
  {Pouquet}}]{Porter+98}
{Porter}, D.~H., {Woodward}, P.~R., \& {Pouquet}, A. 1998, Physics of Fluids,
  10, 237

\bibitem[{{Rathborne} {et~al.}(2009){Rathborne}, {Johnson}, {Jackson}, {Shah},
  \& {Simon}}]{Rathborne+09}
{Rathborne}, J.~M., {Johnson}, A.~M., {Jackson}, J.~M., {Shah}, R.~Y., \&
  {Simon}, R. 2009, \apjs, 182, 131

\bibitem[{{Roman-Duval} {et~al.}(2009){Roman-Duval}, {Jackson}, {Heyer},
  {Johnson}, {Rathborne}, {Shah}, \& {Simon}}]{Roman-Duval+09}
{Roman-Duval}, J., {Jackson}, J.~M., {Heyer}, M., {Johnson}, A., {Rathborne},
  J., {Shah}, R., \& {Simon}, R. 2009, \apj, 699, 1153

\bibitem[{{Roman-Duval} {et~al.}(2010){Roman-Duval}, {Jackson}, {Heyer},
  {Rathborne}, \& {Simon}}]{Roman-Duval+10}
{Roman-Duval}, J., {Jackson}, J.~M., {Heyer}, M., {Rathborne}, J., \& {Simon},
  R. 2010, \apj, 723, 492

\bibitem[{{Salpeter}(1955)}]{Salpeter55}
{Salpeter}, E.~E. 1955, ApJ, 121, 161

\bibitem[{{S{\'a}nchez} {et~al.}(2007){S{\'a}nchez}, {Alfaro}, \&
  {P{\'e}rez}}]{Sanchez+07}
{S{\'a}nchez}, N., {Alfaro}, E.~J., \& {P{\'e}rez}, E. 2007, \apj, 656, 222

\bibitem[{Sanders {et~al.}(1985)Sanders, Scoville, \& Solomon}]{Sanders+85}
Sanders, D.~B., Scoville, N.~Z., \& Solomon, P.~M. 1985, ApJ, 289, 372

\bibitem[{{Schmidt} {et~al.}(2009){Schmidt}, {Federrath}, {Hupp}, {Kern}, \&
  {Niemeyer}}]{Schmidt+09compressive}
{Schmidt}, W., {Federrath}, C., {Hupp}, M., {Kern}, S., \& {Niemeyer}, J.~C.
  2009, \aap, 494, 127

\bibitem[{{Schmidt} {et~al.}(2008){Schmidt}, {Federrath}, \&
  {Klessen}}]{Schmidt+08prl}
{Schmidt}, W., {Federrath}, C., \& {Klessen}, R. 2008, Physical Review Letters,
  101, 194505

\bibitem[{Scoville {et~al.}(1987)Scoville, Yun, Clemens, Sanders, \&
  Waller}]{Scoville+87}
Scoville, N.~Z., Yun, M.~S., Clemens, D.~P., Sanders, D.~B., \& Waller, W.~H.
  1987, \apjs, 63, 821

\bibitem[{{Semenov} {et~al.}(2015){Semenov}, {Kravtsov}, \&
  {Gnedin}}]{Semenov+15}
{Semenov}, V.~A., {Kravtsov}, A.~V., \& {Gnedin}, N.~Y. 2015, ArXiv e-prints

\bibitem[{She \& L{\'e}v{\^e}que(1994)}]{She+Leveque94}
She, Z.-S., \& L{\'e}v{\^e}que, E. 1994, Phys. Rev. Lett., 72, 336

\bibitem[{Solomon {et~al.}(1987)Solomon, Rivolo, Barrett, \&
  Yahil}]{Solomon+87}
Solomon, P.~M., Rivolo, A.~R., Barrett, J.~W., \& Yahil, A.~M. 1987, ApJ, 319,
  730

\bibitem[{{Stilp} {et~al.}(2013){Stilp}, {Dalcanton}, {Warren}, {Skillman},
  {Ott}, \& {Koribalski}}]{Stilp+13_HI}
{Stilp}, A.~M., {Dalcanton}, J.~J., {Warren}, S.~R., {Skillman}, E., {Ott}, J.,
  \& {Koribalski}, B. 2013, \apj, 765, 136

\bibitem[{{Tamburro} {et~al.}(2009){Tamburro}, {Rix}, {Leroy}, {Low}, {Walter},
  {Kennicutt}, {Brinks}, \& {de Blok}}]{Tamburro+09}
{Tamburro}, D., {Rix}, H.-W., {Leroy}, A.~K., {Low}, M.-M.~M., {Walter}, F.,
  {Kennicutt}, R.~C., {Brinks}, E., \& {de Blok}, W.~J.~G. 2009, \aj, 137, 4424

\bibitem[{{Teyssier}(2002)}]{Teyssier02}
{Teyssier}, R. 2002, \aap, 385, 337

\bibitem[{{Walch} {et~al.}(2015){Walch}, {Girichidis}, {Naab}, {Gatto},
  {Glover}, {W{\"u}nsch}, {Klessen}, {Clark}, {Peters}, {Derigs}, \&
  {Baczynski}}]{Walch+15}
{Walch}, S., {Girichidis}, P., {Naab}, T., {Gatto}, A., {Glover}, S.~C.~O.,
  {W{\"u}nsch}, R., {Klessen}, R.~S., {Clark}, P.~C., {Peters}, T., {Derigs},
  D., \& {Baczynski}, C. 2015, \mnras, 454, 238

\bibitem[{{Walch} \& {Naab}(2015)}]{Walch+Naab15SN}
{Walch}, S., \& {Naab}, T. 2015, \mnras, 451, 2757

\bibitem[{{Walch} {et~al.}(2014){Walch}, {Girichidis}, {Naab}, {Gatto},
  {Glover}, {W{\"u}nsch}, {Klessen}, {Clark}, {Peters}, \&
  {Baczynski}}]{Walch+14}
{Walch}, S.~K., {Girichidis}, P., {Naab}, T., {Gatto}, A., {Glover}, S.~C.~O.,
  {W{\"u}nsch}, R., {Klessen}, R.~S., {Clark}, P.~C., {Peters}, T., \&
  {Baczynski}, C. 2014, ArXiv e-prints

\bibitem[{{Williams} \& {McKee}(1997)}]{Williams+McKee97}
{Williams}, J.~P., \& {McKee}, C.~F. 1997, \apj, 476, 166

\bibitem[{{Wolfire} {et~al.}(1995){Wolfire}, {Hollenbach}, {McKee}, {Tielens},
  \& {Bakes}}]{Wolfire+95}
{Wolfire}, M.~G., {Hollenbach}, D., {McKee}, C.~F., {Tielens}, A.~G.~G.~M., \&
  {Bakes}, E.~L.~O. 1995, \apj, 443, 152

\bibitem[{{Wongwathanarat} {et~al.}(2015){Wongwathanarat}, {M{\"u}ller}, \&
  {Janka}}]{Wongwathanarat+15}
{Wongwathanarat}, A., {M{\"u}ller}, E., \& {Janka}, H.-T. 2015, \aap, 577, A48

\end{thebibliography}
\end{document}